\documentclass[longauth]{aa}

\usepackage{graphicx}
\usepackage{txfonts}
\usepackage{longtable}
\usepackage{natbib}
\usepackage{pdflscape}
\usepackage{caption}
\usepackage{color}
\usepackage[switch]{lineno}
\usepackage{hyperref}
\usepackage{threeparttable}
\usepackage{dblfloatfix}    
\newcommand{\dgr}{\ensuremath{^\circ}}

\bibpunct{(}{)}{;}{a}{}{,} 
\begin{document}

\title{
Population study of Galactic supernova remnants at very high $\gamma$-ray energies with H.E.S.S.
}

\author{H.E.S.S. Collaboration
\and H.~Abdalla \inst{\ref{NWU}}
\and A.~Abramowski \inst{\ref{HH}}
\and F.~Aharonian \inst{\ref{MPIK},\ref{DIAS},\ref{NASRA}}
\and F.~Ait~Benkhali \inst{\ref{MPIK}}
\and E.O.~Ang\"uner \inst{\ref{IFJPAN}}
\and M.~Arakawa \inst{\ref{Rikkyo}}
\and M.~Arrieta \inst{\ref{LUTH}}
\and P.~Aubert \inst{\ref{Annecy}}
\and M.~Backes \inst{\ref{UNAM}}
\and A.~Balzer \inst{\ref{GRAPPA}}
\and M.~Barnard \inst{\ref{NWU}}
\and Y.~Becherini \inst{\ref{Linnaeus}}
\and J.~Becker~Tjus \inst{\ref{RUB}}
\and D.~Berge \inst{\ref{GRAPPAHE}}
\and S.~Bernhard \inst{\ref{LFUI}}
\and K.~Bernl\"ohr \inst{\ref{MPIK}}
\and R.~Blackwell \inst{\ref{Adelaide}}
\and M.~B\"ottcher \inst{\ref{NWU}}
\and C.~Boisson \inst{\ref{LUTH}}
\and J.~Bolmont \inst{\ref{LPNHE}}
\and S.~Bonnefoy \inst{\ref{DESY}}
\and P.~Bordas \inst{\ref{MPIK}}
\and J.~Bregeon \inst{\ref{LUPM}}
\and F.~Brun \inst{\ref{CENB}}
\and P.~Brun \inst{\ref{IRFU}}
\and M.~Bryan \inst{\ref{GRAPPA}}
\and M.~B\"{u}chele \inst{\ref{ECAP}}
\and T.~Bulik \inst{\ref{UWarsaw}}
\and M.~Capasso \inst{\ref{IAAT}}
\and S.~Caroff \inst{\ref{LLR}}
\and A.~Carosi \inst{\ref{Annecy}}
\and S.~Casanova \inst{\ref{IFJPAN},\ref{MPIK}}
\and M.~Cerruti \inst{\ref{LPNHE}}
\and N.~Chakraborty \inst{\ref{MPIK}}
\and R.C.G.~Chaves\protect\footnotemark[1] \inst{\ref{LUPM},\ref{CurieChaves}}
\and A.~Chen \inst{\ref{WITS}}
\and J.~Chevalier \inst{\ref{Annecy}}
\and S.~Colafrancesco \inst{\ref{WITS}}
\and B.~Condon \inst{\ref{CENB}}
\and J.~Conrad \inst{\ref{OKC},\ref{FellowConrad}}
\and I.D.~Davids \inst{\ref{UNAM}}
\and J.~Decock \inst{\ref{IRFU}}
\and C.~Deil \inst{\ref{MPIK}}
\and J.~Devin \inst{\ref{LUPM}}
\and P.~deWilt \inst{\ref{Adelaide}}
\and L.~Dirson \inst{\ref{HH}}
\and A.~Djannati-Ata\"i \inst{\ref{APC}}
\and A.~Donath \inst{\ref{MPIK}}
\and L.O'C.~Drury \inst{\ref{DIAS}}
\and K.~Dutson \inst{\ref{Leicester}}
\and J.~Dyks \inst{\ref{NCAC}}
\and T.~Edwards \inst{\ref{MPIK}}
\and K.~Egberts \inst{\ref{UP}}
\and G.~Emery \inst{\ref{LPNHE}}
\and J.-P.~Ernenwein \inst{\ref{CPPM}}
\and S.~Eschbach \inst{\ref{ECAP}}
\and C.~Farnier \inst{\ref{OKC},\ref{Linnaeus}}
\and S.~Fegan \inst{\ref{LLR}}
\and M.V.~Fernandes \inst{\ref{HH}}
\and D.~Fernandez\protect\footnotemark[1] \inst{\ref{LUPM}}
\and A.~Fiasson \inst{\ref{Annecy}}
\and G.~Fontaine \inst{\ref{LLR}}
\and S.~Funk \inst{\ref{ECAP}}
\and M.~F\"u{\ss}ling \inst{\ref{DESY}}
\and S.~Gabici \inst{\ref{APC}}
\and Y.A.~Gallant \inst{\ref{LUPM}}
\and T.~Garrigoux \inst{\ref{NWU}}
\and F.~Gat{\'e} \inst{\ref{Annecy}}
\and G.~Giavitto \inst{\ref{DESY}}
\and B.~Giebels \inst{\ref{LLR}}
\and D.~Glawion \inst{\ref{LSW}}
\and J.F.~Glicenstein \inst{\ref{IRFU}}
\and D.~Gottschall \inst{\ref{IAAT}}
\and M.-H.~Grondin \inst{\ref{CENB}}
\and J.~Hahn\protect\footnotemark[1] \inst{\ref{MPIK}}
\and M.~Haupt \inst{\ref{DESY}}
\and J.~Hawkes \inst{\ref{Adelaide}}
\and G.~Heinzelmann \inst{\ref{HH}}
\and G.~Henri \inst{\ref{Grenoble}}
\and G.~Hermann \inst{\ref{MPIK}}
\and J.A.~Hinton \inst{\ref{MPIK}}
\and W.~Hofmann \inst{\ref{MPIK}}
\and C.~Hoischen \inst{\ref{UP}}
\and T.~L.~Holch \inst{\ref{HUB}}
\and M.~Holler \inst{\ref{LFUI}}
\and D.~Horns \inst{\ref{HH}}
\and A.~Ivascenko \inst{\ref{NWU}}
\and H.~Iwasaki \inst{\ref{Rikkyo}}
\and A.~Jacholkowska \inst{\ref{LPNHE}}
\and M.~Jamrozy \inst{\ref{UJK}}
\and D.~Jankowsky \inst{\ref{ECAP}}
\and F.~Jankowsky \inst{\ref{LSW}}
\and M.~Jingo \inst{\ref{WITS}}
\and L.~Jouvin \inst{\ref{APC}}
\and I.~Jung-Richardt \inst{\ref{ECAP}}
\and M.A.~Kastendieck \inst{\ref{HH}}
\and K.~Katarzy{\'n}ski \inst{\ref{NCUT}}
\and M.~Katsuragawa \inst{\ref{JAXA}}
\and U.~Katz \inst{\ref{ECAP}}
\and D.~Kerszberg \inst{\ref{LPNHE}}
\and D.~Khangulyan \inst{\ref{Rikkyo}}
\and B.~Kh\'elifi \inst{\ref{APC}}
\and J.~King \inst{\ref{MPIK}}
\and S.~Klepser \inst{\ref{DESY}}
\and D.~Klochkov \inst{\ref{IAAT}}
\and W.~Klu\'{z}niak \inst{\ref{NCAC}}
\and Nu.~Komin \inst{\ref{WITS}}
\and K.~Kosack \inst{\ref{IRFU}}
\and S.~Krakau \inst{\ref{RUB}}
\and M.~Kraus \inst{\ref{ECAP}}
\and P.P.~Kr\"uger \inst{\ref{NWU}}
\and H.~Laffon \inst{\ref{CENB}}
\and G.~Lamanna \inst{\ref{Annecy}}
\and J.~Lau \inst{\ref{Adelaide}}
\and J.-P.~Lees \inst{\ref{Annecy}}
\and J.~Lefaucheur \inst{\ref{LUTH}}
\and A.~Lemi\`ere \inst{\ref{APC}}
\and M.~Lemoine-Goumard \inst{\ref{CENB}}
\and J.-P.~Lenain \inst{\ref{LPNHE}}
\and E.~Leser \inst{\ref{UP}}
\and T.~Lohse \inst{\ref{HUB}}
\and M.~Lorentz \inst{\ref{IRFU}}
\and R.~Liu \inst{\ref{MPIK}}
\and R.~L\'opez-Coto \inst{\ref{MPIK}}
\and I.~Lypova \inst{\ref{DESY}}
\and D.~Malyshev \inst{\ref{IAAT}}
\and V.~Marandon\protect\footnotemark[1] \inst{\ref{MPIK}}
\and A.~Marcowith \inst{\ref{LUPM}}
\and C.~Mariaud \inst{\ref{LLR}}
\and R.~Marx \inst{\ref{MPIK}}
\and G.~Maurin \inst{\ref{Annecy}}
\and N.~Maxted \inst{\ref{Adelaide},\ref{MaxtedNowAt}}
\and M.~Mayer \inst{\ref{HUB}}
\and P.J.~Meintjes \inst{\ref{UFS}}
\and M.~Meyer \inst{\ref{OKC},\ref{MeyerNowAt}}
\and A.M.W.~Mitchell \inst{\ref{MPIK}}
\and R.~Moderski \inst{\ref{NCAC}}
\and M.~Mohamed \inst{\ref{LSW}}
\and L.~Mohrmann \inst{\ref{ECAP}}
\and K.~Mor{\aa} \inst{27}
\and E.~Moulin \inst{\ref{IRFU}}
\and T.~Murach \inst{\ref{DESY}}
\and S.~Nakashima  \inst{\ref{JAXA}}
\and M.~de~Naurois \inst{\ref{LLR}}
\and H.~Ndiyavala  \inst{\ref{NWU}}
\and F.~Niederwanger \inst{\ref{LFUI}}
\and J.~Niemiec \inst{\ref{IFJPAN}}
\and L.~Oakes \inst{\ref{HUB}}
\and P.~O'Brien \inst{\ref{Leicester}}
\and H.~Odaka \inst{\ref{JAXA}}
\and S.~Ohm \inst{\ref{DESY}}
\and M.~Ostrowski \inst{\ref{UJK}}
\and I.~Oya \inst{\ref{DESY}}
\and M.~Padovani \inst{\ref{LUPM}}
\and M.~Panter \inst{\ref{MPIK}}
\and R.D.~Parsons \inst{\ref{MPIK}}
\and N.W.~Pekeur \inst{\ref{NWU}}
\and G.~Pelletier \inst{\ref{Grenoble}}
\and C.~Perennes \inst{\ref{LPNHE}}
\and P.-O.~Petrucci \inst{\ref{Grenoble}}
\and B.~Peyaud \inst{\ref{IRFU}}
\and Q.~Piel \inst{\ref{Annecy}}
\and S.~Pita \inst{\ref{APC}}
\and V.~Poireau \inst{\ref{Annecy}}
\and H.~Poon \inst{\ref{MPIK}}
\and D.~Prokhorov \inst{\ref{Linnaeus}}
\and H.~Prokoph \inst{\ref{GRAPPAHE}}
\and G.~P\"uhlhofer \inst{\ref{IAAT}}
\and M.~Punch \inst{\ref{APC},\ref{Linnaeus}}
\and A.~Quirrenbach \inst{\ref{LSW}}
\and S.~Raab \inst{\ref{ECAP}}
\and R.~Rauth \inst{\ref{LFUI}}
\and A.~Reimer \inst{\ref{LFUI}}
\and O.~Reimer \inst{\ref{LFUI}}
\and M.~Renaud\protect\footnotemark[1] \inst{\ref{LUPM}}
\and R.~de~los~Reyes \inst{\ref{MPIK}}
\and F.~Rieger \inst{\ref{MPIK},\ref{FellowRieger}}
\and L.~Rinchiuso \inst{\ref{IRFU}}
\and C.~Romoli \inst{\ref{DIAS}}
\and G.~Rowell \inst{\ref{Adelaide}}
\and B.~Rudak \inst{\ref{NCAC}}
\and C.B.~Rulten \inst{\ref{LUTH}}
\and S.~Safi-Harb \inst{\ref{Winipeg}}
\and V.~Sahakian \inst{\ref{YPI},\ref{NASRA}}
\and S.~Saito \inst{\ref{Rikkyo}}
\and D.A.~Sanchez \inst{\ref{Annecy}}
\and A.~Santangelo \inst{\ref{IAAT}}
\and M.~Sasaki \inst{\ref{ECAP}}
\and R.~Schlickeiser \inst{\ref{RUB}}
\and F.~Sch\"ussler \inst{\ref{IRFU}}
\and A.~Schulz \inst{\ref{DESY}}
\and U.~Schwanke \inst{\ref{HUB}}
\and S.~Schwemmer \inst{\ref{LSW}}
\and M.~Seglar-Arroyo \inst{\ref{IRFU}}
\and M.~Settimo \inst{\ref{LPNHE}}
\and A.S.~Seyffert \inst{\ref{NWU}}
\and N.~Shafi \inst{\ref{WITS}}
\and I.~Shilon \inst{\ref{ECAP}}
\and K.~Shiningayamwe \inst{\ref{UNAM}}
\and R.~Simoni \inst{\ref{GRAPPA}}
\and H.~Sol \inst{\ref{LUTH}}
\and F.~Spanier \inst{\ref{NWU}}
\and M.~Spir-Jacob \inst{\ref{APC}}
\and {\L.}~Stawarz \inst{\ref{UJK}}
\and R.~Steenkamp \inst{\ref{UNAM}}
\and C.~Stegmann \inst{\ref{UP},\ref{DESY}}
\and C.~Steppa \inst{\ref{UP}}
\and I.~Sushch \inst{\ref{NWU}}
\and T.~Takahashi  \inst{\ref{JAXA}}
\and J.-P.~Tavernet \inst{\ref{LPNHE}}
\and T.~Tavernier \inst{\ref{APC}}
\and A.M.~Taylor \inst{\ref{DESY}}
\and R.~Terrier \inst{\ref{APC}}
\and L.~Tibaldo \inst{\ref{MPIK}}
\and D.~Tiziani \inst{\ref{ECAP}}
\and M.~Tluczykont \inst{\ref{HH}}
\and C.~Trichard \inst{\ref{CPPM}}
\and M.~Tsirou \inst{\ref{LUPM}}
\and N.~Tsuji \inst{\ref{Rikkyo}}
\and R.~Tuffs \inst{\ref{MPIK}}
\and Y.~Uchiyama \inst{\ref{Rikkyo}}
\and D.J.~van~der~Walt \inst{\ref{NWU}}
\and C.~van~Eldik \inst{\ref{ECAP}}
\and C.~van~Rensburg \inst{\ref{NWU}}
\and B.~van~Soelen \inst{\ref{UFS}}
\and G.~Vasileiadis \inst{\ref{LUPM}}
\and J.~Veh \inst{\ref{ECAP}}
\and C.~Venter \inst{\ref{NWU}}
\and A.~Viana \inst{\ref{MPIK},\ref{VianaNowAt}}
\and P.~Vincent \inst{\ref{LPNHE}}
\and J.~Vink \inst{\ref{GRAPPA}}
\and F.~Voisin \inst{\ref{Adelaide}}
\and H.J.~V\"olk \inst{\ref{MPIK}}
\and T.~Vuillaume \inst{\ref{Annecy}}
\and Z.~Wadiasingh \inst{\ref{NWU}}
\and S.J.~Wagner \inst{\ref{LSW}}
\and P.~Wagner \inst{\ref{HUB}}
\and R.M.~Wagner \inst{\ref{OKC}}
\and R.~White \inst{\ref{MPIK}}
\and A.~Wierzcholska \inst{\ref{IFJPAN}}
\and P.~Willmann \inst{\ref{ECAP}}
\and A.~W\"ornlein \inst{\ref{ECAP}}
\and D.~Wouters \inst{\ref{IRFU}}
\and R.~Yang \inst{\ref{MPIK}}
\and D.~Zaborov \inst{\ref{LLR}}
\and M.~Zacharias \inst{\ref{NWU}}
\and R.~Zanin \inst{\ref{MPIK}}
\and A.A.~Zdziarski \inst{\ref{NCAC}}
\and A.~Zech \inst{\ref{LUTH}}
\and F.~Zefi \inst{\ref{LLR}}
\and A.~Ziegler \inst{\ref{ECAP}}
\and J.~Zorn \inst{\ref{MPIK}}
\and N.~\.Zywucka \inst{\ref{UJK}}
}

\institute{
Centre for Space Research, North-West University, Potchefstroom 2520, South Africa \label{NWU} \and 
Universit\"at Hamburg, Institut f\"ur Experimentalphysik, Luruper Chaussee 149, D 22761 Hamburg, Germany \label{HH} \and 
Max-Planck-Institut f\"ur Kernphysik, P.O. Box 103980, D 69029 Heidelberg, Germany \label{MPIK} \and 
Dublin Institute for Advanced Studies, 31 Fitzwilliam Place, Dublin 2, Ireland \label{DIAS} \and 
National Academy of Sciences of the Republic of Armenia,  Marshall Baghramian Avenue, 24, 0019 Yerevan, Republic of Armenia  \label{NASRA} \and
Yerevan Physics Institute, 2 Alikhanian Brothers St., 375036 Yerevan, Armenia \label{YPI} \and
Institut f\"ur Physik, Humboldt-Universit\"at zu Berlin, Newtonstr. 15, D 12489 Berlin, Germany \label{HUB} \and
University of Namibia, Department of Physics, Private Bag 13301, Windhoek, Namibia \label{UNAM} \and
GRAPPA, Anton Pannekoek Institute for Astronomy, University of Amsterdam,  Science Park 904, 1098 XH Amsterdam, The Netherlands \label{GRAPPA} \and
Department of Physics and Electrical Engineering, Linnaeus University,  351 95 V\"axj\"o, Sweden \label{Linnaeus} \and
Institut f\"ur Theoretische Physik, Lehrstuhl IV: Weltraum und Astrophysik, Ruhr-Universit\"at Bochum, D 44780 Bochum, Germany \label{RUB} \and
GRAPPA, Anton Pannekoek Institute for Astronomy and Institute of High-Energy Physics, University of Amsterdam,  Science Park 904, 1098 XH Amsterdam, The Netherlands \label{GRAPPAHE} \and
Institut f\"ur Astro- und Teilchenphysik, Leopold-Franzens-Universit\"at Innsbruck, A-6020 Innsbruck, Austria \label{LFUI} \and
School of Physical Sciences, University of Adelaide, Adelaide 5005, Australia \label{Adelaide} \and
LUTH, Observatoire de Paris, PSL Research University, CNRS, Universit\'e Paris Diderot, 5 Place Jules Janssen, 92190 Meudon, France \label{LUTH} \and
Sorbonne Universit\'es, UPMC Universit\'e Paris 06, Universit\'e Paris Diderot, Sorbonne Paris Cit\'e, CNRS, Laboratoire de Physique Nucl\'eaire et de Hautes Energies (LPNHE), 4 place Jussieu, F-75252, Paris Cedex 5, France \label{LPNHE} \and
Laboratoire Univers et Particules de Montpellier, Universit\'e Montpellier, CNRS/IN2P3,  CC 72, Place Eug\`ene Bataillon, F-34095 Montpellier Cedex 5, France \label{LUPM} \and
IRFU, CEA, Universit\'e Paris-Saclay, F-91191 Gif-sur-Yvette, France \label{IRFU} \and
Astronomical Observatory, The University of Warsaw, Al. Ujazdowskie 4, 00-478 Warsaw, Poland \label{UWarsaw} \and
Aix Marseille Universit\'e, CNRS/IN2P3, CPPM, Marseille, France \label{CPPM} \and
Instytut Fizyki J\c{a}drowej PAN, ul. Radzikowskiego 152, 31-342 Krak{\'o}w, Poland \label{IFJPAN} \and
Funded by EU FP7 Marie Curie, grant agreement No. PIEF-GA-2012-332350 \label{CurieChaves}  \and
School of Physics, University of the Witwatersrand, 1 Jan Smuts Avenue, Braamfontein, Johannesburg, 2050 South Africa \label{WITS} \and
Laboratoire d'Annecy-le-Vieux de Physique des Particules, Universit\'{e} Savoie Mont-Blanc, CNRS/IN2P3, F-74941 Annecy-le-Vieux, France \label{Annecy} \and
Landessternwarte, Universit\"at Heidelberg, K\"onigstuhl, D 69117 Heidelberg, Germany \label{LSW} \and
Universit\'e Bordeaux, CNRS/IN2P3, Centre d'\'Etudes Nucl\'eaires de Bordeaux Gradignan, 33175 Gradignan, France \label{CENB} \and
Oskar Klein Centre, Department of Physics, Stockholm University, Albanova University Center, SE-10691 Stockholm, Sweden \label{OKC} \and
Wallenberg Academy Fellow \label{FellowConrad}  \and
Institut f\"ur Astronomie und Astrophysik, Universit\"at T\"ubingen, Sand 1, D 72076 T\"ubingen, Germany \label{IAAT} \and
Laboratoire Leprince-Ringuet, Ecole Polytechnique, CNRS/IN2P3, F-91128 Palaiseau, France \label{LLR} \and
APC, AstroParticule et Cosmologie, Universit\'{e} Paris Diderot, CNRS/IN2P3, CEA/Irfu, Observatoire de Paris, Sorbonne Paris Cit\'{e}, 10, rue Alice Domon et L\'{e}onie Duquet, 75205 Paris Cedex 13, France \label{APC} \and
Univ. Grenoble Alpes, CNRS, IPAG, F-38000 Grenoble, France \label{Grenoble} \and
Department of Physics and Astronomy, The University of Leicester, University Road, Leicester, LE1 7RH, United Kingdom \label{Leicester} \and
Nicolaus Copernicus Astronomical Center, Polish Academy of Sciences, ul. Bartycka 18, 00-716 Warsaw, Poland \label{NCAC} \and
Institut f\"ur Physik und Astronomie, Universit\"at Potsdam,  Karl-Liebknecht-Strasse 24/25, D 14476 Potsdam, Germany \label{UP} \and
Friedrich-Alexander-Universit\"at Erlangen-N\"urnberg, Erlangen Centre for Astroparticle Physics, Erwin-Rommel-Str. 1, D 91058 Erlangen, Germany \label{ECAP} \and
DESY, D-15738 Zeuthen, Germany \label{DESY} \and
Obserwatorium Astronomiczne, Uniwersytet Jagiello{\'n}ski, ul. Orla 171, 30-244 Krak{\'o}w, Poland \label{UJK} \and
Centre for Astronomy, Faculty of Physics, Astronomy and Informatics, Nicolaus Copernicus University,  Grudziadzka 5, 87-100 Torun, Poland \label{NCUT} \and
Department of Physics, University of the Free State,  PO Box 339, Bloemfontein 9300, South Africa \label{UFS} \and
Heisenberg Fellow (DFG), ITA Universit\"at Heidelberg, Germany \label{FellowRieger} \and
Department of Physics, Rikkyo University, 3-34-1 Nishi-Ikebukuro, Toshima-ku, Tokyo 171-8501, Japan \label{Rikkyo} \and
Japan Aerpspace Exploration Agency (JAXA), Institute of Space and Astronautical Science (ISAS), 3-1-1 Yoshinodai, Chuo-ku, Sagamihara, Kanagawa 229-8510,  Japan \label{JAXA} \and
Department of Physics \& Astronomy, University of Manitoba, Winnipeg, MB R3T 2N2, Canada \label{Winipeg} \and
Now at The School of Physics, The University of New South Wales, Sydney, 2052, Australia \label{MaxtedNowAt} \and
Now at Instituto de F\'{i}sica de S\~{a}o Carlos, Universidade de S\~{a}o Paulo, Av. Trabalhador S\~{a}o-carlense, 400 - CEP 13566-590, S\~{a}o Carlos, SP, Brazil \label{VianaNowAt} \and
Now at Kavli Institute for Particle Astrophysics and Cosmology, Department of Physics and SLAC National Accelerator Laboratory, Stanford University, Stanford, California 94305, USA \label{MeyerNowAt}
}

\offprints{H.E.S.S.~collaboration,
\protect\\\email{\href{mailto:contact.hess@hess-experiment.eu}{contact.hess@hess-experiment.eu}};
\protect\\\protect\footnotemark[1] Corresponding authors
}

\titlerunning{Population study of Galactic supernova remnants at very high $\gamma$-ray energies with H.E.S.S.}
\authorrunning{H.E.S.S. Collaboration}

\date{}
\abstract {
Shell-type supernova remnants (SNRs) are considered prime candidates for the acceleration of Galactic cosmic rays (CRs) up to the knee of the CR spectrum at $\mathrm{E} \approx \mathrm{3}\times \mathrm{10}^\mathrm{15}$ eV. Our Milky Way galaxy hosts more than 350 SNRs discovered at radio wavelengths and at high energies, of which 220 fall into the H.E.S.S. Galactic Plane Survey (HGPS) region. Of those, only 50 SNRs are coincident with a H.E.S.S source and in 8 cases the very high-energy (VHE) emission is firmly identified as an SNR. 
The H.E.S.S. GPS provides us with a legacy for SNR population study in VHE $\gamma$-rays and we use this rich data set to extract VHE flux upper limits from all undetected SNRs.
Overall, the derived flux upper limits are not in contradiction with the canonical CR paradigm.
Assuming this paradigm holds true, we can constrain typical ambient density values around shell-type SNRs to $n\leq 7~\textrm{cm}^\textrm{-3}$ and electron-to-proton energy fractions above 10~TeV to $\epsilon_\textrm{ep} \leq 5\times 10^{-3}$.
Furthermore, comparisons of VHE with radio luminosities in non-interacting SNRs reveal a behaviour that is in agreement with the theory of magnetic field amplification at shell-type SNRs.
}


\maketitle

\section{Introduction}
Supernova remnants (SNRs) are considered the most promising candidates for the
origin of Galactic cosmic rays (CRs), a long-standing open problem in astroparticle
physics.  These objects are also a very prominent source class in high-energy
astrophysics, emitting non-thermal radiation in the form of radio waves, X-rays,
and $\gamma$-rays.  According to diffusive shock acceleration (DSA) theory, with
magnetic field amplification (see e.g.~\citet{MalkovDrury}) hadronic CR particles 
such as protons and heavier nuclei can be
accelerated up to PeV energies (the ``knee'' in the CR spectrum) at the
expanding SNR shock front or shell.  When these relativistic particles collide
with other nuclei, for example in the nearby interstellar medium (ISM), they emit
$\gamma$-rays in the very high-energy (VHE;
$0.1 \la E_{\gamma} \la 100$ TeV) band.  Thus, observations with Cherenkov
telescopes, sensitive to VHE $\gamma$-rays, provide a promising avenue to
investigate not only the astrophysics of energetic SNRs themselves but also
their putative connection to the origin of Galactic CRs.  In particular, certain aspects of
DSA theory, such as the efficiency of particle acceleration mechanisms, can be constrained
through such observations.

The High Energy Stereoscopic System (H.E.S.S.) is an array of five imaging
atmospheric Cherenkov telescopes (IACTs) situated in Namibia.  The telescope
array has a field of view of approximately 5\degr\ and can detect $\gamma$-rays above
an energy threshold of $\sim$\,50\,GeV. This array has an energy resolution of
$\sim$\,15\% and a angular resolution of $\sim$\,0.1\degr\ \footnote{The mean point spread function 68\% containment radius is 0.08\degr, see \citet{HGPSForth}.} \citep{Crabpaper}.
The H.E.S.S.\ Galactic Plane Survey \citep[HGPS;][]{HGPSForth} programme has led to the
detection of 78 sources of VHE $\gamma$-rays, of which 8 have been firmly
identified as emission from SNRs, typically by resolving shell-like
morphologies matching those observed at lower energies. These
are \object{RX J1713.7$-$3946} \citep{HGPSForth}, \object{RX J0852.0$-$4622}
\citep{hess_velajr_paper3}, \object{HESS J1731$-$347} \citep{J1731}, \object{RCW 86}
\citep{RCW86-New}, \object{W28} \citep{w28H}, \object{G349.7+0.2} \citep{G349}, \object{W49B} \citep{HGPSForth}, and \object{HESS J1534$-$571}
\citep{HGPSForth}.  One SNR outside of the HGPS region has also been firmly identified,
\object{SN 1006} \citep{SN1006}, bringing the tally to 9.  In addition,
H.E.S.S.\ has detected emission from 8 composite SNRs. For these latter types, it is currently
difficult to determine whether the $\gamma$-rays originate in the interior
pulsar wind nebula (PWN), the surrounding shell, or a combination of these two. In addition to the firm identifications, 16 additional HGPS sources
have also been associated with SNRs based on spatial coincidence, see 
\citet{HGPSForth}.

Using a $\sim$10-year HGPS data set collected between 2004 and 2013, 
we investigate the 
sample of known radio and X-ray SNRs that so far 
have not been detected by IACTs.
To that end, we select a subset of SNRs devoid of any unrelated
VHE emission (the VHE-dark sample, Sect.~\ref{sec-selec}) and derive flux upper limits (Sect.~\ref{sec-res}), 
which we use to test the standard paradigm of the SNRs as the origin of Galactic CRs. 
Assuming hadronic emission, we calculate a constraint on the fraction of the SNR explosion energy that is converted to CR protons in Sect.~\ref{sec-model}. 
We also apply a simple parametric estimate of the inverse-Compton (IC) emission to our results in order to probe the relevant
parameter space in a mixed scenario where both leptonic and hadronic channels contribute to the source emission. Based on our
formalism, we furthermore present expectations on the portion of this parameter space
that will be accessible to the future Cherenkov Telescope Array \citep[CTA;][]{CTA} observatory. 
Finally, in Sect.~\ref{sec-pop}, we compare
the derived flux upper limits to the radio flux densities that have been observed for the investigated source sample
and put these into context of the IACT detections. 

\section{Candidate source selection and data sample}
\label{sec-selec}

We obtained the sample of source candidates from the SNRcat catalogue\footnote{\url{http://www.physics.umanitoba.ca/snr/SNRcat/}\\ Version used here as of 12.14.2015.} \citep{SNRCAT}, which provides
an up-to-date catalogue of SNRs detected from radio to very high energies. The
catalogue comprises SNRs of different morphological
types: shell-type, composite, and filled-centre. In the context of this work, we 
treated only the shell-type and composite SNRs, as filled-centre SNRs
correspond to pulsar wind nebulae, which are discussed in \citet{HGPSForth}. Also, we ignored sources
of uncertain morphology (type `?' in SNRcat).

For our study we used the HGPS data set, which consists of $\sim$ 2700 hours of observations
and features a sensitivity of better than
$\sim 1.5\%$ of the Crab flux in the innermost Galactic regions.

\begin{figure}[h!!!]
  \begin{center}
  \resizebox{\hsize}{!}{\includegraphics[clip=]{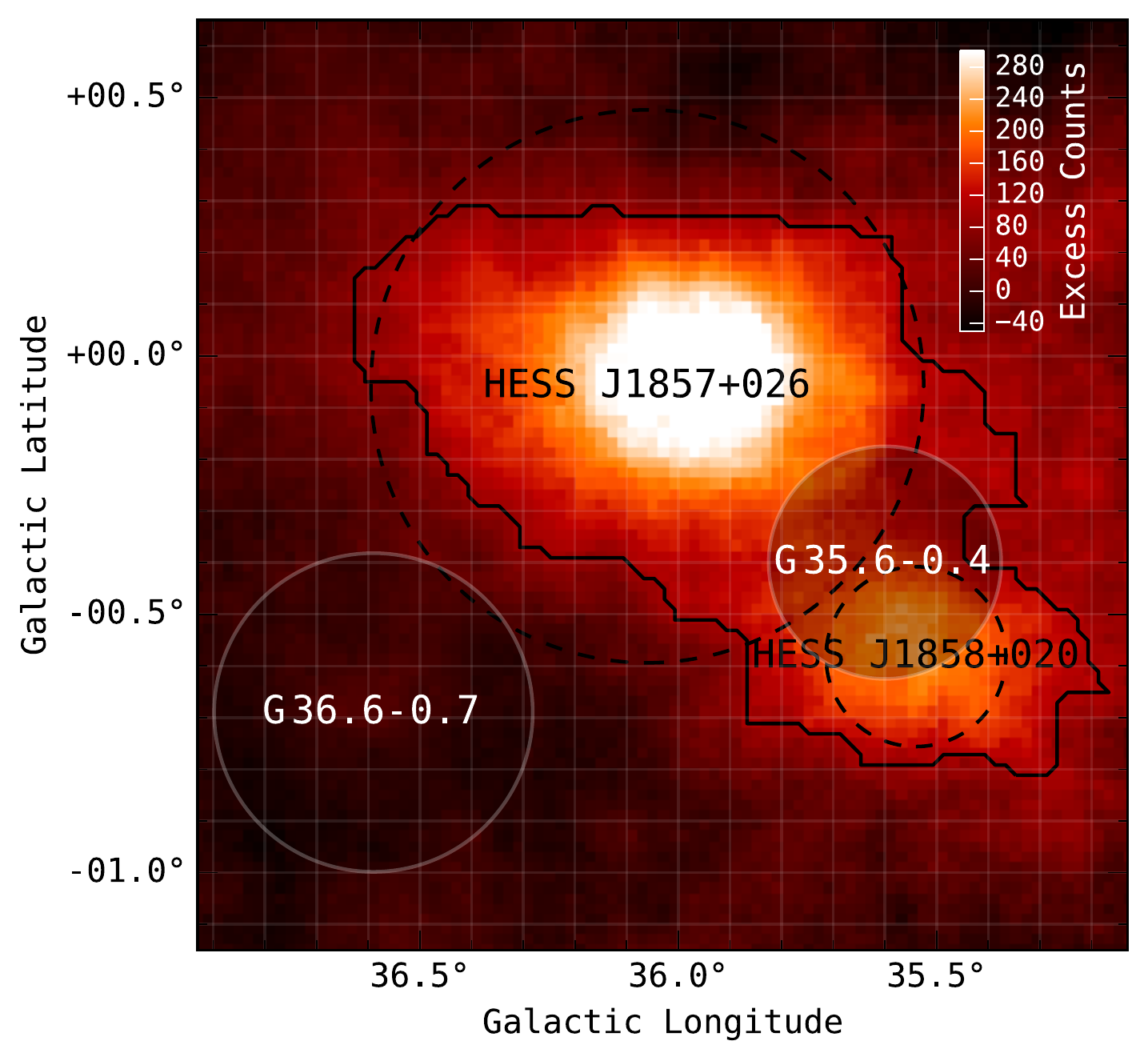}}
  \caption{Illustration of the source selection scheme on the $\gamma$-ray excess image from a given region. The known H.E.S.S. sources HESS~J1857+026 and HESS~J1858+020 with their 80\% flux containment radii are shown with a dashed circle. The de-selection region resulting from the algorithm explained in section \ref{sec-selec} is shown with a solid curve. Because of the overlap with the de-selection region, the SNR~G35.6$-$0.4 is discarded whereas SNR~G36.6$-$0.7 is selected.}
  \label{Figure:Deselection}
  \end{center}
\end{figure}
Of the more than 300 shell-type and composite SNRs listed in SNRcat, 220 fall within
the HGPS region (Galactic longitude from 65\dgr~to 250\dgr, latitude from -3.5\dgr to 3.5\dgr).
However, as this region is populated by almost 80 sources detected by H.E.S.S. (see \citet{HGPSForth}) signal contamination of the analysis regions of the
investigated SNRs is an important consideration. Any unrelated VHE emission in the analysis
regions weaken the derived flux upper limits and complicate
physical interpretation.

In order to obtain a sample of SNRs that does not suffer from this problem, we selected a VHE-dark sample 
from SNRcat consisting of sources that fall outside of regions
of VHE emission. We note that by focussing on SNRs that are not detected at very high energies we are biasing 
ourselves towards low VHE-flux objects. 
Assessing the impact of this bias and a possible correction would require 
population modelling and go beyond the main scope of this paper, which is to derive H.E.S.S. flux upper limits for VHE-undetected SNRs.
The selection method is as follows: We used the HGPS significance map (see \citealt{HGPSForth}) in an iterative way to identify VHE-bright
 regions around HGPS catalogue sources in the Galactic plane. 
More precisely, we chose the significance map with a 0.2$^\circ$ correlation radius, since this radius 
roughly corresponds to the maximum of the source radius distribution for shell-type and
 composite SNRs listed in SNRcat. As starting
points for this iteration, we used all bins\footnote{corresponding to a square of 0.02\dgr~$\times$~0.02\dgr~in Galactic coordinates} of the
significance map that both fall into the 80\% signal containment
radii of the HGPS sources (`R80' in the HGPS catalogue) and have significance values $\geq 4~\sigma$. 
We then saved their respective
neighbouring bins with significances $\geq 4~\sigma$ and used these as starting points for
the next iteration step. The iteration stops when there are no neighbouring bins 
with significances $\geq 4~\sigma$ around the starting points of a given step. 
This procedure results in sets of bins that define contiguous and VHE-bright regions,
in the following referred to as de-selection regions. By construction,
these regions overlap with the circular HGPS source regions, but are in most cases asymmetric
in shape (see Fig.~\ref{Figure:Deselection}).
Such regions are similar to the pre-defined HGPS exclusion regions \citep{HGPSForth}
but smaller in extent to allow for a less conservative compromise between
SNR sample size and signal leakage.

We tested the analysis region (see below) of any candidate object from SNRcat 
for overlap with one of the de-selection regions. If there was at
least one bin in the analysis region belonging to such a region, we discarded the
respective object from the VHE-dark SNR sample. In Fig. \ref{Figure:Deselection},
we illustrate the method.
This procedure results in a sample of 108 SNRs with H.E.S.S.
observations, of which 83 are of shell-type and 25 are of composite morphology.
The latter group includes eight sources with thermal and 15 sources with plerionic characteristics, and
two objects that feature both characteristics.

\section{Analysis}
\label{sec-ana}

In this study, we selected only data of high quality using the
criteria described in \citet{Crabpaper} and the quality cut on
atmospheric transparency conditions developed by
\citet{2014APh....54...25H}. The observation live time of the analysed regions, corrected by the
H.E.S.S. $\gamma$-ray acceptance, spans a range from $\sim$10~min to $\sim$80~h with a
median value of 14.5~h. The majority of data ($\sim$80\%) have been
recorded at average zenith angles smaller than 40\dgr.
Table~\ref{Table:Analysis} lists the analysed source sample and
the corrected observation live time, the averaged zenith angle for the observations
of each source, and the closest H.E.S.S. source.

We analysed the data using the multivariate analysis method described
in \citet{TMVA} with the same analysis configuration used in
\citet{HGPSForth}, TMVA Hard Cuts. For the
background estimation, the Reflected Background method
 \citep{ref:bgmodelling}, was applied. This method is
largely insensitive to acceptance gradients in the cameras and
therefore ideally suited for spectral analysis.  A cross-check was performed using the Model++ Faint Cuts analysis
\citep{Modpp} and yielding compatible results.

The analysis region for every source is given by its position
provided in SNRcat as well as the quoted radii therein and is defined
to be circular. If an SNR is reported with an elliptical shape, we
used the semi-major axis. A margin of $0.1^\circ$ is added to this radius, which
conservatively takes the H.E.S.S. point spread function into account.

For each source, we used all events above the safe energy threshold
in the analysis. The safe energy is defined as the energy above which the energy bias 
is less than 10\% \citep{Crabpaper}. 

We note that the diffuse emission measured in
\citet{Diffuse} was not taken into account in this analysis. This
would result in conservative upper limits, especially for sources close to
the Galactic plane. An attempt to quantify the effect of this
component is described in Sect.~\ref{sec-data} using the maps and
large-scale emission model from \citet{HGPSForth}.

\section{Results}
\label{sec-res}

In Table~\ref{Table:Analysis} we list the significance and upper-limit results
for the individual sources.  We calculate the significance 
using the method proposed by \citet{LiMa}. To obtain the upper limits
on the excess counts above the safe energy, we use the 
profile likelihood method as described in \citet{Rolke} and assumed a confidence 
level of 99\%. We then
express this result as an upper limit on the integrated flux in
the $(1,10)$ TeV interval assuming a power-law source of index 2.3. Such a
value is typical for Galactic sources detected in the VHE range
\citep{HGPSForth}.

\begin{figure}[h!!!]
  \begin{center}
  \resizebox{\hsize}{!}{\includegraphics[clip=]{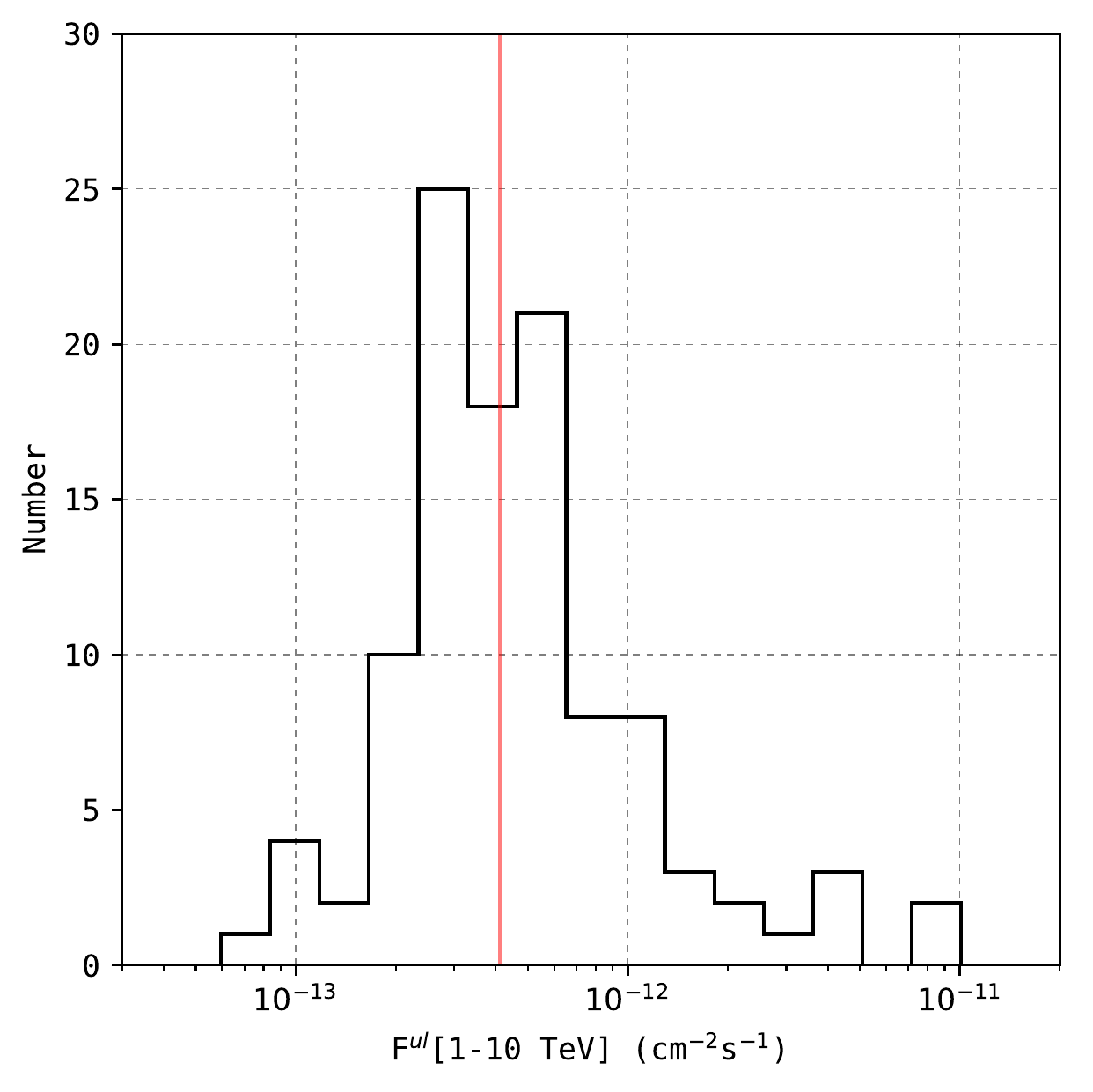}}
  \caption{Distribution of flux upper limits (99\% confidence level) of all investigated SNRs. The red line indicates
          the median value of $\sim$~2\% of the Crab nebula flux.}
  \label{Figure:ULDist}
  \end{center}
\end{figure}
\begin{figure*}
  \begin{center}
  \resizebox{\hsize}{!}{
    \includegraphics[clip=]{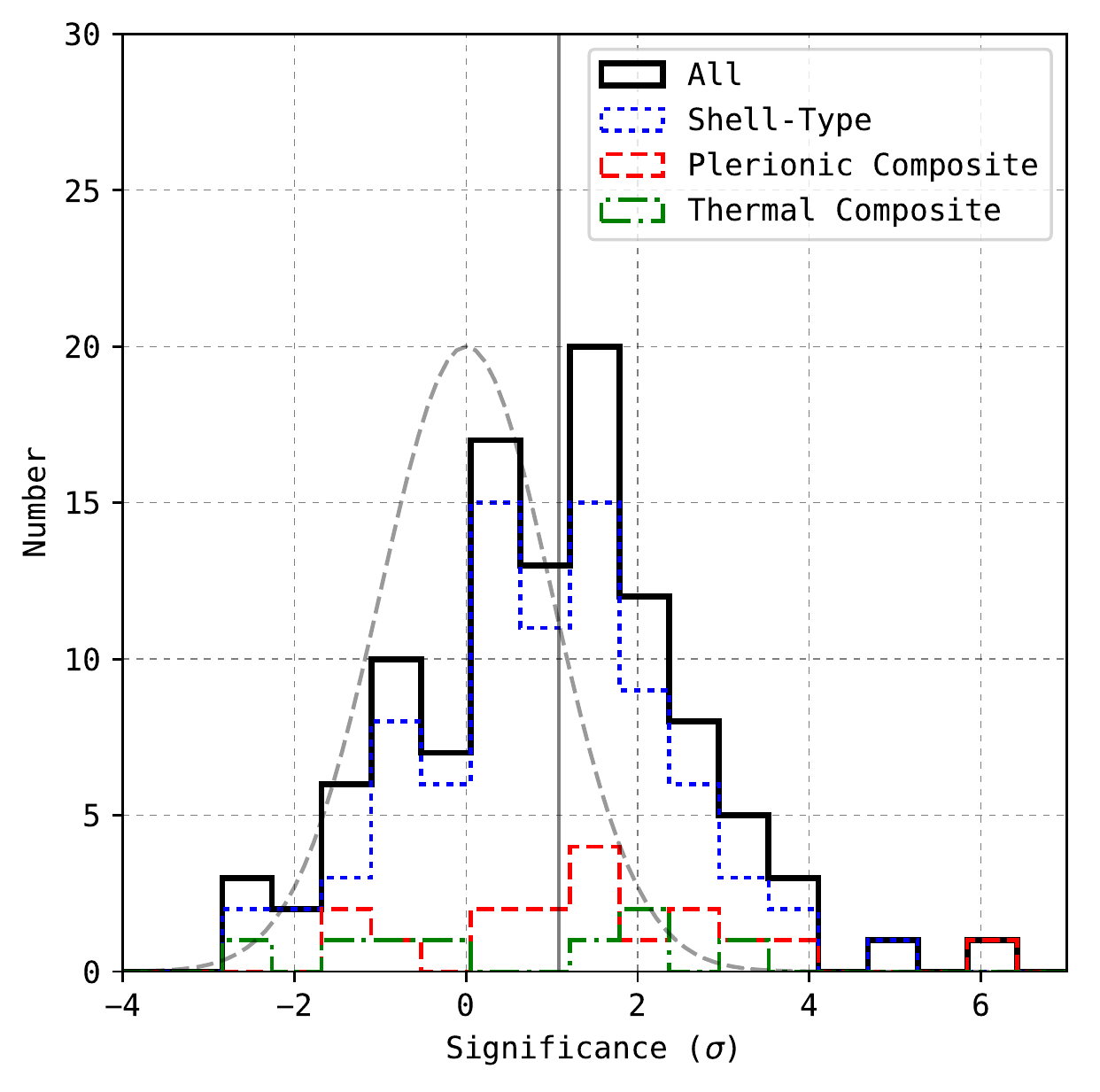}
    \includegraphics[clip=]{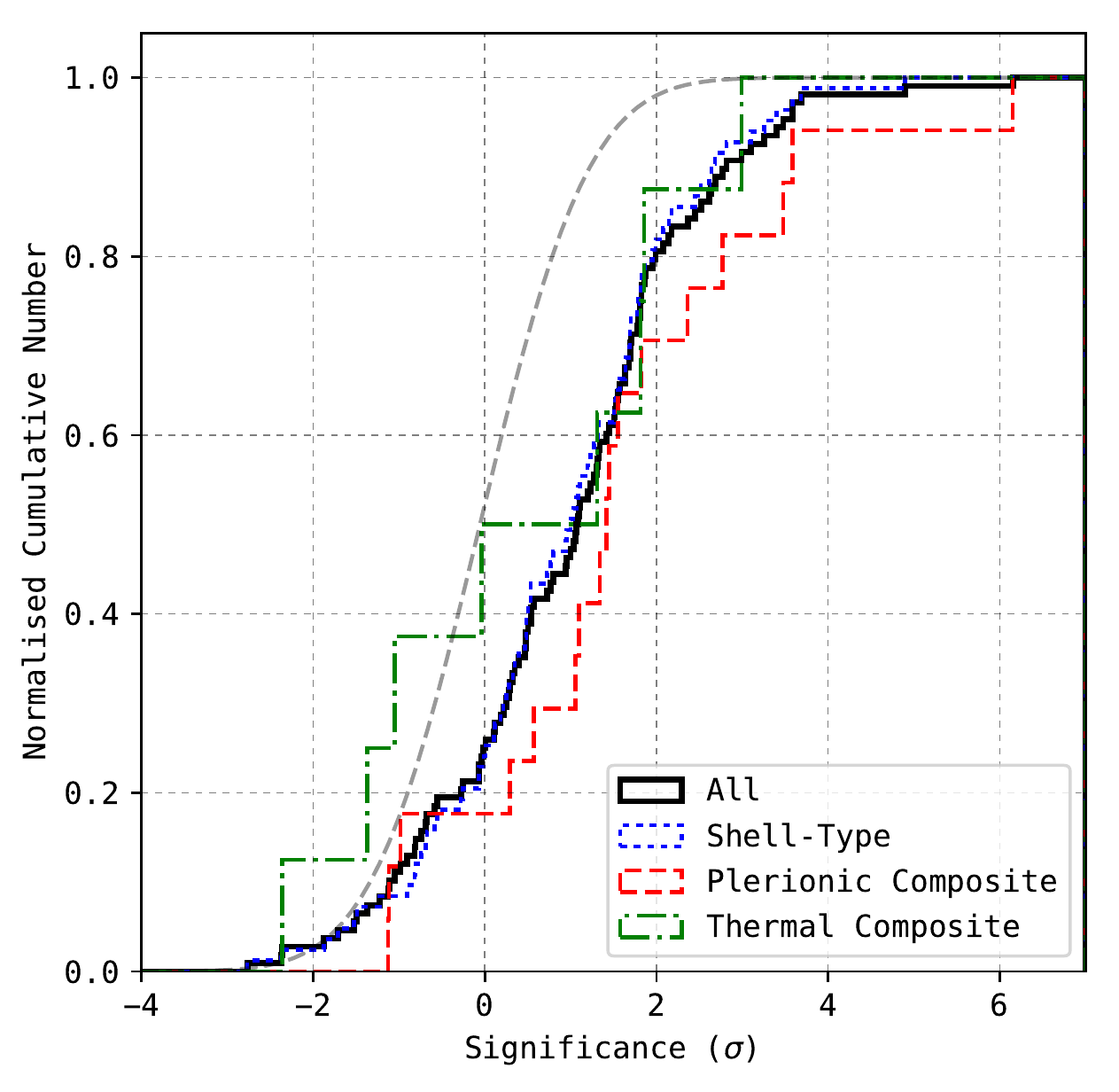}}
  \caption{Left: Significance distribution of the VHE-dark sample of
    SNRs (black, blue, red, and green) and the corresponding median value (grey solid). 
    The grey dashed curve indicates a
    normal Gaussian. Right:
    Cumulative significance distribution of all sources (black),
    those of shell-type (blue), and composite (red and green)
    morphology. The grey dashed line represents a cumulative normal Gaussian.}
  \label{Figure:SigDist}
  \end{center}
\end{figure*}

Fig.~\ref{Figure:ULDist} shows the distribution of upper limits for
the VHE-dark source sample. The total distribution is peaked at the
typical H.E.S.S. sensitivity in the HGPS region of $\sim$2\% of the
Crab nebula flux. The median and variance values of the distribution
of logarithmic flux upper limits are
$M(\mathrm{log}_{10}(F^{ul}/(cm^{-2}s^{-1})) = -12.4$ and $S^2(\mathrm{log}_{10}(F^{ul}/(cm^{-2}s^{-1})) = 0.14$,
respectively.

The significance distribution features median and variance values of 
$M(\sigma) = 1.1$ and $S^2(\sigma) = 2.4$, respectively.
From Table~\ref{Table:Analysis}, one can notice that the $\gamma$-ray excess from the
plerionic composite SNR~G34.7$-$0.4 (W44) shows a significance of
6.2~$\sigma$. However, we are prevented from claiming a detection because this object is a rather large SNR
embedded in a region of high diffuse emission that is not claimed
as a source in the HGPS catalogue (see \citet{HGPSForth}) and
because the signal is below the detection criteria of the
independent cross-check analysis.

We show the significance distribution in the left panel of
Fig. \ref{Figure:SigDist} together with a normal Gaussian, 
corresponding to the expectation in
the absence of any source signal. There is no significant difference in
the shape of the significance distribution with respect to the source
type, as can be seen in the cumulative distributions shown in the right
panel of Fig.~\ref{Figure:SigDist}. All pair-wise comparisons of the various
significance distributions with two-sided Kolmogorov-Smirnov tests result 
in $p$-values larger than 0.42. Therefore we see no indication that the various
samples stem from incompatible underlying distributions.
In particular, we find no indication that the possible additional presence
of a PWN in plerionic composites on average results in higher
significance values in the VHE range.

\section{Discussion}
\label{sec-data}
\subsection{Significance offset}

As shown in Fig. \ref{Figure:SigDist}, the median value of the measured significance
distribution is offset from the normal Gaussian distribution, 
which constitutes the expectation for pure noise (null hypothesis), and a Gaussian
fit to the distribution results in best-fit values for mean and standard deviation of 1.01 and
1.51, respectively. 
A Kolmogorov-Smirnov test of the measured significance distribution against the 
null hypothesis rejects the latter with a p-value of $p=8\times 10^\mathrm{-15}$. 
The origin of this collective excess in positive significance values may arise from several
contributions:
\begin{enumerate}
  \item Low-level signal leakage into the analysis
regions from known VHE $\gamma$-ray sources cannot be completely dismissed.

\begin{figure}[h!!!]
  \begin{center}
  \resizebox{0.9\hsize}{!}{\includegraphics[clip=]{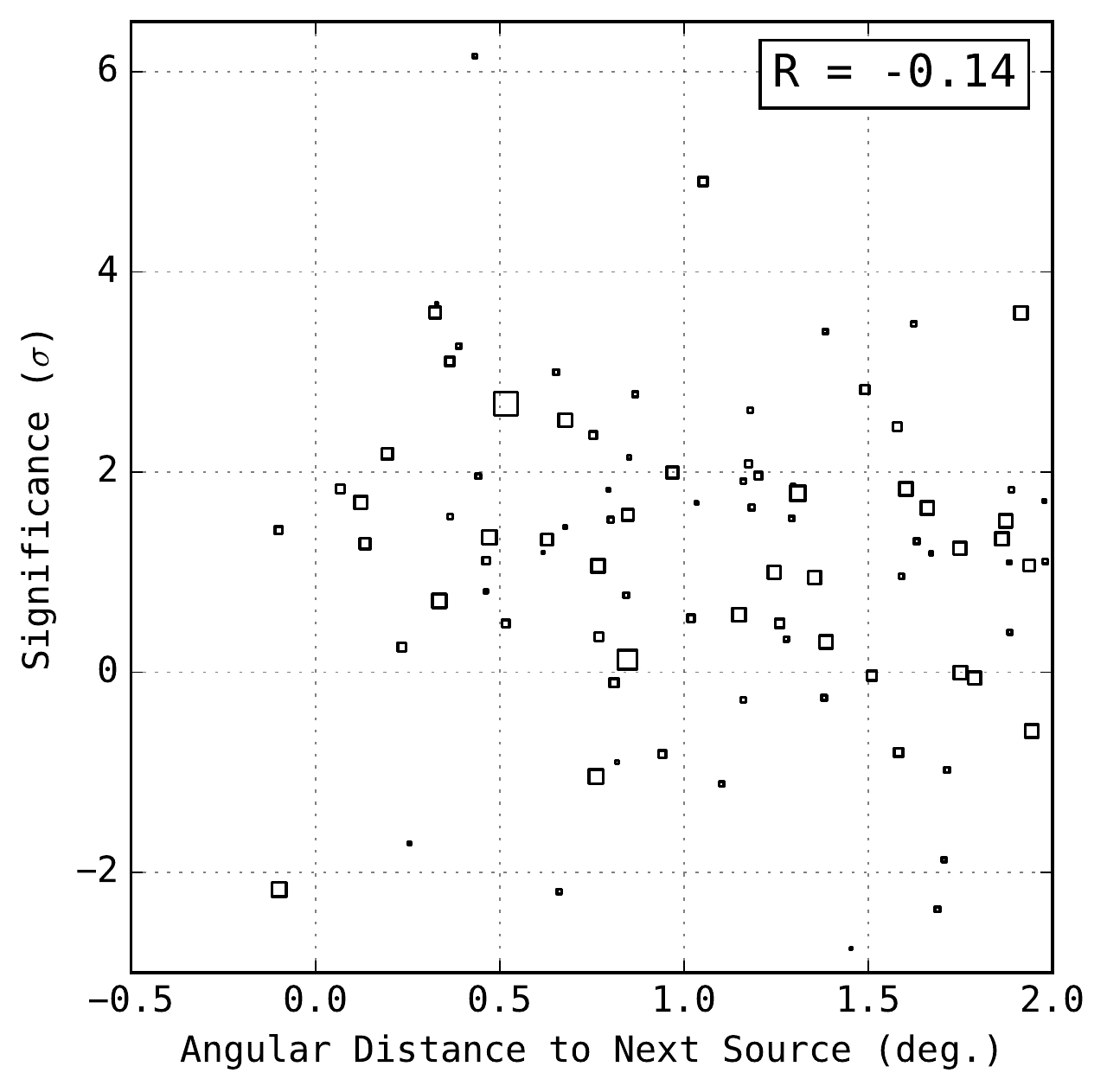}}
  \caption{Source significance values of the VHE-dark source sample vs. the angular distance to the nearest detected
H.E.S.S. source. The angular distance between the edges of our test regions to that of
detected H.E.S.S. sources is shown;  the marker size is proportional to the integrated flux above 1~TeV of the latter. The source region positions and sizes are taken from the HGPS, where they are
defined as the centroids and 80~\% containment radii of symmetric and two-dimensional Gaussian fits to
the integrated flux map, respectively (see \citealt{HGPSForth} for more details).
Negative distance values are possible for asymmetric sources where the Gaussian fit does not
 describe the source morphology well, see e.g. SNR~G36.6$-$0.7 in Fig.~\ref{Figure:Deselection}.}
  \label{Figure:SigvsD}
  \end{center}
\end{figure}

  \item Imperfect background modelling can also lead to spurious excess emission, especially in
   background-dominated observations such as those made by H.E.S.S.
   The background reconstruction, however has been thoroughly studied 
   in several publications (e.g. \citet{ref:bgmodelling} and \citet{NGC253}),
   typically resulting in small systematic uncertainties far below the excess observed here.

  \item Galactic diffuse
emission \citep{Diffuse} in the VHE range
might cause the positive offset in the significance distribution.
While the nature of this emission remains
under discussion, it is believed that it consists both of a diffuse component
of propagating CRs interacting with their environment 
 and a population of unresolved sources (possibly including
SNRs), which emit VHE $\gamma$-rays and a priori are unrelated to the individual SNRs 
investigated in this study.

\begin{figure}[h!!!]
  \begin{center}
  \resizebox{\hsize}{!}{\includegraphics[clip=]{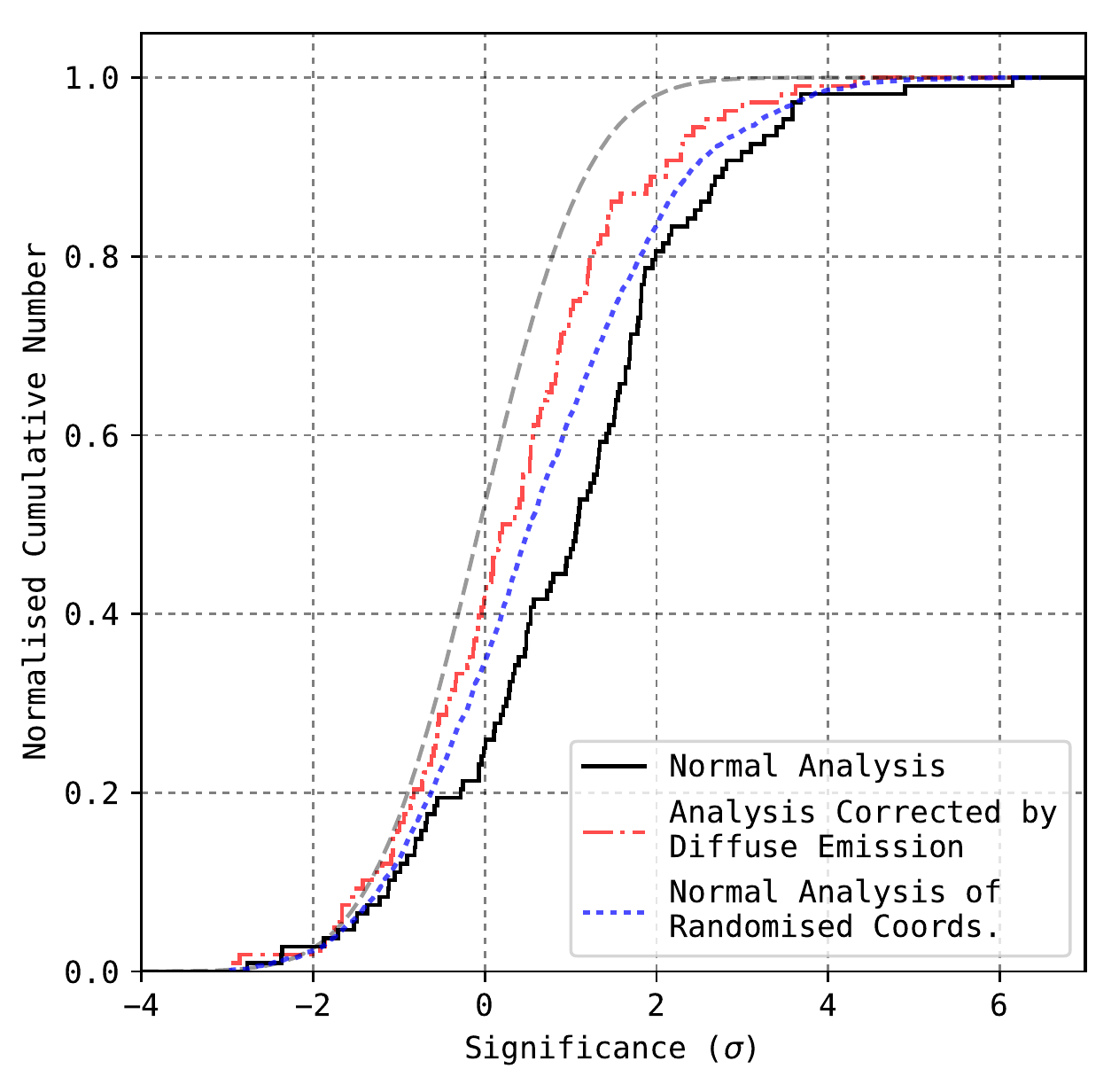}}
  \caption{Cumulated significance distribution of the SNR sample (black) and analogous
distribution corrected for the Galactic diffuse emission component (dash-dotted red). The cumulated 
density function from the analysis of randomised test regions is shown in dotted blue. The grey dashed curve
represents a cumulated normal Gaussian distribution.}
  \label{Figure:SigSubtracted}
  \end{center}
\end{figure}

  \item The significance distribution offset might be the result of
a cumulation of localised emission from the investigated SNR shells, which individually fall below
the HGPS sensitivity. 
\end{enumerate}

The four components are hard to disentangle, and it is possible that 
we impose flux limits on a sum of these contributions. 
However, contribution (1) seems negligible, since no correlation between excess significance
and angular distance to the nearest HGPS source is apparent, as shown in Fig.~\ref{Figure:SigvsD}.

The diffuse emission described in (3) extends out well beyond many regions of strong
 source emission and thus also the de-selection regions. As a result, this emission may
be present in the investigated analysis regions. 
In Fig.~\ref{Figure:SigSubtracted}, we show the cumulative significance distribution of
the investigated SNR sample (compare to the right panel of Fig.~\ref{Figure:SigDist}) together
with the analogous distribution that results if the diffuse emission contribution is accounted for
in the significance calculation. To obtain the latter, we applied the parametric model of the diffuse
emission component presented in \citet{HGPSForth} to the data. It should be mentioned that for this
calculation the significance values were derived directly from the HGPS maps.
A Kolmogorov-Smirnov test of the distribution corrected in this way (red line
in Fig.~\ref{Figure:SigSubtracted}) against a normal (noise-) distribution (grey) results
in a p-value of $p=1.8 \times 10^\mathrm{-2}$, roughly corresponding to a significance value
of $\sim$2.1$\sigma$, which suggests that the parametric 
diffuse emission model can account for a large portion, but not all of the significance offset.

In order to investigate contribution (4), we performed an analysis of randomised SNR analysis regions.
That is, if the
observed significance offset was indeed due to localised, faint emission from SNRs, we
would expect this effect to be absent in a sample of randomised test regions.
We determined the random positions by adding a uniform variate $l \in [-5~,~5)$\degr\ 
to the Galactic longitude value of each real SNR.
We left the Galactic latitude value unchanged so as not to introduce a bias
regarding component (3), which shows a Galactic latitude dependence; see \citet{HGPSForth}.
Also, we added a variate number $s \in [-0.05~,~0.05)$\degr\ to the radius of the test
region, with a lower ceiling of the resulting radius of 0.1\degr. 
Once the random test region is generated, we subjected it to our source selection method 
(see Sec.~\ref{sec-selec}), and additionally tested whether it overlaps with a previously generated region, in 
which case we rejected it. 
If the random test region was rejected, a new test region was generated in the same manner and tested again 
until it passed selection.
We created and analysed 28 randomised samples of the real set of SNRs, 
and each sample yields a set of 108 significance values. From these 28 sets, we calculated 
the average cumulated and normalised significance distribution. We then used this  
distribution as the cumulative density function in a Kolmogorov-Smirnov test against
 the significance distribution of the real SNR sample (blue line in Fig.~\ref{Figure:SigSubtracted}). 
This results in a p-value of $p=5 \times 10^\mathrm{-3}$,
approximately corresponding to a significance of $\sigma \sim 2.6$, which does not
allow us to claim the detection of a cumulated SNR signal.
  
Our results indicate that the observed shift in the significance distribution might be 
the result of the sum of components (3) and (4), although a deeper understanding of 
the large-scale VHE emission along the Galactic plane as well as improved analysis
methods and observation exposure are required to provide definitive answers. Future studies with 
CTA should be able to shed light on this question.

\subsection{Fermi-LAT comparison}

The {\it Fermi}-LAT (Large Area Telescope) collaboration has recently performed a systematic survey in the 1-100 GeV energy band towards known Galactic SNRs \citep{ref:acero16}. Among the 102 source candidates, 30 are classified as likely GeV SNRs, and an additional 14 ``marginally classified'' candidates could be associated with the SNRs. 

\begin{figure}[h!!!]
  \begin{center}
  \resizebox{\hsize}{!}{\includegraphics[clip=]{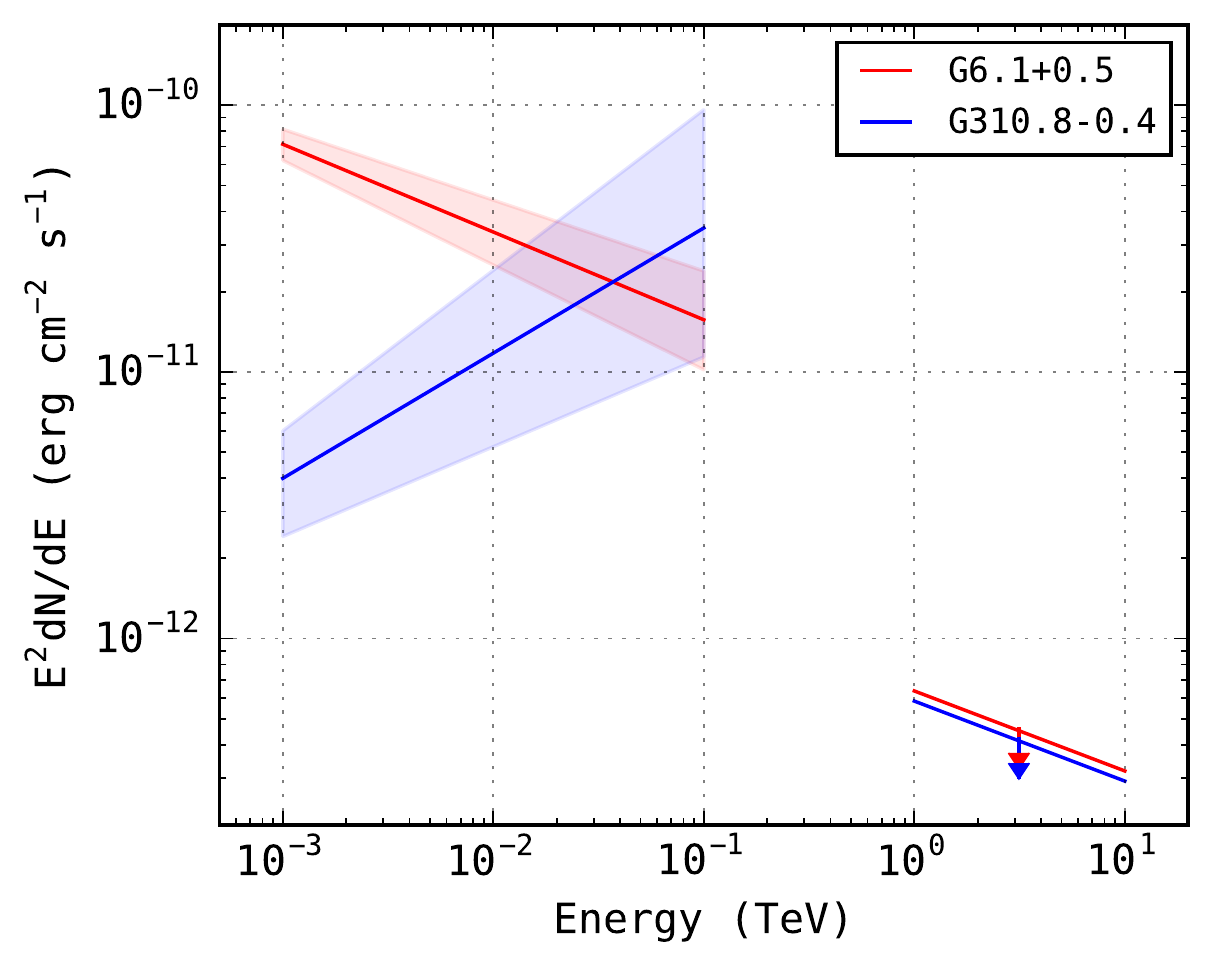}}
  \caption{Gamma-ray spectral energy distributions of G6.1+0.5 and G310.8$-$0.4. The {\it Fermi}-LAT spectra are derived from the spectral parameters reported in the first {\it Fermi}-LAT SNR catalogue. Statistical errors errors are indicated by the shaded bands.}
  \label{Figure:FermiSpec}
  \end{center}
\end{figure}

We compared the H.E.S.S. upper limits for all the GeV SNR candidates in our VHE-dark SNR sample with their extrapolated 1-10 TeV fluxes from the {\it Fermi}-LAT measurements. We note that the coordinates and radii of our analysis regions (see Sect.~\ref{sec-ana}) differ from those found with {\it Fermi}-LAT.   The H.E.S.S. upper limits are only constraining for two sources, namely G6.1+0.5 and G310.8$-$0.4; and hence, these upper limits point towards a spectral steepening in or before the VHE domain. However, these two sources are neither safely nor marginally classified GeV SNR candidates;  the emission from G6.1+0.5 is flagged as doubtful and the GeV extent of G6.1+0.5 and G310.8$-$0.4 (0.64$^\mathrm{\circ}$ and $\sim$ 1$^\mathrm{\circ}$, respectively) are much larger than their radio SNR size (0.1$^\mathrm{\circ}$-~0.15$^\mathrm{\circ}$ and 0.1$^\mathrm{\circ}$, respectively). The  $\gamma$-ray spectra of these two sources are shown in Fig.~\ref{Figure:FermiSpec}.

\subsection{Constraints on the accelerated particles in SNRs}
\label{sec-model}

\begin{figure*}
  \begin{center}
  \resizebox{.72\hsize}{!}{\includegraphics[clip=]{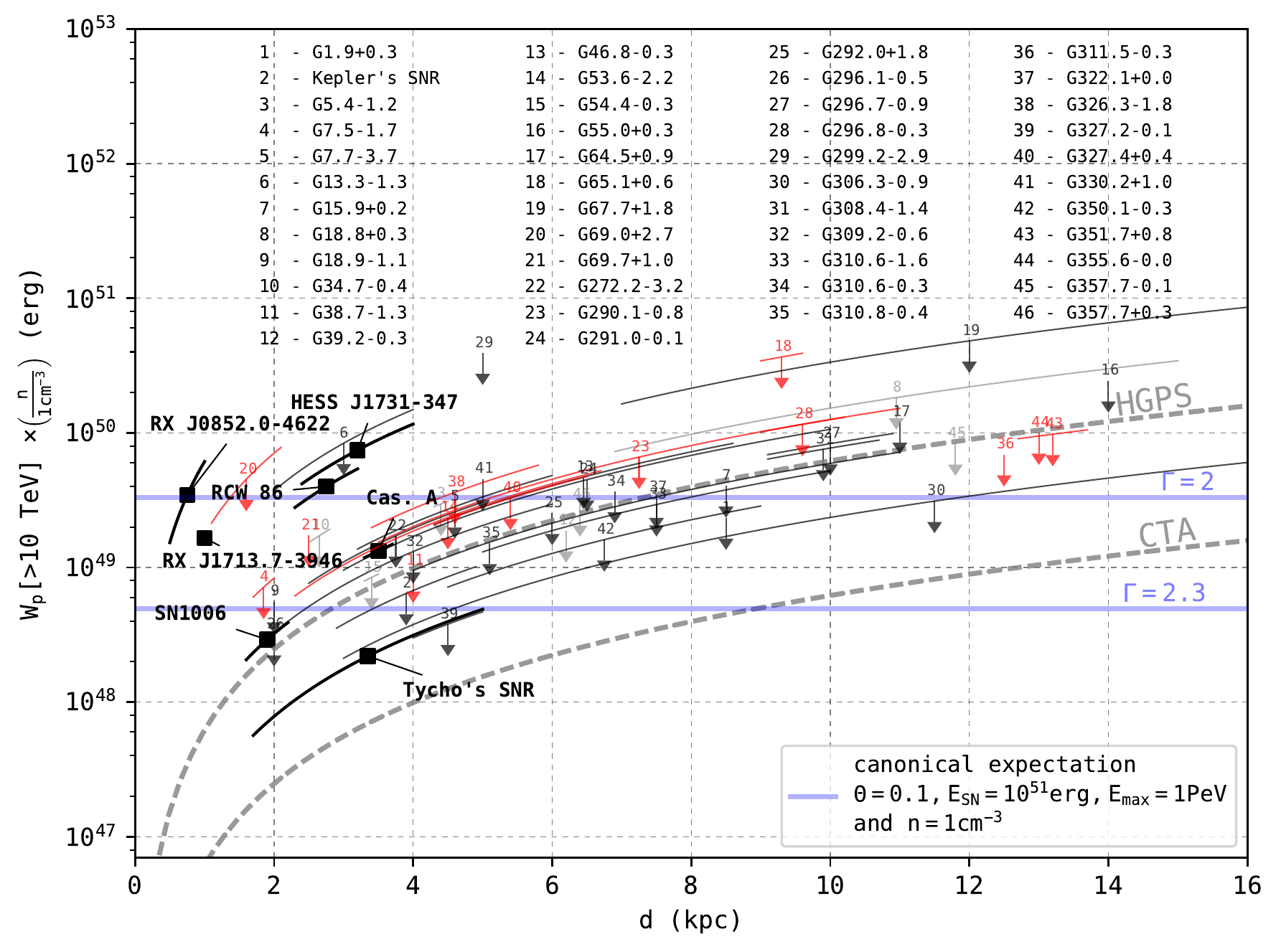}}
  \resizebox{.72\hsize}{!}{\includegraphics[clip=]{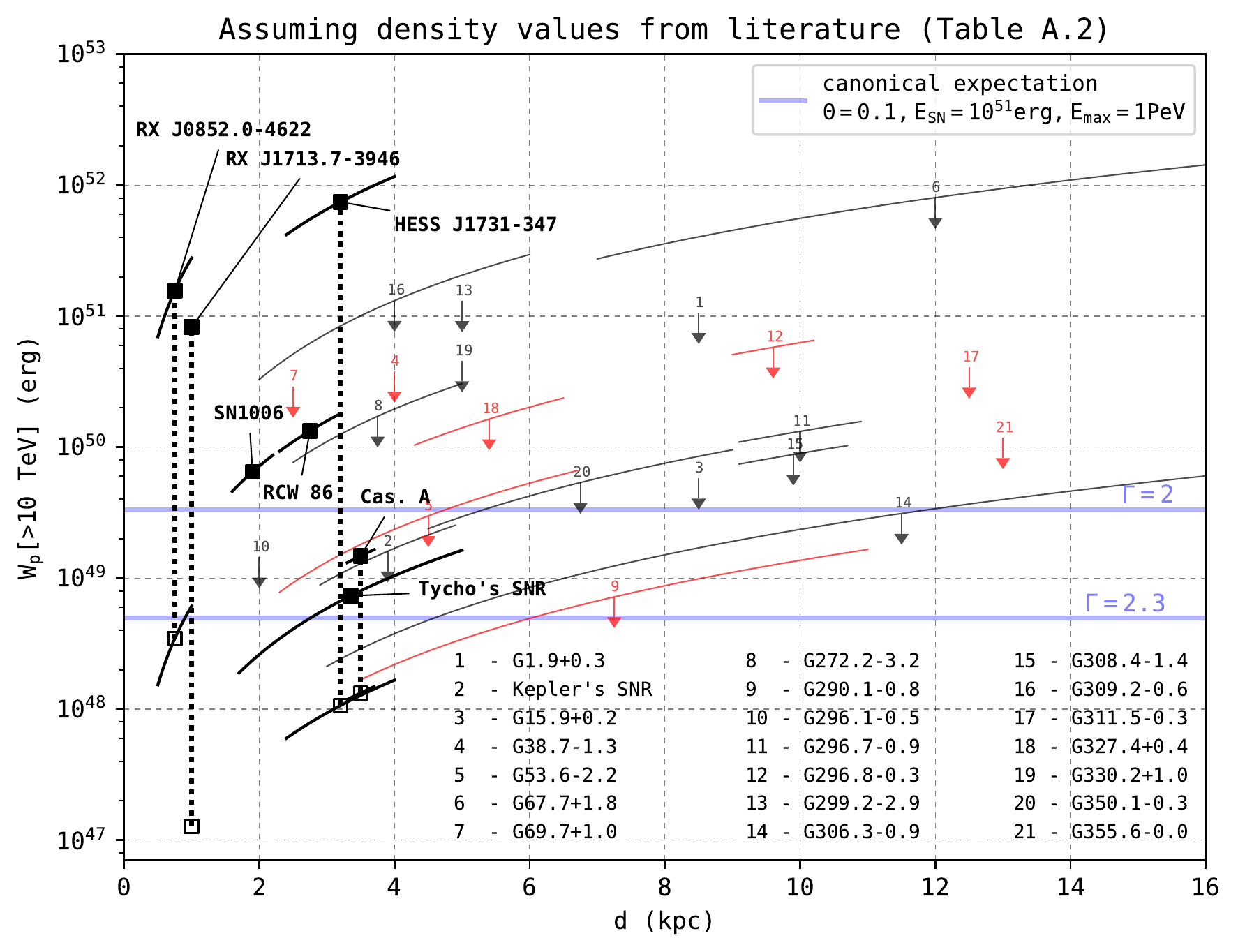}}
  \caption{
{\fontsize{8.3}{8.3}\selectfont Upper limits and measurements on the product of proton energy content above 10~TeV and ambient density (top) and
proton energy content above 10~TeV (bottom) vs. source distance. We assume the ambient density and distance estimates from
the publications listed in Table~\ref{Table:SNRparams} and also show the corresponding values for VHE-detected shell-type SNRs. For the latter, we use the flux normalisations as reported in the publications listed in Table~\ref{Table:SNRspecTable}. In several sources a gas clump correlation of $\gamma$-ray emission was observed. In these cases, filled points indicate the result if the low density values in the rarefied inter-clump medium are used in the calculation, while open points show the result if the high density values inside the clumps are assumed.
We also show the canonical expectation, assuming an acceleration efficiency
of $\Theta$ = 10\%, a SN blast energy of E$_\mathrm{SN}$ = $10^\mathrm{51}$ erg, and a power-
law spectrum with spectral index of $\Gamma$ = 2 (expected from standard first order Fermi acceleration) and $\Gamma$ = 2.3 (corresponding to the average spectral index of HGPS sources) up to $\mathrm{E}_\mathrm{max} = 1$PeV, assuming $n = 1\mathrm{cm}^\mathrm{-3}$ in the top panel. Furthermore, we 
show in the top panel the sensitivity of the HGPS (assuming a mean value of $\sim$1.5\% Crab) and the projected sensitivity of CTA (assuming a ten times higher sensitivity). Red points represent
old sources (>~10~kyrs) that are not expected to accelerate protons to PeV energies,  grey limits in the top panel indicate SNRs that are likely to interact with molecular clouds (see Table~\ref{Table:Analysis}). Values for Kepler's SNR (source 2), G1.9+0.3 (1) and G330.2+1.0 (41 top, 19 bottom) are derived from the limits in \citet{Kepler} and \citet{G1_9}. }
}
  \label{Figure:ECRDist}
  \end{center}
\end{figure*}

By comparing to model expectations,
the derived integral flux upper limits may be used to estimate upper limits
on the energy content in high-energy particles. In the following, we limit our considerations to
inelastic proton-proton collisions and the subsequent $\pi^0$-decay and
inverse-Compton (IC) emission since these processes are believed to be the most relevant $\gamma$-ray emission mechanism in SNRs for the VHE regime.

If the corresponding $\pi^0$ production timescale due to inelastic proton-proton scattering, $\tau_\mathrm{\pi^0}$, and electron cooling time,
$t_\mathrm{IC}$, are known\footnote{The timescales $\tau_\mathrm{\pi^0}$ and $t_\mathrm{IC}$ are
defined as the inverse of the $\pi^0$ production rate from p-p interactions
and the time after which electrons have lost half of their energy due to 
the IC process, respectively.}
and a distance estimate for a source
is given, the presented flux upper limits can be used to place upper limits
on the total energy in electrons and protons at the time of emission.
In the following, we assume that $\gamma$-rays with energies larger than
1~TeV mainly probe particles with energies above 10~TeV in both
the IC and $\pi^0$-decay emission channels.

\begin{figure*}
  \begin{center}
  \resizebox{\hsize}{!}{\includegraphics[clip=]{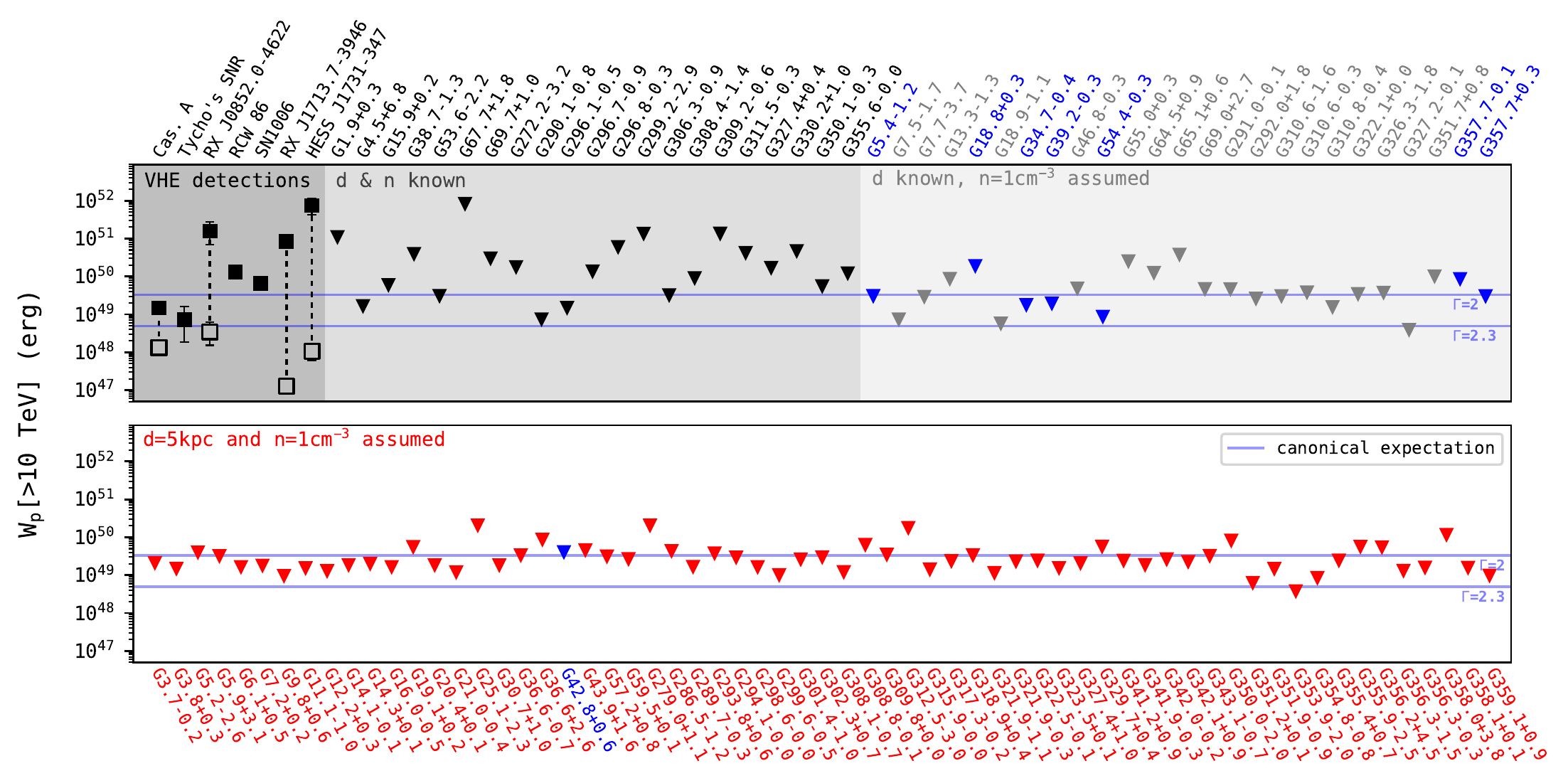}}
  \caption{Sample of upper limits on the proton energy content above 10~TeV (triangles).
Values are derived using the centre of the distance uncertainty interval. Blue triangles indicate SNRs that are likely to interact with molecular clouds  (see Table~\ref{Table:Analysis}). Points are shown for the VHE-detected sample of SNRs, where
the error arising from the distance uncertainty is indicated. In several sources a gas clump correlation of $\gamma$-ray emission was observed. In these cases, filled points indicate the result if the low density values in the rarefied inter-clump medium are used in the calculation, while open points show the result if the high density values inside the clumps are assumed.}
  \label{Figure:ULScat}
  \end{center}
\end{figure*}
Following \citet{Aharonian2004}, we used $\tau_\mathrm{\pi^0} = t_\mathrm{\pi^0} / n$, where 
$n$ is the number density of ambient gas nuclei in units of $1\mathrm{cm}^\mathrm{-3}$ and $t_\mathrm{\pi^0}\approx 5 \times 10^\mathrm{15}\mathrm{s}$.
Unlike in the hadronic case, the IC cooling time depends strongly on the electron energy and  the energy densities of the ambient radiation fields. We adopted a conservative value of the cooling time by assuming the cosmic microwave background (CMB) as the only ambient photon field and using the value for 10~TeV electrons scattering in the Thomson regime\footnote{This approximation introduces a $\sim$14\%~error.}. This results in a cooling time of $t_\mathrm{IC} \approx 10~\mathrm{TeV}/\dot E(10~\mathrm{TeV}) = \frac{3}{4} m_\mathrm{e}^2 /(\sigma_\mathrm{T} c \epsilon_\mathrm{CMB} 10~\mathrm{TeV}) \approx 4\times 10^\mathrm{12} \mathrm{s}$, where $m_\mathrm{e}$ is the electron rest mass, $\sigma_\mathrm{T}$ is the Thomson cross section, $c$ is the speed of light, and $\epsilon_\mathrm{CMB}$ is the energy density of the CMB.

Using these values it is possible to convert the upper limits on the integrated energy flux above 1~TeV into upper limits on the total energy in protons and electrons above 10 TeV at the time of emission. In general, for power-law spectra with $\Gamma \neq 2$ upper limits on the energy flux, $F_\mathrm{E}^\mathrm{ul}$, between energies $E_1$ and $E_2$ are connected to the corresponding integral flux upper limits, $F^\mathrm{ul}$, between $E_1$ and $E_2$ via 

\begin{equation}
F_\mathrm{E}^\mathrm{ul} = \frac{\Gamma - 1}{\Gamma - 2}\times \frac{E_1^\mathrm{2-\Gamma} - E_2^\mathrm{2-\Gamma}}{E_1^\mathrm{1-\Gamma} - E_2^\mathrm{1-\Gamma}}\times F^\mathrm{ul},
\end{equation}
where in our case $E_1 = 1$~TeV, $E_2 = 10$~TeV, and $\Gamma = 2.3$.

Upper limits on the total energy content in electrons can be derived as

\begin{equation}
W_\mathrm{e}^\mathrm{ul}[>10\mathrm{TeV}] \approx F_\mathrm{E}^\mathrm{ul}[>1\mathrm{TeV}] \times t_\mathrm{IC}\times 4\pi d^2.
\end{equation}

The comparable procedure in the case of protons yields
\begin{equation}
\label{wpp}
W_\mathrm{p}^\mathrm{ul}[>10\mathrm{TeV}] \approx F_\mathrm{E}^\mathrm{ul}[>1\mathrm{TeV}] \times \tau_\mathrm{\pi^0} \times 4\pi d^2
\end{equation} or more explicitly, if $n$ is unknown,
\begin{equation}
(W_\mathrm{p}[>10\mathrm{TeV}]\times n)^\mathrm{ul} \approx F_\mathrm{E}^\mathrm{ul}[>1\mathrm{TeV}] \times t_\mathrm{\pi^0} \times 4\pi d^2
,\end{equation} implying that $W_\mathrm{p}[>10\mathrm{TeV}]$ and $n$ are degenerate in this case, such that with this method upper limits on the proton energy
content can only be placed if the corresponding values of the target gas densities are available.

These limits are conservative estimations. The exact proportionality coefficients $c_\mathrm{x}$, defined by $W = c_\mathrm{x} \times F_\mathrm{E} \times t_\mathrm{x} \times 4\pi d^2$ for  $\pi^0$-decay  
are $c_\mathrm{\pi^0}=[0.7,0.7,0.95,0.81]$, using the recent parametrisation of the total inelastic p-p cross section for Geant4, Pythia8, SIBYLL2.1, and QGSJET-I from \citet{Kafexhiu2014}, respectively. For the IC mechanism we obtained $c_\mathrm{IC-CMB}=0.73$ applying the full Klein-Nishina cross section and taking the shape of the electron spectrum into account.

In Table~\ref{Table:Analysis} we list the individual values for $(W_\mathrm{p}\times n)^\mathrm{ul}$ and $W_\mathrm{e}^\mathrm{ul}$. 
The top panel in Fig.~\ref{Figure:ECRDist} shows our limits of $(W_\mathrm{p}\times n)^\mathrm{ul}$
versus the estimated source distance, where we also show the corresponding estimates for  
the VHE-detected sample of isolated shell-type SNRs (see Table~\ref{Table:SNRspecTable} for references). Using the published flux values in case of VHE detections or our calculated flux upper limits for the undetected sources, we calculate the values
according to Eq.~\ref{wpp} using the distance estimates listed in 
Table~\ref{Table:SNRparams}. We see that
the H.E.S.S. upper limits are most constraining for relatively close sources. 
Assuming a hypothetical average ambient density around SNRs of $n\sim 1 \mathrm{cm}^\mathrm{-3}$ and an intrinsic proton power-law spectrum resulting from classical first-order Fermi acceleration with index $\Gamma = 2$,
 a handful of sources within a few 
kpc distance constrain the CR paradigm. This paradigm identifies SNRs as the sources
of Galactic CR assuming that 10\% of the blast energy of $10^\mathrm{51}$erg goes into the acceleration
of CR up to PeV energies. However, ambient density values vary strongly from object to object, and 
in the bottom panel of Fig.~\ref{Figure:ECRDist} we use the
literature estimates listed in Table~\ref{Table:SNRparams} to derive values for $W_\mathrm{p}$. 
The reported density values in this table are in some cases derived from fits to X-ray spectra rather than direct measurements (e.g.
from CO or HI observations).
These values are characterised by the density inside the X-ray emitting bubble (i.e. the inter-clump density), which might not be the relevant quantity
for hadronic high-energy processes. Multi-TeV protons suffer less from energy losses than electrons of similar energy
and are therefore able to propagate further and consequently probe a different environment than X-ray emitting electrons of similar energy. For instance, recent work suggests 
a correlation between TeV $\gamma$-rays brightness and cold $HI$ gas density, for example in the cases of HESS J1731$-$347 (\citealt{Fukuda}) and RX J1713.7$-$3946 (\citealt{Fukui}, \citealt{Sano}). This correlation points to $\gamma$-ray production at dense clumps with number densities well
above 10~cm$^{-3}$ rather than the rarefied inter-clump medium. Therefore, we treat ambient density values that have been derived
from X-ray spectra as lower limits, which results in conservative upper limits on the proton energy content.
In the cases in which studies reveal a correlation between $\gamma$-ray emission and gas clumps, we calculate the proton energy content
for the two scenarios assuming either the low inter-clump medium density or the high value inside the clumps, which are both listed in Table~\ref{Table:SNRparams}.

As can be seen, five of our limits are constraining the canonical
expectation in this case: G290.1$-$0.8 (source 9 in plot), G296.1$-$0.5 (10) as well as G53.6$-$2.2 (5),
G306.3$-$0.9 (14), and G350.1$-$0.3 (20), if its distance is at the lower end of the 
uncertainty interval. The limits become less stringent if the proton spectra are softer. However, 
even if a typical Galactic particle spectral index of $\Gamma = 2.3$ is assumed, inferred from the large number of HGPS sources, 
two sources still constrain the theoretical expectation:
that is G306.3$-$0.9 (14), if the source is situated at the very low end of the distance error
interval, and G290.1$-$0.8 (9). However, the latter is a thermal composite SNR (as is also G53.6$-$2.2), 
estimated
to be of an evolved age (>~10kyrs) and thus unlikely to be a place of efficient
particle acceleration to PeV energies as assumed in the canonical picture (see e.g. \citet{Ptuskin2005}).
Therefore we consider this limit as not constraining. 
Also, the well-studied sources Cassiopeia A, Kepler's SNR, and Tycho 
lie below the canonical estimate if $\Gamma = 2$. However,
the latter two sources require larger indices in order to model
their VHE emission (see \citet{Tycho} and \citet{CasA_new}), and thus in these cases the comparison to
the canonical value is of limited validity.
\begin{figure*}[!b]
  \begin{center}
  \resizebox{0.7\hsize}{!}{\includegraphics[clip=]{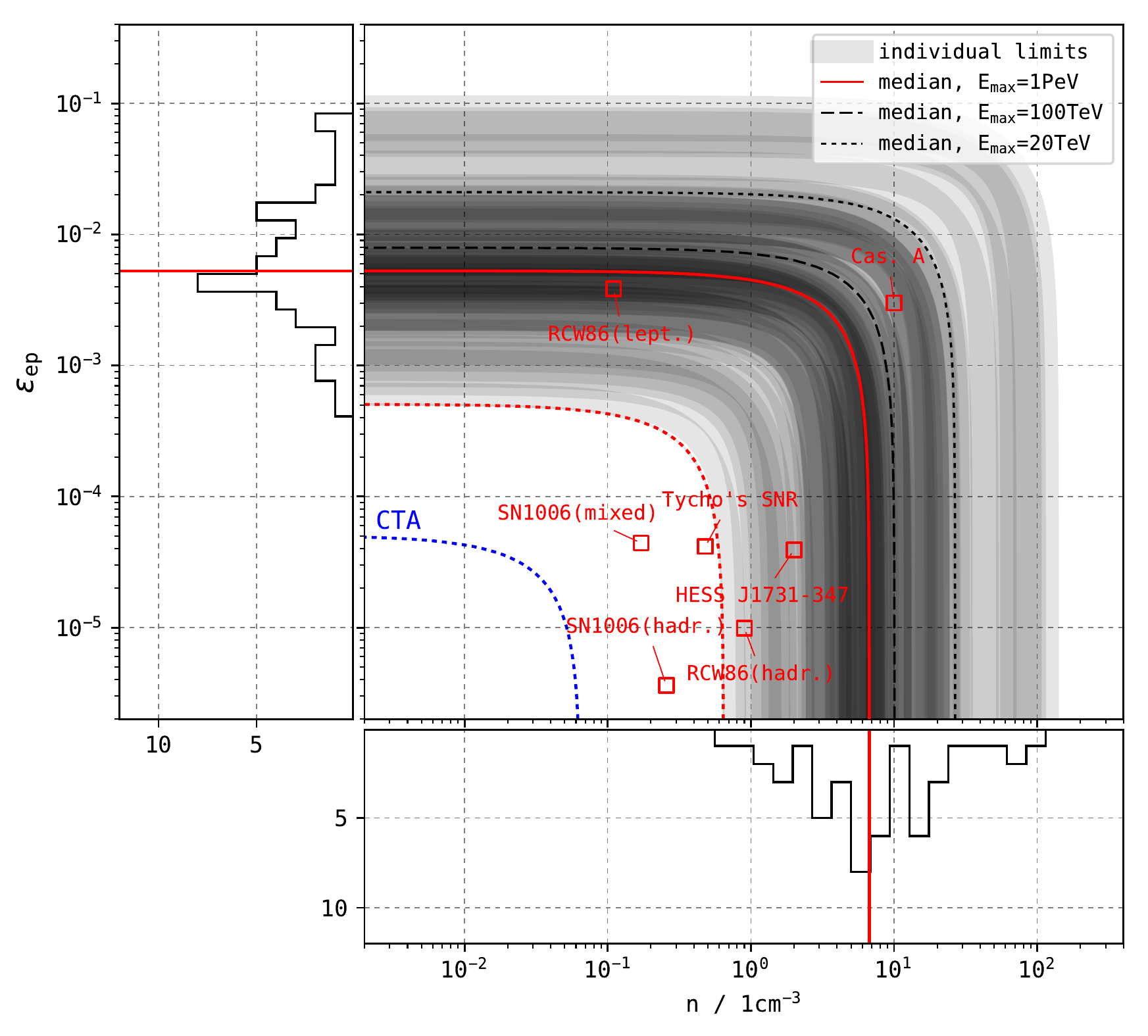}}
  \caption{Constraints on the $\epsilon_\mathrm{ep}$ and $n$ parameter space as imposed by the flux upper limits (grey bands, each limit
           constitutes one band). For each constraint a source-intrinsic power-law spectrum with index $\Gamma=2.3$ was assumed.
           Solid and dotted red lines represent the constraints corresponding to the median and lowest value of the flux
           upper limit distribution, respectively. The blue line indicates the boundary above which CTA might be able to probe the parameter space,
           assuming a ten times higher sensitivity than H.E.S.S. and taking the lowest flux upper limit as a reference.          
           Additionally, values of $\epsilon_\mathrm{ep}$ and $n$ for VHE-detected shell-type SNRs are shown, extracted from the publications in Table~\ref{Table:SNRspecTable}. 
           For better comparison, these values are rescaled to the canonical total CR energy in SNRs of 10$^\mathrm{50}$erg. 
           The dashed and dotted black lines indicate how the median of the distribution shifts if one limits
           the maximum proton energy to 100~TeV and 20~TeV, respectively. The histograms show the parameter limit distributions in the
           extreme hadronic- and leptonic-dominated scenarios.}
           
  \label{Figure:ECRDist2}
  \end{center}
\end{figure*}

In another form of presentation, Fig.~\ref{Figure:ULScat} shows the upper limits on 
$W_\mathrm{p}[>10~\mathrm{TeV}]$ for all objects that have been investigated, assuming a canonical ISM value of $n=1~\mathrm{cm}^{-3}$ as
ambient density and a typical source  distance of $d=5~\mathrm{kpc,}$ where this information is not available. It also shows 
the corresponding values for the VHE-detected SNRs.

The sample shows very high limits, which by
far exceed the canonical expectation. This indicates that
in these cases the sensitivity for the direction of the corresponding sources 
is not sufficient to provide constraining flux upper limits.

In those cases where clump interaction has been observed we see excessive values for
$W_\mathrm{p}$, especially \object{RX J0852.0$-$4622}, \object{RX J1713.7$-$3946} and \object{HESS J1731$-$347}
if the rarefied inter-clump density values are assumed. Adopting a high density 
inside the clumps leads to very low values of $W_\mathrm{p}$ that lie below the canonical 
expectation. Most notably, the derived value for \object{RX J1713.7$-$3946} is 
about 1.5 orders of magnitudes below the canonical expectation for $\Gamma = 2.3$.
This value seems very low as in a hadronic emission scenario a high value for the
cut-off in the proton spectrum of $\sim$90~TeV is required to model the H.E.S.S. 
spectrum (see \citet{RXJ1713New}).

Perhaps the inclusion of $\gamma$-ray emission from 
electrons via the IC mechanism may be a solution to this problem.
Combining both the hadronic and leptonic estimations and introducing the
electron to proton energy fraction above 10~TeV, $\epsilon_\mathrm{ep}$, we can write

\begin{equation}
W_\mathrm{p}^\mathrm{ul}[>10\mathrm{TeV}] = \frac{F_\mathrm{E}^\mathrm{ul}[>1\mathrm{TeV}]\times 4\pi d^2}{n/t_\mathrm{\pi^0} + \epsilon_\mathrm{ep}/t_\mathrm{IC}}.
\end{equation}

In the following we want to determine the portion of the $\{n,\epsilon_\mathrm{ep}\}$ parameter space that is excluded by the flux upper limits, assuming that the accelerated proton distributions in SNRs meet the canonical expectation mentioned in the earlier
paragraphs and follow a power-law distribution with a spectral index of $\Gamma = 2.3$. 

Our upper limits are constraining a parameter set in the $\{n,\epsilon_\mathrm{ep}\}$-plane 
if $W_\mathrm{p}^\mathrm{ul}[>10\mathrm{TeV}]~<~W_\mathrm{p}^\mathrm{th}[>10\mathrm{TeV}]$,
where $W_\mathrm{p}^\mathrm{th}[>10\mathrm{TeV}]$ is the theoretically expected value from the
canonical assumption.
Each individual upper limit then results in a curve in this parameter space above which
the corresponding source would have been detected by H.E.S.S. and is therefore excluding the 
corresponding portion of the parameter space for this object. 
The set of curves is shown in Fig.~\ref{Figure:ECRDist2}.
This figure also shows the distribution of constraints on $\epsilon_\mathrm{ep}$ and $n$ in the asymptotic leptonic- and hadronic-dominated
scenario limits as one-dimensional histograms. The logarithmic medians of those distributions are log$(\epsilon_\mathrm{ep})$ and log$(n/1\mathrm{cm}^\mathrm{-3})$ are -2.28 and 0.83, respectively.
The variance in both cases is 0.26.

In the canonical picture of particle acceleration in SNRs, 
these values constrain hadronic-dominated emission scenarios to 
ambient density values $n \la 7~\mathrm{cm}^\mathrm{-3}$ and leptonic-dominated
emission scenarios to $\epsilon_\mathrm{ep} \la 0.5\%$. The most stringent
upper limit yields $n \la 0.6~\mathrm{cm}^\mathrm{-3}$ and
$\epsilon_\mathrm{ep} \la 0.05\%$.

For comparison, we also show in Fig.~\ref{Figure:ECRDist2} the $\{n,\epsilon_\mathrm{ep}\}$-ntuples for the VHE-detected 
shell-type SNRs. We derive these values from the models in 
the publications listed in Table~\ref{Table:SNRspecTable}. For a valid comparison with the constraints in the
$\{n,\epsilon_\mathrm{ep}\}$ parameter space, we scale the parameters
values by a factor $f = W_\mathrm{p,i}/(10^\mathrm{50}\mathrm{erg})$, where $W_\mathrm{p,i}$
is the total energy in CR protons that has been derived for the sources $i$ in 
the literature because in most cases the latter deviates from the canonical $10^\mathrm{50}$~erg. 
In some publications the authors discuss both hadronic- and leptonic-dominated scenarios, in which
case we show the respective points in the parameter space for both cases.
The majority of the detected sources fall into the part of
the parameter space that would allow for the detection by the HGPS. However, its 
sensitivity would not allow for the detection of the two VHE-faintest SNRs: 
Tycho's SNR, detected by the VERITAS collaboration \citep{Tycho}, and SN 1006. These sources required pointed observations to reach
the exposure necessary for detection.

It should be stressed that the theoretical interpretation of the
presented analysis results is rather simple and does not take into
account the full complexity of the SNRs (which is beyond the scope of
this paper).  Also, many distance and density estimates used in
this study suffer from large uncertainties with factors of a few; see Table~\ref{Table:SNRparams}.

That said, if the assumptions made in our considerations are roughly plausible for the average young to middle-aged SNR, the next generation observatory,  CTA, holds great potential for SNR science.
An improvement in instrument sensitivity by an order of magnitude, as planned with CTA,
will allow it to probe a considerably increased fraction of the parameter space, corresponding to the portion 
of the parameter space above the blue dotted line in Fig.~\ref{Figure:ECRDist2}.
If our theoretical expectations are sound, we can expect CTA to test the SNR paradigm
for ambient densities in the typical ISM range independent of the primary emission mechanism. 
If the $\gamma$-ray emission is dominated by the leptonic channel, even SNRs in rarefied environments such as the interior of bubbles blown by the main-sequence winds of the SNR progenitor stars should be detectable with CTA.

\subsection{Luminosity evolution of SNRs in the radio and VHE bands}

\label{sec-pop}

\begin{figure*}
  \begin{center}
  \resizebox{0.62\hsize}{!}{\includegraphics[clip=]{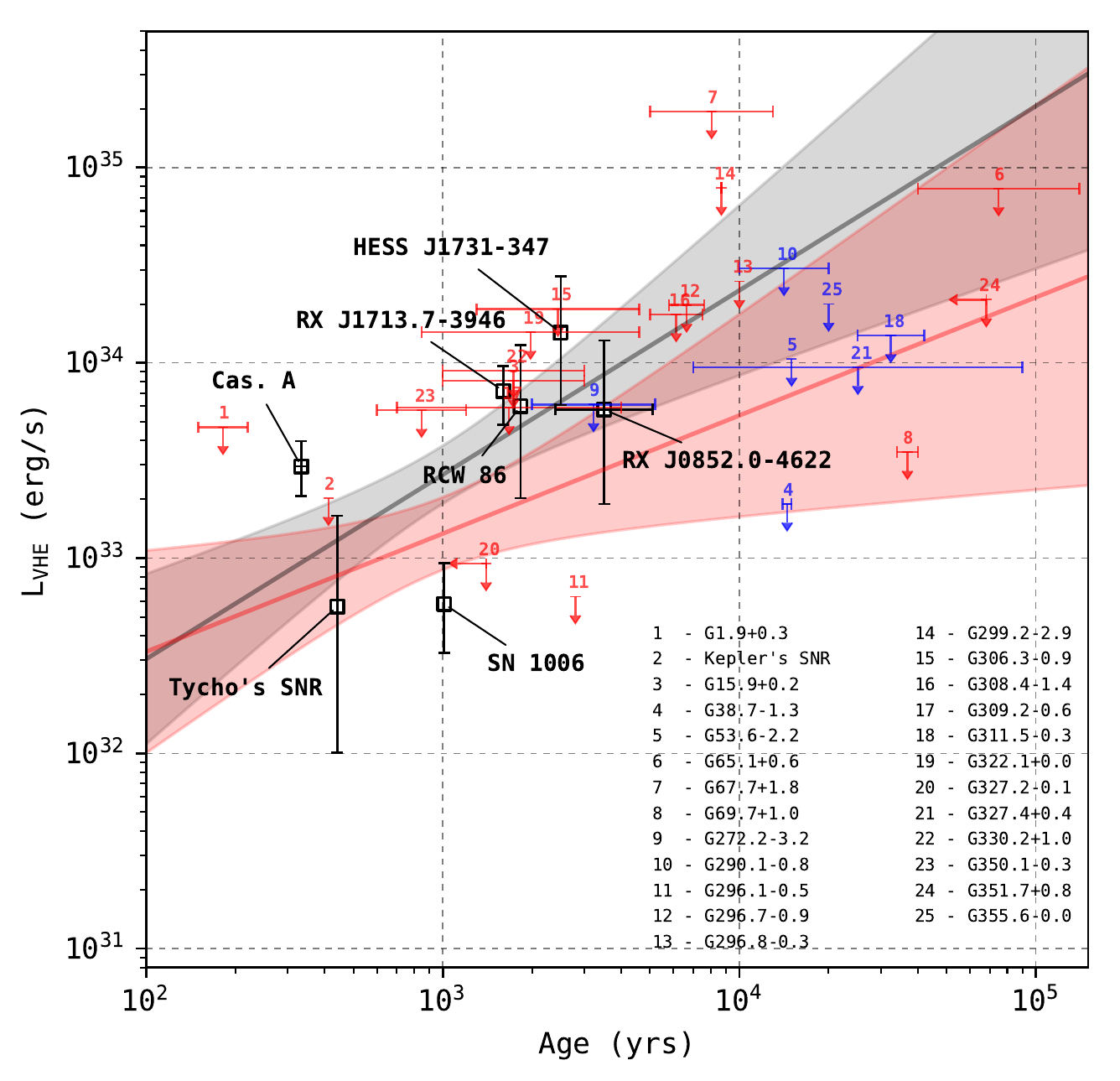}}
  \resizebox{0.62\hsize}{!}{\includegraphics[clip=]{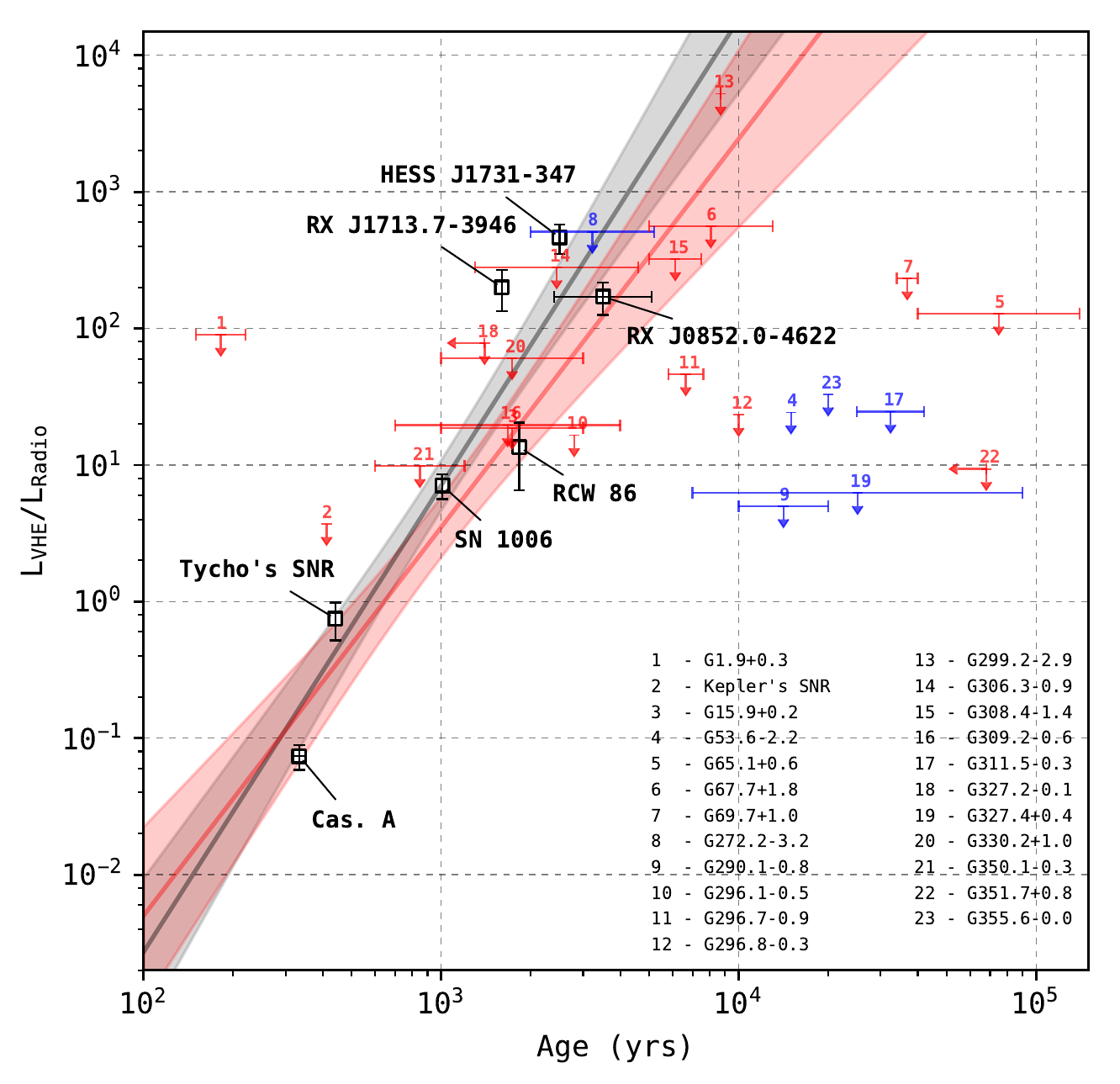}}
  \caption{
  {\fontsize{8.3}{8.3}\selectfont
  Correlations of VHE luminosity between 1~TeV and 10~TeV of $L_\mathrm{VHE}$ (top) and $(L_\mathrm{VHE}/L_\mathrm{radio})$ (bottom)
with source age, where $L_\mathrm{radio}$ is the luminosity at 1~GHz.
Points: detected shell-type SNRs in the VHE regime. 
Arrows: Upper limits. Red: shell-type SNRs. Blue:
thermal composite SNRs. Values for Kepler's SNR, G1.9+0.3 and G330.2+1.0 from \citet{Kepler} and \citet{G1_9}.
Grey lines and uncertainty bands: Best-fit correlations taking only the detected SNRs into account. 
Red lines and uncertainty bands:  
Best-fit correlations when including also those upper limits into the fit where SNRs
 fall into the age interval defined by the sample of VHE-detections  
(<~5.1~kyrs, upper age limit of \object{RX J0852.0$-$4622}).}
}
  \label{Figure:L}
  \end{center}
\end{figure*}

The average sensitivity of the HGPS is at the level of $\sim$2\% of the Crab nebula VHE flux.
There are sky regions of deeper exposure and thus lower values around prominent sources such as the Galactic centre.
Because of the limited sensitivity, we can expect selection effects in the sample of detected SNRs roughly following the relations
$L_\mathrm{VHE}/4\pi d^2 > S$ and $L_\mathrm{VHE}/4\pi d > S$ for point-like and extended sources, respectively. In this work, $S$ is the H.E.S.S. point-like source sensitivity in a given field of view, $d$ is the
distance to the source, and $L_\mathrm{VHE}$ is the source luminosity between 1~TeV and 10~TeV.
By including the sample of flux upper limits on radio SNRs in our considerations,
we can achieve a more complete and less biased view on the VHE emission properties of this source class.
Also, we want to make use of the large amount of radio information on 
SNRs and compare the VHE $\gamma$-ray fluxes to those observed at radio wavelengths.

To that end, we consider $L_\mathrm{VHE}$ and used the radio flux density values at 1~GHz, as provided in the
Green SNR catalogue (\citet{GreenCatalogue}), to calculate the corresponding luminosity $L_\mathrm{Radio}$. We furthermore formally assume a uniform bandwidth of 1~GHz to convert from radio flux density to radio flux. 
The spectral assumptions used in the derivation of the VHE $\gamma$-ray luminosities of the VHE-detected shell-type SNRs are listed in Table~\ref{Table:SNRspecTable}, along with the radio flux densities and age estimates from the SNRcat and Green catalogues.
Finally, we remove SNRs from the sample for which interaction with molecular clouds is established or probable
because in this study we want to investigate the physical processes at isolated SNR shocks. The information about whether a cloud interaction is occurring is also provided by SNRcat. 
However, such information is not available for all SNRs and thus it is possible that
interacting SNRs are still present in the resulting source sample.

In the following, we investigate the data for linear correlations of both the VHE-luminosity ($y=log(L_\mathrm{VHE})$) and the ratio of VHE-to-radio luminosities
($y=log(L_\mathrm{VHE}/L_\mathrm{Radio})$) with source age ($x = log(Age/1kyr)$).The fit results can be found in Table~\ref{Table:FitValues}.

\begin{table}[h!!!]
\begin{threeparttable}
\caption{Fit results from correlation testing.}
\begin{tabular}{l|ccc}
\hline
 & $p$-value & slope & intercept \\
 
\hline
\hline
$L_{VHE}$    & $0.1$ & $0.9\pm 0.4$ & $33.4\pm 0.2$ \\
 \hspace{.5cm}+ limits & $0.1$ & $0.6\pm 0.5$ & $33.1\pm 0.2$  \\
\hline
 
$L_{VHE}/L_{Radio}$ & $2.3\times 10^{-3}$ & $3.4\pm 0.5$ & $0.9\pm 0.2$  \\
 \hspace{.5cm}+ limits & $2.7\times 10^{-2}$ & $2.8\pm 0.6$ & $0.5\pm 0.2$ \\
\hline

\end{tabular}
\label{Table:FitValues}
\end{threeparttable}
\end{table}

In the top panel panel of Fig.~\ref{Figure:L} we show the VHE 
luminosities of shell-type and thermal-composite SNRs as a
function of source age. A linear fit in estimated source age to the data points of those SNRs detected in both the radio and the VHE bands (black points) shows no sign of
correlation (with a $p$-values of $p=0.1$ testing the null-hypothesis of a non-correlation). Also the inclusion of upper limits\footnote{Correlation testing including upper limits was performed using the Cox hazard model and the EM algorithm regression 
provided by the \texttt{asurv} package \citep{asurv},
which is available at \url{http://ascl.net/1406.001}.}
 that fall into the same age interval as the detected SNRs (<~5.1~kyrs) does not change this situation. In this figure, to be conservative, we assumed the largest distances compatible with uncertainties.

There is a large scatter in the data points that partially stems from substantial uncertainties in the distance estimates.
One way to address this problem is to look at the ratio $(L_\mathrm{VHE}/L_\mathrm{radio})$, as it eliminates this uncertainty by construction.
Indeed, the resulting values of the VHE detections show considerably less scatter in the ordinate values around the best-fit linear regression, as
can be seen in the bottom panel of Fig.~\ref{Figure:L}, where we
show $(L_\mathrm{VHE}/L_\mathrm{radio})$ versus source age.
We find evidence for correlation
for the sample of shell-type SNRs that have been detected in both the radio and VHE bands (black) with a p-value of $p=2.3 \times 10^\mathrm{-3}$.
Adding the corresponding upper limit values for SNRs that fall into the same age window as the VHE detections, i.e. with ages <~5.1~kyrs, 
weakens the correlation with age ($p=2.7 \times 10^\mathrm{-2}$). 
An inclusion of the upper limit values for the older
SNRs in the age fit results in a high probability of a non-correlation ($p=0.9$). While it is expected that the
spread in $(L_\mathrm{VHE}/L_\mathrm{radio})$ increases with source age as the diverse environment around
SNRs becomes more and more important for the SNR evolution, the sample 
of upper limit values from evolved ($\ga 10^4$yr) SNRs lies consistently below the extrapolated correlation by some orders of magnitude.
It should be stressed that the correlations found might be different for interacting SNRs, since we excluded these systems
from our test sample.

From our limited data set we find that the increase of VHE luminosity with source age is smaller than that of  $(L_\mathrm{VHE}/L_\mathrm{radio})$,
which implies that in the first several thousand years of SNR evolution the radio-synchrotron emission decreases more 
rapidly than the VHE emission increases. Therefore, although we prefer to look at the ratio
$(L_\mathrm{VHE}/L_\mathrm{radio})$ instead of the individual radio and VHE luminosities to eliminate the large distance uncertainties, 
we note that the observed correlation is mainly driven by the strong decrease in $L_\mathrm{radio}$ with time rather than the relatively constant behaviour of $L_\mathrm{VHE}$.
A time decrease of SNR luminosities at lower energies has been directly observed in Cas A for non-thermal X-ray and radio emission (see \citet{CasA_declineRadio2}, \citet{CasA_declineRadio}, \citet{CasA_Decline_Chandra} and \citet{CasA_Decline_Suzaku}).

At higher ages, there are no more VHE detections of shell-type or thermal composite SNRs,
 while synchrotron emission at radio energies continues.
The latter finding agrees with the theoretical expectation that effective particle acceleration 
to multi-TeV energies occurs mainly in young SNRs but not in more evolved systems. 
The observed behaviour of the younger sources can be interpreted by invoking the notion of magnetic field amplification at SNR shocks. 
In this theory, the amplified B-field is fuelled by a fraction of the shock-generated CR  pressure
 (see e.g. \citet{Bell2004} and \citet{Voelk2005}), and is therefore expected to decrease as the shock 
slows down with increasing source age, which in turn would lead to a decreased overall synchrotron luminosity.

\section{Conclusions}
In this work we investigated a sample of 108 Galactic SNRs, comprised
of sources that have been detected in lower energy bands, for VHE
$\gamma$-ray emission using the H.E.S.S. Phase I data coming from the
HGPS programme \citet{HGPSForth}.  For the first time, upper limits on the
integrated $\gamma$-ray flux between 1 and 10 TeV are provided for
such a large set of Galactic SNRs. We note that the presented upper limits may be useful for continuing
studies, such as in \citet{Cristofari}, where VHE data 
were compared to a SNR population synthesis model to investigate the CR 
standard paradigm.

We paid special attention to the selection of these sources to minimise a possible signal leakage from unrelated
H.E.S.S. detections (see \citet{HGPSForth}) into the analysis regions. 
We found a positive offset of the significance
distribution corresponding to a median value of $1\sigma$. To at
least a large degree, the origin of this offset can be attributed to the
diffuse Galactic TeV emission detected by H.E.S.S. We applied generic
models of the two VHE $\gamma$-ray emission processes believed to be
dominant in SNRs to the data, i.e.  the $\pi^0$-decay in a hadronic
and IC emission in a leptonic emission scenario
to place constraints on the acceleration efficiency in SNRs and
the energy content in electrons and protons above 10~TeV.  Assuming
typical parameters of the ambient gas density and the SN blast energy,
the resulting values do not contradict the standard expecation that
$\sim 10~\%$ of the SN blast energy is converted to CR in SNRs.  We also
investigated the opposite problem assuming that this canonical
paradigm is valid and put constraints on the parameter space spanned
by the ambient gas density around the shock and electron-to-proton
energy fraction.

Overall, the derived flux upper limits are not in contradiction with the canonical CR paradigm.
Assuming this paradigm holds true, we can constrain typical ambient density values around shell-type SNRs to $n\leq 7~\textrm{cm}^\textrm{-3}$ and electron-to-proton energy fractions above 10~TeV to $\epsilon_\textrm{ep} \leq 5\times 10^{-3}$.

Finally, we compared the presented flux upper limits to the flux
measurements of the seven non-interacting shell-type SNRs detected
both in the radio and VHE range. We found evidence of correlation
between the ratio of VHE $\gamma$-ray luminosity to radio luminosity,
$(L_{VHE}/L_{radio})$, and source age.  This correlation can be explained 
by invoking the theory of magnetic field amplification at SNR shocks, 
which accounts for the rapid decrease in radio luminosity by predicting
a declining magnetic field strength as the shock slows down with increasing source age. 

Further development in
the SNR population from the observational point of view should be achieved
with the next generation instrument: the CTA observatory. In this
work we have also estimated the performance of this future
observatory to probe the ambient gas density and the SN blast energy
parameter space. The results suggest that this instrument will be an
important leap forward in the investigation of the Galactic SNRs and
will likely be able to confirm or invalidate the CR paradigm.


\begin{acknowledgements}
The support of the Namibian authorities and of the University of Namibia in facilitating the construction and operation of H.E.S.S. is gratefully acknowledged, as is the support by the German Ministry for Education and Research (BMBF), the Max Planck Society, the German Research Foundation (DFG), the Alexander von Humboldt Foundation, the Deutsche Forschungsgemeinschaft, the French Ministry for Research, the CNRS-IN2P3 and the Astroparticle Interdisciplinary Programme of the CNRS, the U.K. Science and Technology Facilities Council (STFC), the IPNP of the Charles University, the Czech Science Foundation, the Polish National Science Centre, the South African Department of Science and Technology and National Research Foundation, the University of Namibia, the National Commission on Research, Science \& Technology of Namibia (NCRST), the Innsbruck University, the Austrian Science Fund (FWF), and the Austrian Federal Ministry for Science, Research and Economy, the University of Adelaide and the Australian Research Council, the Japan Society for the Promotion of Science, and the University of Amsterdam.
We appreciate the excellent work of the technical support staff in Berlin, Durham, Hamburg, Heidelberg, Palaiseau, Paris, Saclay, and in Namibia in the construction and operation of the equipment. This work benefited from services provided by the H.E.S.S. Virtual Organisation, supported by the national resource providers of the EGI Federation.

\end{acknowledgements}

\bibliographystyle{aa}
\bibliography{SNRPop}

\begin{thebibliography}{79}
\expandafter\ifx\csname natexlab\endcsname\relax\def\natexlab#1{#1}\fi

\bibitem[{{Abdalla} {et~al.}(2016{\natexlab{a}}){Abdalla}, {Abdalla},
  {Abramowski}, {Aharonian}, {Ait Benkhali}, {Akhperjanian}, {Andersson},
  {Ang{\"u}ner}, \& et~al.}]{RXJ1713New}
{Abdalla}, H., {Abdalla}, H., {Abramowski}, A., {et~al.} 2016{\natexlab{a}},
  {H.E.S.S. observations of RX J1713.7-3946 with improved angular and spectral
  resolution; evidence for gamma-ray emission extending beyond the X-ray
  emitting shell}, accepted to \aap , arXiv pre-print: 1609.08671

\bibitem[{{Abdalla} {et~al.}(2017){Abdalla}, {Abramowski}, {Aharonian}, {Ait
  Benkhali}, \& {Akhperjanian}}]{HGPSForth}
{Abdalla}, H., {Abramowski}, A., {Aharonian}, F., {Ait Benkhali}, F., \&
  {Akhperjanian}, A. 2017, {The H.E.S.S. Galactic plane survey}, \aap\ in
  preparation

\bibitem[{{Abdalla} {et~al.}(2016{\natexlab{b}}){Abdalla}, {Abramowski},
  {Aharonian}, {Ait Benkhali}, {Akhperjanian}, {Andersson}, {Ang{\"u}ner},
  {Arakawa}, {Arrieta}, \& et~al.}]{hess_velajr_paper3}
{Abdalla}, H., {Abramowski}, A., {Aharonian}, F., {et~al.} 2016{\natexlab{b}},
  {Deeper H.E.S.S. Observations of Vela Junior (RX J0852.0-4622): Morphology
  Studies and Resolved Spectroscopy}, accepted to \aap , arXiv pre-print:
  1611.01863

\bibitem[{{Abramowski} {et~al.}(2011){Abramowski}, {Acero}, {Aharonian},
  {Akhperjanian}, {Anton}, {Balzer}, {Barnacka}, {Barres de Almeida},
  {Becherini}, {Becker}, {Behera}, {Bernl{\"o}hr}, {Bochow}, {Boisson},
  {Bolmont}, {Bordas}, {Brucker}, {Brun}, {Brun}, {Bulik}, {B{\"u}sching},
  {Carrigan}, {Casanova}, {Cerruti}, {Chadwick}, {Charbonnier}, {Chaves},
  {Cheesebrough}, {Chounet}, {Clapson}, {Coignet}, {Cologna}, {Conrad},
  {Dalton}, {Daniel}, {Davids}, {Degrange}, {Deil}, {Dickinson},
  {Djannati-Ata{\"i}}, {Domainko}, {Drury}, {Dubois}, {Dubus}, {Dutson},
  {Dyks}, {Dyrda}, {Egberts}, {Eger}, {Espigat}, {Fallon}, {Farnier}, {Fegan},
  {Feinstein}, {Fernandes}, {Fiasson}, {Fontaine}, {F{\"o}rster},
  {F{\"u}{\ss}ling}, {Gallant}, {Gast}, {G{\'e}rard}, {Gerbig}, {Giebels},
  {Glicenstein}, {Gl{\"u}ck}, {Goret}, {G{\"o}ring}, {H{\"a}ffner}, {Hague},
  {Hampf}, {Hauser}, {Heinz}, {Heinzelmann}, {Henri}, {Hermann}, {Hinton},
  {Hoffmann}, {Hofmann}, {Hofverberg}, {Holler}, {Horns}, {Jacholkowska}, {de
  Jager}, {Jahn}, {Jamrozy}, {Jung}, {Kastendieck}, {Katarzy{\'n}ski}, {Katz},
  {Kaufmann}, {Keogh}, {Khangulyan}, {Kh{\'e}lifi}, {Klochkov}, {Klu{\'z}niak},
  {Kneiske}, {Komin}, {Kosack}, {Kossakowski}, {Laffon}, {Lamanna}, {Lennarz},
  {Lohse}, {Lopatin}, {Lu}, {Marandon}, {Marcowith}, {Masbou}, {Maurin},
  {Maxted}, {McComb}, {Medina}, {M{\'e}hault}, {Moderski}, {Moulin}, {Naumann},
  {Naumann-Godo}, {de Naurois}, {Nedbal}, {Nekrassov}, {Nguyen}, {Nicholas},
  {Niemiec}, {Nolan}, {Ohm}, {de O{\~n}a Wilhelmi}, {Opitz}, {Ostrowski},
  {Oya}, {Panter}, {Paz Arribas}, {Pedaletti}, {Pelletier}, {Petrucci}, {Pita},
  {P{\"u}hlhofer}, {Punch}, {Quirrenbach}, {Raue}, {Rayner}, {Reimer},
  {Reimer}, {Renaud}, {de los Reyes}, {Rieger}, {Ripken}, {Rob}, {Rosier-Lees},
  {Rowell}, {Rudak}, {Rulten}, {Ruppel}, {Ryde}, {Sahakian}, {Santangelo},
  {Schlickeiser}, {Sch{\"o}ck}, {Schulz}, {Schwanke}, {Schwarzburg},
  {Schwemmer}, {Sikora}, {Skilton}, {Sol}, {Spengler}, {Stawarz}, {Steenkamp},
  {Stegmann}, {Stinzing}, {Stycz}, {Sushch}, {Szostek}, {Tavernet}, {Terrier},
  {Tluczykont}, {Valerius}, {van Eldik}, {Vasileiadis}, {Venter}, {Vialle},
  {Viana}, {Vincent}, {V{\"o}lk}, {Volpe}, {Vorobiov}, {Vorster}, {Wagner},
  {Ward}, {White}, {Wierzcholska}, {Zacharias}, {Zajczyk}, {Zdziarski}, {Zech},
  \& {Zechlin}}]{J1731}
{Abramowski}, A., {Acero}, F., {Aharonian}, F., {et~al.} 2011, \aap, 531, A81

\bibitem[{{Abramowski} {et~al.}(2012){Abramowski}, {Acero}, {Aharonian},
  {Akhperjanian}, {Anton}, {Balzer}, {Barnacka}, {Becherini}, {Becker},
  {Bernl{\"o}hr}, {Birsin}, {Biteau}, {Bochow}, {Boisson}, {Bolmont}, {Bordas},
  {Brucker}, {Brun}, {Brun}, {Bulik}, {B{\"u}sching}, {Carrigan}, {Casanova},
  {Cerruti}, {Chadwick}, {Charbonnier}, {Chaves}, {Cheesebrough}, {Cologna},
  {Conrad}, {Couturier}, {Dalton}, {Daniel}, {Davids}, {Degrange}, {Deil},
  {Dickinson}, {Djannati-Ata{\"i}}, {Domainko}, {Drury}, {Dubus}, {Dutson},
  {Dyks}, {Dyrda}, {Egberts}, {Eger}, {Espigat}, {Fallon}, {Fegan},
  {Feinstein}, {Fernandes}, {Fiasson}, {Fontaine}, {F{\"o}rster},
  {F{\"u}{\ss}ling}, {Gajdus}, {Gallant}, {Garrigoux}, {Gast}, {G{\'e}rard},
  {Giebels}, {Glicenstein}, {Gl{\"u}ck}, {G{\"o}ring}, {Grondin},
  {H{\"a}ffner}, {Hague}, {Hahn}, {Hampf}, {Harris}, {Hauser}, {Heinz},
  {Heinzelmann}, {Henri}, {Hermann}, {Hillert}, {Hinton}, {Hofmann},
  {Hofverberg}, {Holler}, {Horns}, {Jacholkowska}, {Jahn}, {Jamrozy}, {Jung},
  {Kastendieck}, {Katarzy{\'n}ski}, {Katz}, {Kaufmann}, {Kh{\'e}lifi},
  {Klochkov}, {Klu{\'z}niak}, {Kneiske}, {Komin}, {Kosack}, {Kossakowski},
  {Krayzel}, {Laffon}, {Lamanna}, {Lenain}, {Lennarz}, {Lohse}, {Lopatin},
  {Lu}, {Marandon}, {Marcowith}, {Masbou}, {Maurin}, {Maxted}, {Mayer},
  {McComb}, {Medina}, {M{\'e}hault}, {Moderski}, {Mohamed}, {Moulin},
  {Naumann}, {Naumann-Godo}, {de Naurois}, {Nedbal}, {Nekrassov}, {Nguyen},
  {Nicholas}, {Niemiec}, {Nolan}, {Ohm}, {de O{\~n}a Wilhelmi}, {Opitz},
  {Ostrowski}, {Oya}, {Panter}, {Paz Arribas}, {Pekeur}, {Pelletier}, {Perez},
  {Petrucci}, {Peyaud}, {Pita}, {P{\"u}hlhofer}, {Punch}, {Quirrenbach},
  {Raue}, {Reimer}, {Reimer}, {Renaud}, {de los Reyes}, {Rieger}, {Ripken},
  {Rob}, {Rosier-Lees}, {Rowell}, {Rudak}, {Rulten}, {Sahakian}, {Sanchez},
  {Santangelo}, {Schlickeiser}, {Schulz}, {Schwanke}, {Schwarzburg},
  {Schwemmer}, {Sheidaei}, {Skilton}, {Sol}, {Spengler}, {Stawarz},
  {Steenkamp}, {Stegmann}, {Stinzing}, {Stycz}, {Sushch}, {Szostek},
  {Tavernet}, {Terrier}, {Tluczykont}, {Valerius}, {van Eldik}, {Vasileiadis},
  {Venter}, {Viana}, {Vincent}, {V{\"o}lk}, {Volpe}, {Vorobiov}, {Vorster},
  {Wagner}, {Ward}, {White}, {Wierzcholska}, {Zacharias}, {Zajczyk},
  {Zdziarski}, {Zech}, {Zechlin}, \& {H.~E.~S.~S.~Collaboration}}]{NGC253}
{Abramowski}, A., {Acero}, F., {Aharonian}, F., {et~al.} 2012, \apj, 757, 158

\bibitem[{{Abramowski} {et~al.}(2014{\natexlab{a}}){Abramowski}, {Aharonian},
  {Ait Benkhali}, {Akhperjanian}, {Ang{\"u}ner}, {Backes}, {Balenderan},
  {Balzer}, {Barnacka}, {Becherini}, \& et~al.}]{Diffuse}
{Abramowski}, A., {Aharonian}, F., {Ait Benkhali}, F., {et~al.}
  2014{\natexlab{a}}, \prd, 90, 122007

\bibitem[{{Abramowski} {et~al.}(2015){Abramowski}, {Aharonian}, {Ait Benkhali},
  {Akhperjanian}, {Ang{\"u}ner}, {Backes}, {Balenderan}, {Balzer}, {Barnacka},
  \& et~al.}]{G349}
{Abramowski}, A., {Aharonian}, F., {Ait Benkhali}, F., {et~al.} 2015, \aap,
  574, A100

\bibitem[{{Abramowski} {et~al.}(2016){Abramowski}, {Aharonian}, {Ait Benkhali},
  {Akhperjanian}, {Ang{\"u}ner}, {Backes}, {Balzer}, {Becherini}, {Becker
  Tjus}, \& et~al.}]{RCW86-New}
{Abramowski}, A., {Aharonian}, F., {Ait Benkhali}, F., {et~al.} 2016, {Detailed
  spectral and morphological analysis of the shell type SNR RCW 86}, accepted
  to \aap , arXiv pre-print: 1601.04461

\bibitem[{{Abramowski} {et~al.}(2014{\natexlab{b}}){Abramowski}, {Aharonian},
  {Benkhali}, {Akhperjanian}, {Ang{\"u}ner}, {Anton}, {Balenderan}, {Balzer},
  {Barnacka}, \& et~al.}]{G1_9}
{Abramowski}, A., {Aharonian}, F., {Benkhali}, F.~A., {et~al.}
  2014{\natexlab{b}}, \mnras, 441, 790

\bibitem[{{Acciari} {et~al.}(2011){Acciari}, {Aliu}, {Arlen}, {Aune},
  {Beilicke}, {Benbow}, {Bradbury}, {Buckley}, {Bugaev}, {Byrum}, {Cannon},
  {Cesarini}, {Ciupik}, {Collins-Hughes}, {Cui}, {Dickherber}, {Duke},
  {Errando}, {Finley}, {Finnegan}, {Fortson}, {Furniss}, {Galante}, {Gall},
  {Gillanders}, {Godambe}, {Griffin}, {Grube}, {Guenette}, {Gyuk}, {Hanna},
  {Holder}, {Hughes}, {Hui}, {Humensky}, {Kaaret}, {Karlsson}, {Kertzman},
  {Kieda}, {Krawczynski}, {Krennrich}, {Lang}, {LeBohec}, {Madhavan}, {Maier},
  {Majumdar}, {McArthur}, {McCann}, {Moriarty}, {Mukherjee}, {Ong}, {Orr},
  {Otte}, {Pandel}, {Park}, {Perkins}, {Pohl}, {Quinn}, {Ragan}, {Reyes},
  {Reynolds}, {Roache}, {Rose}, {Saxon}, {Schroedter}, {Sembroski}, {Senturk},
  {Slane}, {Smith}, {Te{\v s}i{\'c}}, {Theiling}, {Thibadeau}, {Tsurusaki},
  {Varlotta}, {Vassiliev}, {Vincent}, {Vivier}, {Wakely}, {Ward}, {Weekes},
  {Weinstein}, {Weisgarber}, {Williams}, {Wood}, \& {Zitzer}}]{Tycho}
{Acciari}, V.~A., {Aliu}, E., {Arlen}, T., {et~al.} 2011, \apjl, 730, L20

\bibitem[{{Acero} {et~al.}(2016){Acero}, {Ackermann}, {Ajello}, {Baldini},
  {Ballet}, {Barbiellini}, {Bastieri}, {Bellazzini}, {Bissaldi}, {Blandford},
  {Bloom}, {Bonino}, {Bottacini}, {Brandt}, {Bregeon}, {Bruel}, {Buehler},
  {Buson}, {Caliandro}, {Cameron}, {Caputo}, {Caragiulo}, {Caraveo},
  {Casandjian}, {Cavazzuti}, {Cecchi}, {Chekhtman}, {Chiang}, {Chiaro},
  {Ciprini}, {Claus}, {Cohen}, {Cohen-Tanugi}, {Cominsky}, {Condon}, {Conrad},
  {Cutini}, {D'Ammando}, {de Angelis}, {de Palma}, {Desiante}, {Digel}, {Di
  Venere}, {Drell}, {Drlica-Wagner}, {Favuzzi}, {Ferrara}, {Franckowiak},
  {Fukazawa}, {Funk}, {Fusco}, {Gargano}, {Gasparrini}, {Giglietto}, {Giommi},
  {Giordano}, {Giroletti}, {Glanzman}, {Godfrey}, {Gomez-Vargas}, {Grenier},
  {Grondin}, {Guillemot}, {Guiriec}, {Gustafsson}, {Hadasch}, {Harding},
  {Hayashida}, {Hays}, {Hewitt}, {Hill}, {Horan}, {Hou}, {Iafrate}, {Jogler},
  {J{\'o}hannesson}, {Johnson}, {Kamae}, {Katagiri}, {Kataoka}, {Katsuta},
  {Kerr}, {Kn{\"o}dlseder}, {Kocevski}, {Kuss}, {Laffon}, {Lande}, {Larsson},
  {Latronico}, {Lemoine-Goumard}, {Li}, {Li}, {Longo}, {Loparco}, {Lovellette},
  {Lubrano}, {Magill}, {Maldera}, {Marelli}, {Mayer}, {Mazziotta}, {Michelson},
  {Mitthumsiri}, {Mizuno}, {Moiseev}, {Monzani}, {Moretti}, {Morselli},
  {Moskalenko}, {Murgia}, {Nemmen}, {Nuss}, {Ohsugi}, {Omodei}, {Orienti},
  {Orlando}, {Ormes}, {Paneque}, {Perkins}, {Pesce-Rollins}, {Petrosian},
  {Piron}, {Pivato}, {Porter}, {Rain{\`o}}, {Rando}, {Razzano}, {Razzaque},
  {Reimer}, {Reimer}, {Renaud}, {Reposeur}, {Rousseau}, {Saz Parkinson},
  {Schmid}, {Schulz}, {Sgr{\`o}}, {Siskind}, {Spada}, {Spandre}, {Spinelli},
  {Strong}, {Suson}, {Tajima}, {Takahashi}, {Tanaka}, {Thayer}, {Thompson},
  {Tibaldo}, {Tibolla}, {Torres}, {Tosti}, {Troja}, {Uchiyama}, {Vianello},
  {Wells}, {Wood}, {Wood}, {Yassine}, {den Hartog}, \& {Zimmer}}]{ref:acero16}
{Acero}, F., {Ackermann}, M., {Ajello}, M., {et~al.} 2016, \apjs, 224, 8

\bibitem[{{Acero} {et~al.}(2010){Acero}, {Aharonian}, {Akhperjanian}, {Anton},
  {Barres de Almeida}, {Bazer-Bachi}, {Becherini}, {Behera}, {Beilicke},
  {Bernl{\"o}hr}, {Bochow}, {Boisson}, {Bolmont}, {Borrel}, {Brucker}, {Brun},
  {Brun}, {B{\"u}hler}, {Bulik}, {B{\"u}sching}, {Boutelier}, {Chadwick},
  {Charbonnier}, {Chaves}, {Cheesebrough}, {Conrad}, {Chounet}, {Clapson},
  {Coignet}, {Dalton}, {Daniel}, {Davids}, {Degrange}, {Deil}, {Dickinson},
  {Djannati-Ata{\"i}}, {Domainko}, {O'C.~Drury}, {Dubois}, {Dubus}, {Dyks},
  {Dyrda}, {Egberts}, {Eger}, {Espigat}, {Fallon}, {Farnier}, {Fegan},
  {Feinstein}, {Fiasson}, {F{\"o}rster}, {Fontaine}, {F{\"u}{\ss}ling},
  {Gabici}, {Gallant}, {G{\'e}rard}, {Gerbig}, {Giebels}, {Glicenstein},
  {Gl{\"u}ck}, {Goret}, {G{\"o}ring}, {Hauser}, {Hauser}, {Heinz},
  {Heinzelmann}, {Henri}, {Hermann}, {Hinton}, {Hoffmann}, {Hofmann},
  {Hofverberg}, {Holleran}, {Hoppe}, {Horns}, {Jacholkowska}, {de Jager},
  {Jahn}, {Jung}, {Katarzy{\'n}ski}, {Katz}, {Kaufmann}, {Kerschhaggl},
  {Khangulyan}, {Kh{\'e}lifi}, {Keogh}, {Klochkov}, {Klu{\'z}niak}, {Kneiske},
  {Komin}, {Kosack}, {Kossakowski}, {Lamanna}, {Lemoine-Goumard}, {Lenain},
  {Lohse}, {Marandon}, {Marcowith}, {Masbou}, {Maurin}, {McComb}, {Medina},
  {M{\'e}hault}, {Moderski}, {Moulin}, {Naumann-Godo}, {de Naurois}, {Nedbal},
  {Nekrassov}, {Nicholas}, {Niemiec}, {Nolan}, {Ohm}, {Olive}, {de O{\~n}a
  Wilhelmi}, {Orford}, {Ostrowski}, {Panter}, {Paz Arribas}, {Pedaletti},
  {Pelletier}, {Petrucci}, {Pita}, {P{\"u}hlhofer}, {Punch}, {Quirrenbach},
  {Raubenheimer}, {Raue}, {Rayner}, {Reimer}, {Renaud}, {de Los Reyes},
  {Rieger}, {Ripken}, {Rob}, {Rosier-Lees}, {Rowell}, {Rudak}, {Rulten},
  {Ruppel}, {Ryde}, {Sahakian}, {Santangelo}, {Schlickeiser}, {Sch{\"o}ck},
  {Sch{\"o}nwald}, {Schwanke}, {Schwarzburg}, {Schwemmer}, {Shalchi}, {Sushch},
  {Sikora}, {Skilton}, {Sol}, {Stawarz}, {Steenkamp}, {Stegmann}, {Stinzing},
  {Superina}, {Szostek}, {Tam}, {Tavernet}, {Terrier}, {Tibolla}, {Tluczykont},
  {van Eldik}, {Vasileiadis}, {Venter}, {Venter}, {Vialle}, {Vincent}, {Vink},
  {Vivier}, {V{\"o}lk}, {Volpe}, {Vorobiov}, {Wagner}, {Ward}, {Zdziarski},
  {Zech}, \& {H.E.S.S.~Collaboration}}]{SN1006}
{Acero}, F., {Aharonian}, F., {Akhperjanian}, A.~G., {et~al.} 2010, \aap, 516,
  A62

\bibitem[{{Acero} {et~al.}(2009){Acero}, {Ballet}, {Decourchelle},
  {Lemoine-Goumard}, {Ortega}, {Giacani}, {Dubner}, \&
  {Cassam-Chena{\"i}}}]{RXJ1713_Acero}
{Acero}, F., {Ballet}, J., {Decourchelle}, A., {et~al.} 2009, \aap, 505, 157

\bibitem[{{Actis} {et~al.}(2011){Actis}, {Agnetta}, {Aharonian},
  {Akhperjanian}, {Aleksi{\'c}}, {Aliu}, {Allan}, {Allekotte}, {Antico},
  {Antonelli}, \& et~al.}]{CTA}
{Actis}, M., {Agnetta}, G., {Aharonian}, F., {et~al.} 2011, Experimental
  Astronomy, 32, 193

\bibitem[{{Aharonian} {et~al.}(2006){Aharonian}, {A. G. Akhperjanian}, {A. R.
  Bazer-Bachi}, {M. Beilicke}, {W. Benbow}, {D. Berge}, {K. Bernl\"ohr}, {C.
  Boisson}, {O. Bolz}, {V. Borrel}, {I. Braun}, {F. Breitling}, {A. M. Brown},
  {R. B\"uhler}, {I. B\"usching}, {S. Carrigan}, {P. M. Chadwick}, {L.-M.
  Chounet}, {R. Cornils}, {L. Costamante}, {B. Degrange}, {H. J. Dickinson},
  {A. Djannati-Ata\"{\i}}, {L. O'C. Drury}, {G. Dubus}, {K. Egberts}, {D.
  Emmanoulopoulos}, {P. Espigat}, {F. Feinstein}, {E. Ferrero}, {A. Fiasson},
  {G. Fontaine}, {Seb. Funk}, {S. Funk}, {Y. A. Gallant}, {B. Giebels}, {J. F.
  Glicenstein}, {P. Goret}, {C. Hadjichristidis}, {D. Hauser}, {M. Hauser}, {G.
  Heinzelmann}, {G. Henri}, {G. Hermann}, {J. A. Hinton}, {W. Hofmann}, {M.
  Holleran}, {D. Horns}, {A. Jacholkowska}, {O. C. de Jager}, {B. Kh\'elifi},
  {Nu. Komin}, {A. Konopelko}, {K. Kosack}, {I. J. Latham}, {R. Le Gallou}, {A.
  Lemi\`ere}, {M. Lemoine-Goumard}, {T. Lohse}, {J. M. Martin}, {O.
  Martineau-Huynh}, {A. Marcowith}, {C. Masterson}, {T. J. L. McComb}, {M. de
  Naurois}, {D. Nedbal}, {S. J. Nolan}, {A. Noutsos}, {K. J. Orford}, {J. L.
  Osborne}, {M. Ouchrif}, {M. Panter}, {G. Pelletier}, {S. Pita}, {G.
  P\"uhlhofer}, {M. Punch}, {B. C. Raubenheimer}, {M. Raue}, {S. M. Rayner},
  {A. Reimer}, {O. Reimer}, {J. Ripken}, {L. Rob}, {L. Rolland}, {G. Rowell},
  {V. Sahakian}, {L. Saug\'e}, {S. Schlenker}, {R. Schlickeiser}, {U.
  Schwanke}, {H. Sol}, {D. Spangler}, {F. Spanier}, {R. Steenkamp}, {C.
  Stegmann}, {G. Superina}, {J.-P. Tavernet}, {R. Terrier}, {C. G. Th\'eoret},
  {M. Tluczykont}, {C. van Eldik}, {G. Vasileiadis}, {C. Venter}, {P. Vincent},
  {H. J. V\"olk}, {S. J. Wagner}, \& {M. Ward}}]{Crabpaper}
{Aharonian}, {A. G. Akhperjanian}, {A. R. Bazer-Bachi}, {et~al.} 2006, A\&A,
  457, 899

\bibitem[{{Aharonian} {et~al.}(2008{\natexlab{a}}){Aharonian}, {Akhperjanian},
  {Barres de Almeida}, {Bazer-Bachi}, {Behera}, {Beilicke}, {Benbow}, {Berge},
  {Bernl{\"o}hr}, {Boisson}, {Bolz}, {Borrel}, {Braun}, {Brion}, {Brucker},
  {B{\"u}hler}, {Bulik}, {B{\"u}sching}, {Boutelier}, {Carrigan}, {Chadwick},
  {Chounet}, {Clapson}, {Coignet}, {Cornils}, {Costamante}, {Dalton},
  {Degrange}, {Dickinson}, {Djannati-Ata{\"i}}, {Domainko}, {O'C.~Drury},
  {Dubois}, {Dubus}, {Dyks}, {Egberts}, {Emmanoulopoulos}, {Espigat},
  {Farnier}, {Feinstein}, {Fiasson}, {F{\"o}rster}, {Fontaine},
  {F{\"u}{\ss}ling}, {Gallant}, {Giebels}, {Glicenstein}, {Gl{\"u}ck}, {Goret},
  {Hadjichristidis}, {Hauser}, {Hauser}, {Heinzelmann}, {Henri}, {Hermann},
  {Hinton}, {Hoffmann}, {Hofmann}, {Holleran}, {Hoppe}, {Horns},
  {Jacholkowska}, {de Jager}, {Jung}, {Katarzy{\'n}ski}, {Kendziorra},
  {Kerschhaggl}, {Kh{\'e}lifi}, {Keogh}, {Komin}, {Kosack}, {Lamanna},
  {Latham}, {Lemoine-Goumard}, {Lenain}, {Lohse}, {Martin}, {Martineau-Huynh},
  {Marcowith}, {Masterson}, {Maurin}, {McComb}, {Moderski}, {Moulin},
  {Naumann-Godo}, {de Naurois}, {Nedbal}, {Nekrassov}, {Nolan}, {Ohm}, {Olive},
  {de O{\~n}a Wilhelmi}, {Orford}, {Osborne}, {Ostrowski}, {Panter},
  {Pedaletti}, {Pelletier}, {Petrucci}, {Pita}, {P{\"u}hlhofer}, {Punch},
  {Quirrenbach}, {Raubenheimer}, {Raue}, {Rayner}, {Renaud}, {Ripken}, {Rob},
  {Rosier-Lees}, {Rowell}, {Rudak}, {Ruppel}, {Sahakian}, {Santangelo},
  {Schlickeiser}, {Sch{\"o}ck}, {Schr{\"o}der}, {Schwanke}, {Schwarzburg},
  {Schwemmer}, {Shalchi}, {Sol}, {Spangler}, {Stawarz}, {Steenkamp},
  {Stegmann}, {Superina}, {Tam}, {Tavernet}, {Terrier}, {van Eldik},
  {Vasileiadis}, {Venter}, {Vialle}, {Vincent}, {Vivier}, {V{\"o}lk}, {Volpe},
  {Wagner}, {Ward}, {Zdziarski}, \& {Zech}}]{Kepler}
{Aharonian}, F., {Akhperjanian}, A.~G., {Barres de Almeida}, U., {et~al.}
  2008{\natexlab{a}}, \aap, 488, 219

\bibitem[{{Aharonian} {et~al.}(2008{\natexlab{b}}){Aharonian}, {Akhperjanian},
  {Bazer-Bachi}, {Behera}, {Beilicke}, {Benbow}, {Berge}, {Bernl{\"o}hr},
  {Boisson}, {Bolz}, {Borrel}, {Braun}, {Brion}, {Brown}, {B{\"u}hler},
  {Bulik}, {B{\"u}sching}, {Boutelier}, {Carrigan}, {Chadwick}, {Chounet},
  {Clapson}, {Coignet}, {Cornils}, {Costamante}, {Degrange}, {Dickinson},
  {Djannati-Ata{\"i}}, {Domainko}, {O'C.~Drury}, {Dubus}, {Dyks}, {Egberts},
  {Emmanoulopoulos}, {Espigat}, {Farnier}, {Feinstein}, {Fiasson},
  {F{\"o}rster}, {Fontaine}, {Fukui}, {Funk}, {Funk}, {F{\"u}{\ss}ling},
  {Gallant}, {Giebels}, {Glicenstein}, {Gl{\"u}ck}, {Goret}, {Hadjichristidis},
  {Hauser}, {Hauser}, {Heinzelmann}, {Henri}, {Hermann}, {Hinton}, {Hoffmann},
  {Hofmann}, {Holleran}, {Hoppe}, {Horns}, {Jacholkowska}, {de Jager},
  {Kendziorra}, {Kerschhaggl}, {Kh{\'e}lifi}, {Komin}, {Kosack}, {Lamanna},
  {Latham}, {Le Gallou}, {Lemi{\`e}re}, {Lemoine-Goumard}, {Lenain}, {Lohse},
  {Martin}, {Martineau-Huynh}, {Marcowith}, {Masterson}, {Maurin}, {McComb},
  {Moderski}, {Moriguchi}, {Moulin}, {de Naurois}, {Nedbal}, {Nolan}, {Olive},
  {Orford}, {Osborne}, {Ostrowski}, {Panter}, {Pedaletti}, {Pelletier},
  {Petrucci}, {Pita}, {P{\"u}hlhofer}, {Punch}, {Ranchon}, {Raubenheimer},
  {Raue}, {Rayner}, {Reimer}, {Renaud}, {Ripken}, {Rob}, {Rolland},
  {Rosier-Lees}, {Rowell}, {Rudak}, {Ruppel}, {Sahakian}, {Santangelo},
  {Saug{\'e}}, {Schlenker}, {Schlickeiser}, {Schr{\"o}der}, {Schwanke},
  {Schwarzburg}, {Schwemmer}, {Shalchi}, {Sol}, {Spangler}, {Stawarz},
  {Steenkamp}, {Stegmann}, {Superina}, {Takeuchi}, {Tam}, {Tavernet},
  {Terrier}, {van Eldik}, {Vasileiadis}, {Venter}, {Vialle}, {Vincent},
  {Vivier}, {V{\"o}lk}, {Volpe}, {Wagner}, \& {Ward}}]{w28H}
{Aharonian}, F., {Akhperjanian}, A.~G., {Bazer-Bachi}, A.~R., {et~al.}
  2008{\natexlab{b}}, \aap, 481, 401

\bibitem[{{Aharonian}(2004)}]{Aharonian2004}
{Aharonian}, F.~A. 2004, {Very high energy cosmic gamma radiation : a crucial
  window on the extreme Universe}

\bibitem[{{Ahnen} {et~al.}(2017){Ahnen}, {Ansoldi}, {Antonelli}, {Arcaro},
  {Babi{\'c}}, {Banerjee}, {Bangale}, {Barres de Almeida}, {Barrio}, {Becerra
  Gonz{\'a}lez}, {Bednarek}, {Bernardini}, {Berti}, {Bhattacharyya},
  {Biasuzzi}, {Biland}, {Blanch}, {Bonnefoy}, {Bonnoli}, {Carosi}, {Carosi},
  {Chatterjee}, {Colak}, {Colin}, {Colombo}, {Contreras}, {Cortina}, {Covino},
  {Cumani}, {Da Vela}, {Dazzi}, {De Angelis}, {De Lotto}, {de O{\~n}a
  Wilhelmi}, {Di Pierro}, {Doert}, {Dom{\'{\i}}nguez}, {Dominis Prester},
  {Dorner}, {Doro}, {Einecke}, {Eisenacher Glawion}, {Elsaesser},
  {Engelkemeier}, {Fallah Ramazani}, {Fern{\'a}ndez-Barral}, {Fidalgo},
  {Fonseca}, {Font}, {Fruck}, {Galindo}, {Garc{\'{\i}}a L{\'o}pez},
  {Garczarczyk}, {Gaug}, {Giammaria}, {Godinovi{\'c}}, {Gora}, {Guberman},
  {Hadasch}, {Hahn}, {Hassan}, {Hayashida}, {Herrera}, {Hose}, {Hrupec},
  {Inada}, {Ishio}, {Konno}, {Kubo}, {Kushida}, {Kuve{\v z}di{\'c}}, {Lelas},
  {Lindfors}, {Lombardi}, {Longo}, {L{\'o}pez}, {Maggio}, {Majumdar},
  {Makariev}, {Maneva}, {Manganaro}, {Mannheim}, {Maraschi}, {Mariotti},
  {Mart{\'{\i}}nez}, {Mazin}, {Menzel}, {Minev}, {Mirzoyan}, {Moralejo},
  {Moreno}, {Moretti}, {Neustroev}, {Niedzwiecki}, {Nievas Rosillo}, {Nilsson},
  {Ninci}, {Nishijima}, {Noda}, {Nogu{\'e}s}, {Paiano}, {Palacio}, {Paneque},
  {Paoletti}, {Paredes}, {Pedaletti}, {Peresano}, {Perri}, {Persic}, {Prada
  Moroni}, {Prandini}, {Puljak}, {Garcia}, {Reichardt}, {Rhode}, {Rib{\'o}},
  {Rico}, {Righi}, {Saito}, {Satalecka}, {Schroeder}, {Schweizer}, {Shore},
  {Sitarek}, {{\v S}nidari{\'c}}, {Sobczynska}, {Stamerra}, {Strzys},
  {Suri{\'c}}, {Takalo}, {Tavecchio}, {Temnikov}, {Terzi{\'c}}, {Tescaro},
  {Teshima}, {Torres-Alb{\`a}}, {Treves}, {Vanzo}, {Vazquez Acosta}, {Vovk},
  {Ward}, {Will}, \& {Zari{\'c}}}]{CasA_new}
{Ahnen}, M.~L., {Ansoldi}, S., {Antonelli}, L.~A., {et~al.} 2017, ArXiv
  e-print: 1707.01583

\bibitem[{{Allen} {et~al.}(2015){Allen}, {Chow}, {DeLaney}, {Filipovi{\'c}},
  {Houck}, {Pannuti}, \& {Stage}}]{VelaJr-Allen}
{Allen}, G.~E., {Chow}, K., {DeLaney}, T., {et~al.} 2015, \apj, 798, 82

\bibitem[{{Auchettl} {et~al.}(2015){Auchettl}, {Slane}, {Castro}, {Foster}, \&
  {Smith}}]{G290-1-Auchettl}
{Auchettl}, K., {Slane}, P., {Castro}, D., {Foster}, A.~R., \& {Smith}, R.~K.
  2015, \apj, 810, 43

\bibitem[{{Bell}(2004)}]{Bell2004}
{Bell}, A.~R. 2004, \mnras, 353, 550

\bibitem[{{Berge} {et~al.}(2007){Berge}, {Funk}, \& {Hinton}}]{ref:bgmodelling}
{Berge}, D., {Funk}, S., \& {Hinton}, J. 2007, \aap, 466, 1219

\bibitem[{{Broersen} {et~al.}(2014){Broersen}, {Chiotellis}, {Vink}, \&
  {Bamba}}]{RCW86-Broersen}
{Broersen}, S., {Chiotellis}, A., {Vink}, J., \& {Bamba}, A. 2014, \mnras, 441,
  3040

\bibitem[{{Broersen} \& {Vink}(2015)}]{G53-6-Broersen}
{Broersen}, S. \& {Vink}, J. 2015, \mnras, 446, 3885

\bibitem[{{Busser} {et~al.}(1996){Busser}, {Egger}, \&
  {Aschenbach}}]{G299-2-Busser}
{Busser}, J.-U., {Egger}, R., \& {Aschenbach}, B. 1996, \aap, 310, L1

\bibitem[{{Carlton} {et~al.}(2011){Carlton}, {Borkowski}, {Reynolds}, {Hwang},
  {Petre}, {Green}, {Krishnamurthy}, \& {Willett}}]{G1-9-Carlton}
{Carlton}, A.~K., {Borkowski}, K.~J., {Reynolds}, S.~P., {et~al.} 2011, \apjl,
  737, L22

\bibitem[{{Cassam-Chena{\"i}} {et~al.}(2004){Cassam-Chena{\"i}},
  {Decourchelle}, {Ballet}, {Sauvageot}, {Dubner}, \&
  {Giacani}}]{RXJ1713-Cassam}
{Cassam-Chena{\"i}}, G., {Decourchelle}, A., {Ballet}, J., {et~al.} 2004, \aap,
  427, 199

\bibitem[{{Castro} {et~al.}(2011){Castro}, {Slane}, {Gaensler}, {Hughes}, \&
  {Patnaude}}]{G296-1-Castro}
{Castro}, D., {Slane}, P.~O., {Gaensler}, B.~M., {Hughes}, J.~P., \&
  {Patnaude}, D.~J. 2011, \apj, 734, 86

\bibitem[{{Chen} {et~al.}(2008){Chen}, {Seward}, {Sun}, \& {Li}}]{G327-4-Chen}
{Chen}, Y., {Seward}, F.~D., {Sun}, M., \& {Li}, J.-t. 2008, \apj, 676, 1040

\bibitem[{{Cristofari} {et~al.}(2013){Cristofari}, {Gabici}, {Casanova},
  {Terrier}, \& {Parizot}}]{Cristofari}
{Cristofari}, P., {Gabici}, S., {Casanova}, S., {Terrier}, R., \& {Parizot}, E.
  2013, \mnras, 434, 2748

\bibitem[{{de Naurois} \& {Rolland}(2009)}]{Modpp}
{de Naurois}, M. \& {Rolland}, L. 2009, Astroparticle Physics, 32, 231

\bibitem[{{DeLaney} \& {Rudnick}(2003)}]{CasA-DeLaney}
{DeLaney}, T. \& {Rudnick}, L. 2003, \apj, 589, 818

\bibitem[{{Dent} {et~al.}(1974){Dent}, {Aller}, \&
  {Olsen}}]{CasA_declineRadio2}
{Dent}, W.~A., {Aller}, H.~D., \& {Olsen}, E.~T. 1974, \apjl, 188, L11

\bibitem[{{Ferrand} \& {Safi-Harb}(2012)}]{SNRCAT}
{Ferrand}, G. \& {Safi-Harb}, S. 2012, Advances in Space Research, 49, 1313

\bibitem[{{Fukuda} {et~al.}(2014){Fukuda}, {Yoshiike}, {Sano}, {Torii},
  {Yamamoto}, {Acero}, \& {Fukui}}]{Fukuda}
{Fukuda}, T., {Yoshiike}, S., {Sano}, H., {et~al.} 2014, \apj, 788, 94

\bibitem[{{Fukui} {et~al.}(2012){Fukui}, {Sano}, {Sato}, {Torii}, {Horachi},
  {Hayakawa}, {McClure-Griffiths}, {Rowell}, {Inoue}, {Inutsuka}, {Kawamura},
  {Yamamoto}, {Okuda}, {Mizuno}, {Onishi}, {Mizuno}, \& {Ogawa}}]{Fukui}
{Fukui}, Y., {Sano}, H., {Sato}, J., {et~al.} 2012, \apj, 746, 82

\bibitem[{{Green}(2014)}]{GreenCatalogue}
{Green}, D.~A. 2014, {`A Catalogue of Galactic Supernova Remnants (2014 May
  version)', Cavendish Laboratory, Cambridge, United Kingdom (available at
  "http://www.mrao.cam.ac.uk/surveys/snrs/")}

\bibitem[{{Hahn} {et~al.}(2014){Hahn}, {de los Reyes}, {Bernl{\"o}hr},
  {Kr{\"u}ger}, {Lo}, {Chadwick}, {Daniel}, {Deil}, {Gast}, {Kosack}, \&
  {Marandon}}]{2014APh....54...25H}
{Hahn}, J., {de los Reyes}, R., {Bernl{\"o}hr}, K., {et~al.} 2014,
  Astroparticle Physics, 54, 25

\bibitem[{{Huang} {et~al.}(2014){Huang}, {Wu}, {Hui}, {Seo}, {Trepl}, \&
  {Kong}}]{G38-7-Huang}
{Huang}, R.~H.~H., {Wu}, J.~H.~K., {Hui}, C.~Y., {et~al.} 2014, \apj, 785, 118

\bibitem[{{Hui} \& {Becker}(2009)}]{G67-7-Hui}
{Hui}, C.~Y. \& {Becker}, W. 2009, \aap, 494, 1005

\bibitem[{{Hui} {et~al.}(2012){Hui}, {Seo}, {Huang}, {Trepl}, {Woo}, {Lu},
  {Kong}, \& {Walter}}]{G308-4-Hui}
{Hui}, C.~Y., {Seo}, K.~A., {Huang}, R.~H.~H., {et~al.} 2012, \apj, 750, 7

\bibitem[{{Hwang} \& {Laming}(2003)}]{Hwang}
{Hwang}, U. \& {Laming}, J.~M. 2003, \apj, 597, 362

\bibitem[{{Kafexhiu} {et~al.}(2014){Kafexhiu}, {Aharonian}, {Taylor}, \&
  {Vila}}]{Kafexhiu2014}
{Kafexhiu}, E., {Aharonian}, F., {Taylor}, A.~M., \& {Vila}, G.~S. 2014, \prd,
  90, 123014

\bibitem[{{Lavalley} {et~al.}(1992){Lavalley}, {Isobe}, \& {Feigelson}}]{asurv}
{Lavalley}, M., {Isobe}, T., \& {Feigelson}, E. 1992, 25, 245

\bibitem[{{Lee} {et~al.}(2014){Lee}, {Park}, {Hughes}, \& {Slane}}]{CasA-Lee}
{Lee}, J.-J., {Park}, S., {Hughes}, J.~P., \& {Slane}, P.~O. 2014, \apj, 789, 7

\bibitem[{{Li} \& {Ma}(1983)}]{LiMa}
{Li}, T.-P. \& {Ma}, Y.-Q. 1983, \apj, 272, 317

\bibitem[{{Long} {et~al.}(1991){Long}, {Blair}, {Matsui}, \&
  {White}}]{G53-6-Long}
{Long}, K.~S., {Blair}, W.~P., {Matsui}, Y., \& {White}, R.~L. 1991, \apj, 373,
  567

\bibitem[{{Lovchinsky} {et~al.}(2011){Lovchinsky}, {Slane}, {Gaensler},
  {Hughes}, {Ng}, {Lazendic}, {Gelfand}, \& {Brogan}}]{G350-1-Lovchinsky}
{Lovchinsky}, I., {Slane}, P., {Gaensler}, B.~M., {et~al.} 2011, \apj, 731, 70

\bibitem[{{Malkov} \& {Drury}(2001)}]{MalkovDrury}
{Malkov}, M.~A. \& {Drury}, L.~O. 2001, Reports on Progress in Physics, 64, 429

\bibitem[{{Minami} {et~al.}(2013){Minami}, {Ota}, {Yamauchi}, \&
  {Koyama}}]{G355-6-Minami}
{Minami}, S., {Ota}, N., {Yamauchi}, S., \& {Koyama}, K. 2013, \pasj, 65

\bibitem[{{Moriguchi} {et~al.}(2001){Moriguchi}, {Yamaguchi}, {Onishi},
  {Mizuno}, \& {Fukui}}]{Moriguchi}
{Moriguchi}, Y., {Yamaguchi}, N., {Onishi}, T., {Mizuno}, A., \& {Fukui}, Y.
  2001, \pasj, 53, 1025

\bibitem[{{Ohm} {et~al.}(2009){Ohm}, {van Eldik}, \& {Egberts}}]{TMVA}
{Ohm}, S., {van Eldik}, C., \& {Egberts}, K. 2009, Astroparticle Physics, 31,
  383

\bibitem[{{Pannuti} {et~al.}(2014){Pannuti}, {Rho}, {Heinke}, \&
  {Moffitt}}]{G311-5-Pannuti}
{Pannuti}, T.~G., {Rho}, J., {Heinke}, C.~O., \& {Moffitt}, W.~P. 2014, \aj,
  147, 55

\bibitem[{{Park}(2015)}]{Tycho_Park}
{Park}, N. e.~a. 2015, ArXiv e-print: 1508.07068

\bibitem[{{Park} {et~al.}(2009){Park}, {Kargaltsev}, {Pavlov}, {Mori}, {Slane},
  {Hughes}, {Burrows}, \& {Garmire}}]{330-2-Park}
{Park}, S., {Kargaltsev}, O., {Pavlov}, G.~G., {et~al.} 2009, \apj, 695, 431

\bibitem[{{Patnaude} {et~al.}(2011){Patnaude}, {Vink}, {Laming}, \&
  {Fesen}}]{CasA_Decline_Chandra}
{Patnaude}, D.~J., {Vink}, J., {Laming}, J.~M., \& {Fesen}, R.~A. 2011, \apjl,
  729, L28

\bibitem[{{Prinz} \& {Becker}(2012)}]{G308-4-Hui-Prinz}
{Prinz}, T. \& {Becker}, W. 2012, \aap, 544, A7

\bibitem[{{Prinz} \& {Becker}(2013)}]{G296-7-Prinz}
{Prinz}, T. \& {Becker}, W. 2013, \aap, 550, A33

\bibitem[{{Ptuskin} \& {Zirakashvili}(2005)}]{Ptuskin2005}
{Ptuskin}, V.~S. \& {Zirakashvili}, V.~N. 2005, \aap, 429, 755

\bibitem[{{Rakowski} {et~al.}(2006){Rakowski}, {Badenes}, {Gaensler},
  {Gelfand}, {Hughes}, \& {Slane}}]{G337-2-Rakowski}
{Rakowski}, C.~E., {Badenes}, C., {Gaensler}, B.~M., {et~al.} 2006, \apj, 646,
  982

\bibitem[{{Rakowski} {et~al.}(2001){Rakowski}, {Hughes}, \&
  {Slane}}]{G309-2-Rakowski}
{Rakowski}, C.~E., {Hughes}, J.~P., \& {Slane}, P. 2001, \apj, 548, 258

\bibitem[{{Reynolds} {et~al.}(2013){Reynolds}, {Loi}, {Murphy}, {Miller},
  {Maitra}, {G{\"u}ltekin}, {Gehrels}, {Kennea}, {Siegel}, {Gelbord}, {Kuin},
  {Moss}, {Reeves}, {Robbins}, {Gaensler}, {Reis}, \&
  {Petre}}]{G306-3-Reynolds}
{Reynolds}, M.~T., {Loi}, S.~T., {Murphy}, T., {et~al.} 2013, \apj, 766, 112

\bibitem[{{Reynolds} {et~al.}(2008){Reynolds}, {Borkowski}, {Green}, {Hwang},
  {Harrus}, \& {Petre}}]{G1-9-Reynolds}
{Reynolds}, S.~P., {Borkowski}, K.~J., {Green}, D.~A., {et~al.} 2008, \apjl,
  680, L41

\bibitem[{{Reynolds} {et~al.}(2006){Reynolds}, {Borkowski}, {Hwang}, {Harrus},
  {Petre}, \& {Dubner}}]{G15-9-Reynolds}
{Reynolds}, S.~P., {Borkowski}, K.~J., {Hwang}, U., {et~al.} 2006, \apjl, 652,
  L45

\bibitem[{Rolke {et~al.}(2005)Rolke, L\'opez, \& Conrad}]{Rolke}
Rolke, W.~A., L\'opez, A.~M., \& Conrad, J. 2005, Nuclear Instruments and
  Methods in Physics Research Section A: Accelerators, Spectrometers, Detectors
  and Associated Equipment, 551, 493

\bibitem[{{S{\'a}nchez-Ayaso} {et~al.}(2012){S{\'a}nchez-Ayaso}, {Combi},
  {Albacete Colombo}, {L{\'o}pez-Santiago}, {Mart{\'{\i}}}, \&
  {Mu{\~n}oz-Arjonilla}}]{G296-8-Sanchez}
{S{\'a}nchez-Ayaso}, E., {Combi}, J.~A., {Albacete Colombo}, J.~F., {et~al.}
  2012, \apss, 337, 573

\bibitem[{{S{\'a}nchez-Ayaso} {et~al.}(2013){S{\'a}nchez-Ayaso}, {Combi},
  {Bocchino}, {Albacete-Colombo}, {L{\'o}pez-Santiago}, {Mart{\'{\i}}}, \&
  {Castro}}]{G272-2-Sanchez}
{S{\'a}nchez-Ayaso}, E., {Combi}, J.~A., {Bocchino}, F., {et~al.} 2013, \aap,
  552, A52

\bibitem[{{Sano} {et~al.}(2015){Sano}, {Fukuda}, {Yoshiike}, {Sato}, {Horachi},
  {Kuwahara}, {Torii}, {Hayakawa}, {Tanaka}, {Matsumoto}, {Inoue}, {Yamazaki},
  {Inutsuka}, {Kawamura}, {Yamamoto}, {Okuda}, {Tachihara}, {Mizuno}, {Onishi},
  {Mizuno}, {Acero}, \& {Fukui}}]{Sano}
{Sano}, H., {Fukuda}, T., {Yoshiike}, S., {et~al.} 2015, \apj, 799, 175

\bibitem[{Sato {et~al.}(2017)Sato, Maeda, Bamba, Katsuda, Ohira, Yamazaki,
  Masai, Matsumoto, Sawada, Terada, Hughes, \& Ishida}]{CasA_Decline_Suzaku}
Sato, T., Maeda, Y., Bamba, A., {et~al.} 2017, The Astrophysical Journal, 836,
  225

\bibitem[{{Slane} {et~al.}(2014){Slane}, {Lee}, {Ellison}, {Patnaude},
  {Hughes}, {Eriksen}, {Castro}, \& {Nagataki}}]{Tycho-Slane}
{Slane}, P., {Lee}, S.-H., {Ellison}, D.~C., {et~al.} 2014, \apj, 783, 33

\bibitem[{{Tian} {et~al.}(2007){Tian}, {Haverkorn}, \& {Zhang}}]{351-7-Tian}
{Tian}, W.~W., {Haverkorn}, M., \& {Zhang}, H.~Y. 2007, \mnras, 378, 1283

\bibitem[{{Vink}(2008)}]{Kepler-Vink}
{Vink}, J. 2008, \apj, 689, 231

\bibitem[{Vinyaikin(2014)}]{CasA_declineRadio}
Vinyaikin, E.~N. 2014, Astronomy Reports, 58, 626

\bibitem[{{V{\"o}lk} {et~al.}(2005){V{\"o}lk}, {Berezhko}, \&
  {Ksenofontov}}]{Voelk2005}
{V{\"o}lk}, H.~J., {Berezhko}, E.~G., \& {Ksenofontov}, L.~T. 2005, \aap, 433,
  229

\bibitem[{{V{\"o}lk} {et~al.}(2002){V{\"o}lk}, {Berezhko}, {Ksenofontov}, \&
  {Rowell}}]{Tycho-Voelk}
{V{\"o}lk}, H.~J., {Berezhko}, E.~G., {Ksenofontov}, L.~T., \& {Rowell}, G.~P.
  2002, \aap, 396, 649

\bibitem[{{Winkler} {et~al.}(2014){Winkler}, {Williams}, {Reynolds}, {Petre},
  {Long}, {Katsuda}, \& {Hwang}}]{SN1006-Winkler}
{Winkler}, P.~F., {Williams}, B.~J., {Reynolds}, S.~P., {et~al.} 2014, \apj,
  781, 65

\bibitem[{{Yoshita} {et~al.}(2000){Yoshita}, {Miyata}, \&
  {Tsunemi}}]{G69-7-Yoshita}
{Yoshita}, K., {Miyata}, E., \& {Tsunemi}, H. 2000, \pasj, 52, 867

\bibitem[{{Zoglauer} {et~al.}(2015){Zoglauer}, {Reynolds}, {An}, {Boggs},
  {Christensen}, {Craig}, {Fryer}, {Grefenstette}, {Harrison}, {Hailey},
  {Krivonos}, {Madsen}, {Miyasaka}, {Stern}, \& {Zhang}}]{G1-9-Zoglauer}
{Zoglauer}, A., {Reynolds}, S.~P., {An}, H., {et~al.} 2015, \apj, 798, 98

\end{thebibliography}

\onecolumn
\begin{landscape}
\begin{center}
{\fontsize{9.2}{8.7}\selectfont
 \begin{longtable}{l | c c c c c c c c | c c c c}
 \captionsetup{width=1.2\textwidth}
\hline
Source & $l$ & $b$ & $R_{\mathrm{ON}}$ & Type & Distance & Live time & Av. zenith & Closest detection & $F^\mathrm{ul}$[1-10 TeV] & \multicolumn{1}{c}{$\sigma$} &  ($W_\mathrm{p}\times n$)$^\mathrm{ul}$[>10 TeV] & $W^\mathrm{ul} _\mathrm{e}$[>10 TeV] \\
 &  (\dgr) & (\dgr) & (\dgr) & & (kpc) & (h) & (\dgr) & / Angular distance (\dgr) & ($\times 10^{-13}$cm$^{-2}$s$^{-1}$) & & ($\times 10^{48}$erg)& ($\times 10^{45}$erg) \\
\hline
\endfirsthead

\hline
Source & $l$ & $b$ & $R_{\mathrm{ON}}$  & Type & Distance & Live time & Av. zenith & Closest detection & $F^\mathrm{ul}$[1-10 TeV] & \multicolumn{1}{c}{$\sigma$} &  ($W_\mathrm{p}\times n$)$^\mathrm{ul}$[>10 TeV] & $W^\mathrm{ul} _\mathrm{e}$[>10 TeV] \\
&  (\dgr) & (\dgr) &  (\dgr) & & (kpc) & (h) & (\dgr) & / Angular distance (\dgr) & ($\times 10^{-13}$cm$^{-2}$s$^{-1}$) & & ($\times 10^{48}$erg)& ($\times 10^{45}$erg) \\

\hline
\endhead

G3.7$-$0.2 & 3.78 & -0.28 & 0.22 & S & - & 24.3 & 20.3 & HESS J1747-248 / 2.0 & 3.7 & 2.8 & - & - \\ 
G3.8+0.3 & 3.81 & 0.39 & 0.25 & S? & - & 37.6 & 20.8 & HESS J1747-248 / 1.3 & 2.6 & 1.7 & - & - \\ 
G5.2$-$2.6 & 5.20 & -2.60 & 0.25 & S & - & 1.5 & 17.5 & HESS J1800-240 / 2.3 & 7.1 & -0.8 & - & - \\ 
G5.4$-$1.2 & 5.35 & -1.13 & 0.39 & C(p)?* & 4.4 & 16.7 & 18.5 & HESS J1800-240 / 0.9 & 7.0 & 1.8 & 29.6 & 23.3\\ 
G5.9+3.1 & 5.90 & 3.13 & 0.27 & S & - & 7.5 & 21.1 & HESS J1747-248 / 2.5 & 5.7 & 0.2 & - & - \\ 
G6.1+0.5 & 6.10 & 0.53 & 0.25 & S & - & 17.7 & 19.1 & HESS J1800-240 / 1.0 & 2.9 & 0.3 & - & - \\ 
G7.2+0.2 & 7.20 & 0.20 & 0.20 & S & - & 19.1 & 18.9 & HESS J1804-216 / 1.2 & 3.2 & 2.5 & - & - \\ 
G7.5$-$1.7 & 7.54 & -1.90 & 0.51 & C(t\&p) & 1.85 & 8.3 & 18.2 & HESS J1804-216 / 2.0 & 9.5 & 0.6 & 7.1 & 5.6\\ 
G7.7$-$3.7 & 7.75 & -3.77 & 0.28 & S & 4.6 & 2.4 & 8.7 & HESS J1804-216 / 3.7 & 6.1 & 0.0 & 28.3 & 22.3\\ 
G9.8+0.6 & 9.75 & 0.57 & 0.20 & S & - & 39.0 & 17.0 & HESS J1808-204 / 0.9 & 1.7 & 1.2 & - & - \\ 
G11.1$-$1.0 & 11.17 & -1.04 & 0.25 & S & - & 35.4 & 16.0 & HESS J1809-193 / 1.0 & 2.7 & 1.7 & - & - \\ 
G12.2+0.3 & 12.26 & 0.30 & 0.15 & S & - & 23.0 & 18.0 & HESS J1813-178 / 0.6 & 2.3 & 2.2 & - & - \\ 
G13.3$-$1.3 & 13.32 & -1.30 & 0.68 & S? & 3.0 & 3.2 & 16.3 & HESS J1813-178 / 1.4 & 42.6 & 3.6 & 83.8 & 65.9\\ 
G14.1$-$0.1 & 14.19 & 0.11 & 0.15 & S & - & 7.3 & 18.5 & HESS J1818-154 / 1.2 & 3.3 & 1.6 & - & - \\ 
G14.3+0.1 & 14.30 & 0.14 & 0.14 & S & - & 7.5 & 18.9 & HESS J1818-154 / 1.1 & 3.6 & 2.0 & - & - \\ 
G15.1$-$1.6 & 15.11 & -1.61 & 0.35 & S & - & 31.9 & 19.0 & HESS J1818-154 / 1.8 & 1.1 & -2.8 & - & - \\ 
G15.9+0.2 & 15.88 & 0.20 & 0.16 & S? & 8.5 & 43.2 & 20.8 & HESS J1818-154 / 0.5 & 2.6 & 3.7 & 40.5 & 31.8\\ 
G16.0$-$0.5 & 16.05 & -0.55 & 0.23 & S & - & 40.9 & 20.4 & HESS J1818-154 / 1.0 & 2.9 & 2.7 & - & - \\ 
G18.8+0.3 & 18.80 & 0.35 & 0.24 & S* & 10.95 & 14.0 & 23.4 & HESS J1826-130 / 0.8 & 7.0 & 3.3 & 183.1 & 143.9\\ 
G18.9$-$1.1 & 18.95 & -1.18 & 0.38 & C(p) & 2.0 & 22.7 & 23.1 & HESS J1826-130 / 0.9 & 6.4 & 1.6 & 5.6 & 4.4\\ 
G19.1+0.2 & 19.15 & 0.27 & 0.33 & S & - & 6.0 & 23.9 & HESS J1826-130 / 0.9 & 9.8 & 2.0 & - & - \\ 
G20.4+0.1 & 20.47 & 0.16 & 0.17 & S & - & 14.4 & 18.2 & HESS J1828-099 / 1.0 & 3.3 & 2.1 & - & - \\ 
G21.0$-$0.4 & 21.04 & -0.47 & 0.17 & S & - & 18.4 & 20.4 & HESS J1833-105 / 0.6 & 2.1 & 0.8 & - & - \\ 
G25.1$-$2.3 & 25.10 & -2.25 & 0.77 & S & - & 1.8 & 27.9 & HESS J1837-069 / 2.2 & 36.5 & 0.1 & - & - \\ 
G30.7+1.0 & 30.72 & 0.95 & 0.30 & S? & - & 19.9 & 27.3 & HESS J1848-018 / 1.2 & 3.2 & 0.5 & - & - \\ 
G34.7$-$0.4 & 34.67 & -0.39 & 0.39 & C(t\&p)* & 2.65 & 23.6 & 30.7 & HESS J1858+020 / 0.9 & 11.2 & 6.2 & 17.2 & 13.5\\ 
G36.6$-$0.7 & 36.59 & -0.69 & 0.31 & S? & - & 11.1 & 34.7 & HESS J1857+026 / 0.8 & 5.9 & 1.3 & - & - \\ 
G36.6+2.6 & 36.58 & 2.60 & 0.24 & S & - & 0.2 & 37.3 & HESS J1857+026 / 2.7 & 15.4 & -1.5 & - & - \\ 
G38.7$-$1.3 & 38.74 & -1.41 & 0.37 & C(t) & 4.0 & 18.5 & 37.5 & HESS J1908+063 / 1.9 & 2.7 & -1.0 & 9.5 & 7.5\\ 
G39.2$-$0.3 & 39.24 & -0.32 & 0.17 & C(p)* & 6.2 & 22.9 & 38.1 & HESS J1908+063 / 1.4 & 2.2 & 1.3 & 18.8 & 14.8\\ 
G42.8+0.6 & 42.82 & 0.64 & 0.30 & S* & - & 27.8 & 41.0 & HESS J1911+090 / 0.9 & 7.2 & 4.9 & - & - \\ 
G43.9+1.6 & 43.91 & 1.61 & 0.60 & S? & - & 17.3 & 40.1 & HESS J1912+101 / 1.8 & 8.1 & 0.4 & - & - \\ 
G46.8$-$0.3 & 46.77 & -0.30 & 0.24 & S & 6.45 & 15.7 & 38.9 & HESS J1923+141 / 2.3 & 5.2 & 2.5 & 47.0 & 37.0\\ 
G53.6$-$2.2 & 53.63 & -2.26 & 0.38 & C(t) & 4.5 & 3.8 & 47.7 & HESS J1930+188 / 2.6 & 5.3 & -2.4 & 23.6 & 18.5\\ 
G54.4$-$0.3 & 54.47 & -0.29 & 0.43 & S* & 3.4 & 10.5 & 47.4 & HESS J1930+188 / 0.7 & 3.3 & -1.7 & 8.4 & 6.6\\ 
G55.0+0.3 & 55.11 & 0.42 & 0.27 & S & 14.0 & 13.9 & 47.6 & HESS J1930+188 / 1.1 & 5.7 & 1.8 & 244.7 & 192.3\\ 
G57.2+0.8 & 57.30 & 0.83 & 0.20 & S? & - & 8.9 & 49.3 & HESS J1943+213 / 2.2 & 5.5 & 1.7 & - & - \\ 
G59.5+0.1 & 59.58 & 0.12 & 0.23 & S & - & 8.9 & 50.1 & HESS J1943+213 / 2.3 & 4.7 & 0.5 & - & - \\ 
G64.5+0.9 & 64.52 & 0.95 & 0.17 & S & 11.0 & 8.0 & 53.5 & HESS J1943+213 / 7.1 & 4.6 & 0.3 & 120.7 & 94.8\\ 
G65.1+0.6 & 65.27 & 0.30 & 0.85 & S & 9.3 & 5.9 & 53.4 & HESS J1943+213 / 7.7 & 19.4 & -0.7 & 366.4 & 287.9\\ 
G67.7+1.8 & 67.74 & 1.82 & 0.23 & S & 12.0 & 1.5 & 55.2 & HESS J1943+213 / 10.4 & 15.3 & 0.1 & 482.7 & 379.3\\ 
G69.0+2.7 & 68.84 & 2.78 & 0.77 & C(p) & 1.6 & 0.5 & 56.3 & HESS J1943+213 / 11.8 & 80.9 & -1.1 & 45.3 & 35.6\\ 
G69.7+1.0 & 69.69 & 1.00 & 0.23 & S & 2.5 & 0.9 & 57.0 & HESS J1943+213 / 12.1 & 12.8 & -1.2 & 17.4 & 13.7\\ 
G272.2$-$3.2 & 272.22 & -3.18 & 0.23 & C(t) & 3.75 & 1.9 & 34.8 & HESS J0852-463 / 6.2 & 5.6 & -1.4 & 17.2 & 13.5\\ 
G279.0+1.1 & 278.63 & 1.22 & 0.89 & S & - & 2.7 & 30.8 & HESS J1023-575 / 5.8 & 36.4 & 1.0 & - & - \\ 
G286.5$-$1.2 & 286.57 & -1.21 & 0.32 & S? & - & 25.4 & 39.5 & HESS J1026-582 / 1.9 & 7.7 & 3.4 & - & - \\ 
G289.7$-$0.3 & 289.69 & -0.29 & 0.25 & S & - & 11.7 & 39.3 & HESS J1119-614 / 2.5 & 2.9 & -0.5 & - & - \\ 
G290.1$-$0.8 & 290.15 & -0.78 & 0.26 & C(t) & 7.25 & 11.1 & 39.8 & HESS J1119-614 / 2.0 & 5.8 & 1.3 & 66.2 & 52.0\\ 
G291.0$-$0.1 & 291.02 & -0.08 & 0.23 & C(p) & 6.5 & 20.6 & 39.5 & HESS J1119-614 / 1.2 & 4.9 & 2.8 & 44.9 & 35.3\\ 
G292.0+1.8 & 292.03 & 1.75 & 0.20 & C(p) & 6.0 & 17.9 & 38.5 & HESS J1119-614 / 2.3 & 3.2 & 1.1 & 25.3 & 19.9\\ 
G293.8+0.6 & 293.77 & 0.60 & 0.27 & C(p)? & - & 22.4 & 39.9 & HESS J1119-614 / 2.0 & 6.7 & 3.5 & - & - \\ 
G294.1$-$0.0 & 294.12 & -0.06 & 0.43 & S & - & 16.3 & 40.8 & HESS J1119-614 / 2.0 & 5.2 & -0.1 & - & - \\ 
G296.1$-$0.5 & 296.05 & -0.50 & 0.41 & S & 2.0 & 11.3 & 42.8 & HESS J1119-614 / 3.9 & 3.6 & -1.5 & 3.2 & 2.5\\ 
G296.7$-$0.9 & 296.66 & -0.92 & 0.19 & S & 10.0 & 9.9 & 43.1 & HESS J1119-614 / 4.5 & 3.8 & 0.5 & 83.2 & 65.4\\ 
G296.8$-$0.3 & 296.88 & -0.34 & 0.27 & S & 9.6 & 15.6 & 42.2 & HESS J1119-614 / 4.8 & 5.7 & 1.7 & 115.8 & 91.0\\ 
G298.6$-$0.0 & 298.61 & -0.06 & 0.20 & S & - & 14.5 & 41.9 & HESS J1303-631 / 5.6 & 2.9 & 0.7 & - & - \\ 
G299.2$-$2.9 & 299.18 & -2.89 & 0.25 & S & 5.0 & 0.2 & 47.6 & HESS J1302-638 / 5.3 & 71.8 & 0.2 & 392.7 & 308.5\\ 
G299.6$-$0.5 & 299.59 & -0.47 & 0.21 & S & - & 17.3 & 42.3 & HESS J1302-638 / 4.6 & 1.8 & -0.7 & - & - \\ 
G301.4$-$1.0 & 301.44 & -0.98 & 0.41 & S & - & 10.3 & 42.3 & HESS J1302-638 / 2.7 & 4.6 & -0.7 & - & - \\ 
G302.3+0.7 & 302.29 & 0.73 & 0.24 & S & - & 11.8 & 40.8 & HESS J1303-631 / 2.2 & 5.3 & 1.2 & - & - \\ 
G306.3$-$0.9 & 306.31 & -0.89 & 0.12 & S & 11.5 & 24.8 & 43.7 & HESS J1302-638 / 2.1 & 1.1 & -0.1 & 31.2 & 24.5\\ 
G308.1$-$0.7 & 308.13 & -0.66 & 0.21 & S & - & 14.7 & 42.8 & HESS J1356-645 / 2.5 & 2.1 & -0.6 & - & - \\ 
G308.4$-$1.4 & 308.44 & -1.38 & 0.18 & S & 9.9 & 12.6 & 43.1 & HESS J1356-645 / 1.8 & 3.5 & 1.0 & 75.7 & 59.4\\ 
G308.8$-$0.1 & 308.81 & -0.10 & 0.35 & C(p)? & - & 13.5 & 42.3 & HESS J1356-645 / 2.6 & 11.1 & 3.6 & - & - \\ 
G309.2$-$0.6 & 309.16 & -0.70 & 0.23 & S & 4.0 & 16.4 & 42.6 & HESS J1356-645 / 1.9 & 3.8 & 0.9 & 13.2 & 10.3\\ 
G309.8+0.0 & 309.79 & -0.00 & 0.31 & S & - & 12.6 & 41.9 & HESS J1356-645 / 2.5 & 6.2 & 1.3 & - & - \\ 
G310.6$-$1.6 & 310.60 & -1.60 & 0.11 & C(p) & 7.5 & 11.4 & 42.7 & HESS J1356-645 / 1.2 & 2.4 & 1.1 & 29.4 & 23.1\\ 
G310.6$-$0.3 & 310.62 & -0.28 & 0.17 & S & 6.9 & 12.2 & 41.8 & HESS J1356-645 / 2.4 & 3.5 & 1.5 & 36.8 & 28.9\\ 
G310.8$-$0.4 & 310.81 & -0.46 & 0.20 & S & 5.1 & 12.0 & 41.8 & HESS J1356-645 / 2.3 & 2.7 & 0.0 & 15.1 & 11.9\\ 
G311.5$-$0.3 & 311.53 & -0.34 & 0.14 & C(t)? & 12.5 & 10.3 & 41.9 & HESS J1418-609 / 1.8 & 2.0 & 0.0 & 69.1 & 54.3\\ 
G312.5$-$3.0 & 312.49 & -3.00 & 0.27 & S & - & 0.7 & 46.0 & HESS J1356-645 / 2.7 & 31.2 & 0.5 & - & - \\ 
G315.9$-$0.0 & 315.86 & -0.03 & 0.31 & C(p) & - & 19.1 & 39.2 & HESS J1427-608 / 1.4 & 2.5 & -1.1 & - & - \\ 
G316.3$-$0.0 & 316.29 & -0.01 & 0.34 & S & - & 20.4 & 38.8 & HESS J1427-608 / 1.9 & 4.8 & 0.5 & - & - \\ 
G317.3$-$0.2 & 317.31 & -0.24 & 0.19 & S & - & 21.7 & 38.0 & HESS J1457-593 / 1.1 & 4.2 & 3.1 & - & - \\ 
G318.9+0.4 & 318.91 & 0.39 & 0.35 & C(p)? & - & 19.3 & 36.7 & HESS J1503-582 / 0.7 & 6.0 & 1.4 & - & - \\ 
G321.9$-$1.1 & 321.89 & -1.07 & 0.33 & S & - & 26.3 & 37.5 & HESS J1514-591 / 1.6 & 2.0 & -0.8 & - & - \\ 
G321.9$-$0.3 & 321.90 & -0.30 & 0.36 & S & - & 20.1 & 37.1 & HESS J1514-591 / 1.8 & 4.1 & 0.5 & - & - \\ 
G322.1+0.0 & 322.12 & 0.04 & 0.17 & S & 7.5 & 22.2 & 37.2 & HESS J1534-571 / 1.9 & 2.7 & 2.1 & 33.4 & 26.2\\ 
G322.5$-$0.1 & 322.46 & -0.11 & 0.23 & C(p)? & - & 15.3 & 38.2 & HESS J1534-571 / 1.5 & 4.3 & 2.4 & - & - \\ 
G323.5+0.1 & 323.49 & 0.11 & 0.21 & S & - & 21.1 & 37.6 & HESS J1534-571 / 1.1 & 2.7 & 1.1 & - & - \\ 
G326.3$-$1.8 & 326.30 & -1.76 & 0.42 & C(p) & 4.6 & 14.0 & 34.9 & HESS J1554-550 / 1.1 & 7.8 & 1.5 & 36.2 & 28.4\\ 
G327.2$-$0.1 & 327.24 & -0.13 & 0.13 & S & 4.5 & 21.7 & 33.7 & HESS J1554-550 / 1.0 & 0.9 & -0.9 & 3.8 & 3.0\\ 
G327.4+0.4 & 327.25 & 0.49 & 0.28 & C(t) & 5.4 & 17.3 & 33.5 & HESS J1554-550 / 1.6 & 5.1 & 1.9 & 32.8 & 25.8\\ 
G327.4+1.0 & 327.37 & 1.01 & 0.22 & S & - & 15.3 & 33.2 & HESS J1554-550 / 2.1 & 3.7 & 1.1 & - & - \\ 
G329.7+0.4 & 329.72 & 0.41 & 0.43 & S & - & 9.1 & 32.5 & HESS J1614-518 / 2.0 & 10.0 & 1.8 & - & - \\ 
G341.2+0.9 & 341.19 & 0.86 & 0.28 & C(p) & - & 11.3 & 26.4 & HESS J1646-458 / 2.5 & 4.3 & 0.3 & - & - \\ 
G341.9$-$0.3 & 341.86 & -0.32 & 0.16 & S & - & 13.9 & 26.0 & HESS J1708-443 / 2.3 & 3.3 & 1.8 & - & - \\ 
G342.0$-$0.2 & 341.94 & -0.21 & 0.20 & S & - & 8.3 & 23.3 & HESS J1702-420 / 2.3 & 4.7 & 1.6 & - & - \\ 
G342.1+0.9 & 342.10 & 0.89 & 0.18 & S & - & 7.5 & 23.8 & HESS J1702-420 / 2.4 & 4.0 & 1.1 & - & - \\ 
G343.1$-$0.7 & 343.08 & -0.59 & 0.33 & S & - & 14.6 & 23.9 & HESS J1702-420 / 1.2 & 5.7 & 1.3 & - & - \\ 
G350.0$-$2.0 & 349.92 & -2.05 & 0.47 & S & - & 2.6 & 18.4 & HESS J1718-385 / 1.9 & 14.4 & 0.3 & - & - \\ 
G350.1$-$0.3 & 349.72 & 0.25 & 0.13 & S? & 6.75 & 21.0 & 28.1 & HESS J1718-374 / 0.1 & 1.6 & 0.8 & 16.1 & 12.7\\ 
G351.2+0.1 & 351.27 & 0.16 & 0.16 & C(p)? & - & 19.2 & 20.0 & HESS J1718-374 / 1.6 & 1.1 & -1.0 & - & - \\ 
G351.7+0.8 & 351.70 & 0.82 & 0.25 & S & 13.2 & 16.3 & 19.6 & HESS J1729-345 / 1.9 & 2.6 & -0.3 & 97.8 & 76.8\\ 
G351.9$-$0.9 & 351.92 & -0.96 & 0.20 & S & - & 23.0 & 20.4 & HESS J1731-347 / 1.6 & 2.6 & 1.5 & - & - \\ 
G353.9$-$2.0 & 353.94 & -2.09 & 0.21 & S & - & 29.2 & 18.7 & HESS J1731-347 / 1.5 & 0.7 & -2.4 & - & - \\ 
G354.8$-$0.8 & 354.87 & -0.78 & 0.26 & S & - & 34.9 & 18.1 & HESS J1731-347 / 1.3 & 1.5 & -0.3 & - & - \\ 
G355.4+0.7 & 355.40 & 0.73 & 0.31 & S & - & 16.9 & 15.8 & HESS J1729-345 / 2.1 & 4.3 & 1.0 & - & - \\ 
G355.6$-$0.0 & 355.69 & -0.08 & 0.17 & C(t) & 13.0 & 18.6 & 15.8 & HESS J1731-347 / 2.2 & 2.7 & 1.8 & 100.1 & 78.6\\ 
G355.9$-$2.5 & 355.95 & -2.54 & 0.21 & S & - & 2.7 & 17.0 & HESS J1746-308 / 2.9 & 10.0 & 1.8 & - & - \\ 
G356.2+4.5 & 356.22 & 4.46 & 0.31 & S & - & 5.4 & 10.9 & HESS J1741-302 / 4.9 & 9.6 & 1.9 & - & - \\ 
G356.3$-$1.5 & 356.31 & -1.50 & 0.27 & S & - & 4.5 & 14.5 & HESS J1746-308 / 2.2 & 2.3 & -1.9 & - & - \\ 
G356.3$-$0.3 & 356.30 & -0.35 & 0.19 & S & - & 8.9 & 14.8 & HESS J1741-302 / 2.0 & 2.8 & 0.4 & - & - \\ 
G357.7$-$0.1 & 357.69 & -0.12 & 0.17 & C(t)?* & 11.8 & 32.6 & 20.4 & HESS J1741-302 / 0.6 & 2.7 & 3.0 & 83.2 & 65.4\\ 
G357.7+0.3 & 357.67 & 0.35 & 0.30 & S* & 6.4 & 47.3 & 19.2 & HESS J1741-302 / 0.7 & 3.3 & 1.5 & 29.3 & 23.0\\ 
G358.0+3.8 & 357.96 & 3.80 & 0.42 & S & - & 2.9 & 11.5 & HESS J1741-302 / 3.8 & 20.6 & 2.6 & - & - \\ 
G358.1+0.1 & 358.12 & 1.04 & 0.27 & S & - & 51.1 & 20.7 & HESS J1741-302 / 1.0 & 2.8 & 1.6 & - & - \\ 
G359.1+0.9 & 359.10 & 0.99 & 0.20 & S & - & 78.0 & 20.7 & HESS J1741-302 / 1.2 & 1.7 & 2.6 & - & - \\ 

\hline

 \caption*{Table 2: Results overview. The SNR morphology type is indicated by S for shell-type and C for mixed morphology SNRs. The type of the latter is indicated as thermal (t), plerionic (p), or both (t\&p). If molecular cloud interaction is probable or certain (information taken from SNRcat), this is indicated by an asterisk. For sources with an observational live time $\la 1$h, instrumental effects can impact the result and should be used with care. R$_{\mathrm{ON}}$ are the radii of the circular analysis regions. Distance assumptions correspond to the arithmetic mean of the density range limits reported in SNRcat.}
\label{Table:Analysis}
\end{longtable}

}
\end{center}

\begin{table}
\begin{threeparttable}
\caption{Literature estimates on source ages, distances, and ambient densities. Some of the ambient density values are not direct
measurements but are derived from fits to the X-ray spectra. In some cases correlations with high density gas clumps are observed or inferred. The corresponding densities inside the clumps are denoted by an asterisk.}
\begin{tabular}{l|ccc|r}
\hline
Source & Distance & Age & Density & References\\
 & (kpc) & (kyrs) & ($\mathrm{cm}^{-3}$)  &  \\
\hline
Cassiopeia A & 3.3 - 3.7& 0.316 - 0.352 & 0.6 - 1.2 / $\sim$10* &  \cite{CasA-DeLaney}, \cite{CasA-Lee}, \cite{Hwang}\\ 
Tycho's SNR & 1.7 - 5& 0.441 & 0.3  &  \cite{Tycho-Slane}, \cite{Tycho-Voelk}\\ 
RX J0852.0$-$4622 & 0.5 - 1& 2.4 - 5.1 & 0.022 / $\sim$10* & \cite{VelaJr-Allen}, \cite{Moriguchi}\\ 
RCW 86 & 2.3 - 3.2& 1.83 & 0.3  &  \cite{RCW86-Broersen}\\ 
SN 1006 & 1.6 - 2.2& 1.01 & 0.045 &  \cite{SN1006-Winkler}\\ 
RX J1713.7$-$3946 & 1& 1.6 & <0.02 / $\sim$130* &  \cite{RXJ1713-Cassam}, \cite{RXJ1713_Acero}, \cite{Fukui}\\ 
H.E.S.S. J1731$-$347 & 2.4 - 4& 2.5 & <0.01 / $\sim$40-100* &  \cite{J1731}, \cite{Fukuda}\\ 
Kepler's SNR & 2.9 - 4.9& 0.412 & 0.4 - 5 &  \cite{Kepler-Vink}\\ 
G1.9+0.3 & 8.5& 0.15 - 0.22 & 0.022 - 0.04 &  \cite{G1-9-Reynolds}, \cite{G1-9-Carlton}, \cite{G1-9-Zoglauer}\\ 
G15.9+0.2 & 8.5& 1 - 3 & 0.7 &  \cite{G15-9-Reynolds}\\ 
G38.7$-$1.3 & 4& 14 - 15 & 0.025 - 0.034 &  \cite{G38-7-Huang}\\ 
G53.6$-$2.2 & 2.3 - 6.7& 15 & 0.79 - 1.36 &  \cite{G53-6-Long}, \cite{G53-6-Broersen}\\ 
G67.7+1.8 & 7 - 17& 5 - 13 & 0.06 - 0.1 &  \cite{G67-7-Hui}\\ 
G69.7+1.0 & 2.5& 34 - 40 & 0.06 &  \cite{G69-7-Yoshita}\\ 
G272.2$-$3.2 & 2.5 - 5& 2 - 5.2 & 0.1 &  \cite{G272-2-Sanchez}\\ 
G290.1$-$0.8 & 3.5 - 11& 10 - 20 & 9.2 &  \cite{G290-1-Auchettl}\\ 
G296.1$-$0.5 & 2& 2.8 & 0.22 &  \cite{G296-1-Castro}\\ 
G296.7$-$0.9 & 9.1 - 10.9& 5.8 - 7.6 & 0.63 - 0.91 &  \cite{G296-7-Prinz}\\ 
G296.8$-$0.3 & 9& 10 & 0.2 &  \cite{G296-8-Sanchez}\\ 
G299.2$-$2.9 & 5& 8.7 & 0.3 &  \cite{G299-2-Busser}\\ 
G306.3$-$0.9 & 8& 1.3 - 4.6 & 1 - 49.1 &  \cite{G306-3-Reynolds}\\ 
G308.4$-$1.4 & 9.1 - 10.7& 5 - 7.5 & 0.86 - 1.06 &  \cite{G308-4-Hui}, \cite{G308-4-Hui-Prinz}\\ 
G309.2$-$0.6 & 2 - 6& 0.7 - 4 & 0.01 - 0.05 &  \cite{G309-2-Rakowski}\\ 
G311.5$-$0.3 & 12.5& 25 - 42 & 0.17 &  \cite{G311-5-Pannuti}\\ 
G327.4+0.4 & 4.3 - 6.5& 7 - 90 & 0.2 - 0.4 &  \cite{G327-4-Chen}\\ 
G330.2+1.0 & >5& 1 - 3 & 0.1 &  \cite{330-2-Park}\\ 
G337.2$-$0.7 & 2 - 9.3& 0.75 - 3.5 & 0.6 &  \cite{G337-2-Rakowski}\\ 
G350.1$-$0.3 & 4.5 - 9& 0.6 - 1.2 & 0.3 &  \cite{G350-1-Lovchinsky}\\ 
G351.7+0.8 & 12.7 - 13.7& <68 & <0.4 &  \cite{351-7-Tian}\\ 
G355.6$-$0.0 & 13& 20 & 0.85 &  \cite{G355-6-Minami}\\ 

\hline
\end{tabular}
\label{Table:SNRparams}
\end{threeparttable}
\end{table}

\begin{table}
\begin{threeparttable}
\caption{Spectral parameters of the shell-type SNRs detected in both the radio 
and VHE bands. Except for the value for RX J1713-3946, all radio flux density 
values are taken from Green's catalogue \cite{GreenCatalogue}. Also given are
the references to the models applied in Sec.~\ref{sec-data}.}
\begin{tabular}{l|c|cr|cccr}
\hline
SNR & Radio flux density & Integrated flux $F_{1-10\mathrm{TeV}}$ & Flux references & Model references\\
  & (@1GHz, Jy) & ($10^{-12} \mathrm{cm}^{-2} \mathrm{s}^{-1}$) & & \\
\hline
\hline
Cassiopeia A & 2720 & 0.58 $\pm$ 0.12\tnote{a}& \cite{CasA_new} &  \cite{CasA_new}\\
Tycho's SNR &  56 &  0.11 $\pm$ 0.04& \cite{Tycho_Park} & \cite{Tycho-Slane}\\
RX J0852.0$-$4622 &  50 &  22.8 $\pm$ 6.1 & \cite{hess_velajr_paper3} & \cite{hess_velajr_paper3}\\
RCW 86 &  49 &  1.82 $\pm$ 0.94 & \cite{RCW86-New} & \cite{RCW86-New}\\
SN 1006 & 19 &  $\sim$0.37 $\pm$ 0.08\tnote{b} & \cite{SN1006} & \cite{SN1006}\\
RX J1713.7$-$3946 &  $\sim$30\tnote{c}  & 16.4 $\pm$ 5.4 &\cite{RXJ1713New} & \cite{RXJ1713New}\\
H.E.S.S. J1731$-$347 &  2.5 & 3.37 $\pm$ 0.82  & \cite{J1731} & \cite{J1731}\\
\hline
\end{tabular}
\begin{tablenotes}
\item[a] assuming a systematic flux error of 20$\%$
\item[b] sum of north-east and south-west regions
\item[c] value from \citep{RXJ1713_Acero} extrapolated to 1GHz
\end{tablenotes}
\label{Table:SNRspecTable}
\end{threeparttable}
\end{table}

\end{landscape}

\begin{figure*}
  \begin{center}
  \resizebox{1\hsize}{!}{\includegraphics[clip=]{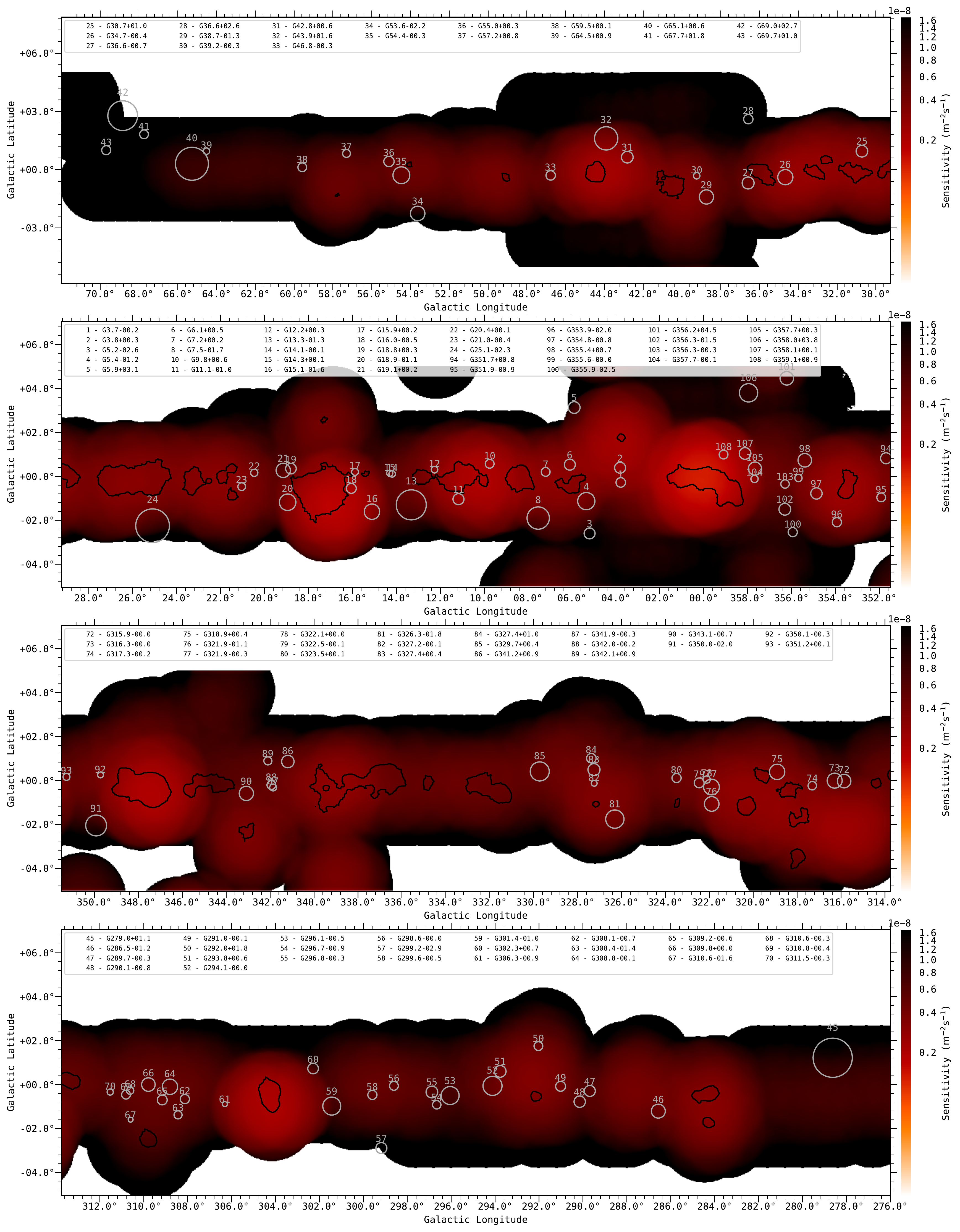}}
  \caption{
  {\fontsize{8.3}{8.3}\selectfont
 HGPS Sensitivity map (for a correlation radius of 0.2\dgr) overlaid with the analysis regions
of our source sample (grey) and the de-selection regions (black); see also Sect.~\ref{sec-selec}.
}}
  \label{Figure:DS_map}
  \end{center}
\end{figure*}
\end{document}